\documentclass[iop]{emulateapj}
\usepackage{graphicx}

\shorttitle{HTTP catalog}
\shortauthors{Sabbi et al.}

\begin{document}
\tracingall


\title{Hubble Tarantula Treasury Project. III. Photometric Catalog and Resulting Constraints on the Progression of Star Formation in the 30~Doradus Region$^{**}$}


\author{E. Sabbi\altaffilmark{1}, 
D.J. Lennon\altaffilmark{2}, 
J. Anderson\altaffilmark{1}, 
M. Cignoni\altaffilmark{1},
R.P. van der Marel\altaffilmark{1}, 
D. Zaritsky\altaffilmark{3},
G. de Marchi\altaffilmark{4},
N. Panagia\altaffilmark{1,5,6},
D.A. Gouliermis\altaffilmark{7,8},
E.K. Grebel\altaffilmark{9},
J.S. Gallager III\altaffilmark{10},
L.J. Smith\altaffilmark{11},
H. Sana\altaffilmark{1},
A. Aloisi\altaffilmark{1},
M. Tosi\altaffilmark{12},
C.J. Evans\altaffilmark{13},
H. Arab\altaffilmark{1},
M. Boyer\altaffilmark{14,15},
S.E. de Mink\altaffilmark{16},
K. Gordon\altaffilmark{1},
A.M. Koekemoer\altaffilmark{1},
S.S. Larsen\altaffilmark{17},
J.E. Ryon\altaffilmark{10},
P. Zeidler\altaffilmark{9}
}

\email{sabbi@stsci.edu}

\altaffiltext{1}{Space Telescope Science Institute, 3700 San Martin Drive, Baltimore, MD, 21218, USA }
\altaffiltext{2}{ESA - European Space Astronomy Center, Apdo. de Correo 78, 28691 Associate Villanueva de la Ca\~{n}ada, Madrid, Spain}
\altaffiltext{3}{Steward Observatory, University of Arizona, 933 North Cherry Avenue, Tucson, AZ 85721, USA}
\altaffiltext{4}{Space Science Department, European Space Agency, Keplerlaan 1, 2200 AG Noordwijk, The Netherlands}
\altaffiltext{5}{Istituto Nazionale di Astrofisica, Osservatorio Astrofisico di Catania, Via Santa Sofia 78, 95123 Catania, Italy}
\altaffiltext{6}{Supernova Limited, OYV 131, Northsound Road, Virgin Gorda, British Virgin Islands}
\altaffiltext{7}{Zentrum f\"ur Astronomie der Universit\"at Heidelberg, Institut f\"ur Theoretische Astrophysik, Albert-Ueberle-Str. 2, 69120 Heidelberg, Germany} 
\altaffiltext{8}{Max-Planck-Institut f\"{u}r Astronomie, K\"{o}nigstuhl 17, 69117 Heidelberg, Germany}
\altaffiltext{9}{Astronomisches Rechen-Institut, Zentrum f\"ur Astronomie der Universit\"at Heidelberg, M\"onchhofstr.\ 12--14,
69120 Heidelberg, Germany}
\altaffiltext{10}{Department of Astronomy, University of Wisconsin, 475 North Charter Street, Madison, WI 53706, USA}
\altaffiltext{11}{ESA/STScI, 3700 San Martin Drive, Baltimore, MD, 21218, USA}
\altaffiltext{12}{Istituto Nazionale di Astrofisica, Osservatorio Astronomico di Bologna, Via Ranzani 1, I-40127 Bologna, Italy}
\altaffiltext{13}{UK Astronomy Technology Center, Royal Observatory Edinburgh, Blackford Hill, Edinburgh, EH9 3HJ, UK} 
\altaffiltext{14}{Observational Cosmology Lab, Code 665, NASA, Goddard Space Flight Center, Greenbelt, MD, 20771, USA}
\altaffiltext{15}{Oak Ridge Associate Universities (ORAU), Oak Ridge, TN 37831, USA}
\altaffiltext{16}{Astronomical Institute ''Anton Pannekoek'', University of Amsterdam, P.O. Box 94249, NL-1090 GE Amsterdam, the Netherlands}
\altaffiltext{17}{Department of Astrophysics / IMAPP, Radboud University Nijmegen, PO Box 9010, 6500 GL Nijmegen, The Netherlands}

\altaffiltext{**}{Based on observations with the NASA/ESA Hubble Space Telescope, obtained at the Space Telescope Science Institute, which is operated by AURA Inc., under NASA contract NAS 5-26555}


\begin{abstract}

We present and describe the astro-photometric catalog of more than 800,000 sources found in the Hubble Tarantula Treasury Project (HTTP). HTTP is a Hubble Space Telescope (HST) Treasury program designed to image the entire 30 Doradus region down to the sub-solar ($\sim 0.5\, {\rm M}_\odot$) mass regime using the Wide Field Camera 3 (WFC3) and the Advanced Camera for Surveys (ACS). We observed 30 Doradus in the near ultraviolet (F275W, F336W), optical (F555W, F658N, F775W), and near infrared (F110W, F160W) wavelengths. The stellar photometry was measured using  point-spread function (PSF) fitting across all the bands simultaneously. The relative astrometric accuracy of the catalog is 0.4 mas. The astro-photometric catalog, results from artificial star experiments and the mosaics for all the filters are available for download. Color-magnitude diagrams are presented showing the spatial distributions and ages of stars within 30 Dor as well as in the surrounding fields. HTTP provides the first rich and statistically significant sample of intermediate and low mass pre-main sequence candidates and allows us to trace how star formation has been developing through the region. The depth and high spatial resolution of our analysis highlight the dual role of stellar feedback in quenching and triggering star formation on the giant H{\sc ii} region scale.  Our results are consistent with stellar sub-clustering in a partially filled gaseous nebula that is offset towards our side of the Large Magellanic Cloud.
\end{abstract}





\section{Introduction}
\label{Intro}

30 Doradus (a.k.a. the Tarantula Nebula) in the Large Magellanic Cloud (LMC) is the most powerful source of H$\alpha$ emission in the Local Group \citep[$f({H\alpha})\sim 1.3\times 10^{-8}\, {\rm erg\, cm^{-2}\, s^{-1}}$,][]{kennicutt86}. Covering an area of $\sim 40,000\, {\rm pc^2}$, 30 Doradus is the closest extragalactic giant H{\sc ii} region and is comparable in size to the unresolved luminous H{\sc ii} complexes observed in distant galaxies \citep{oey03, hunt09}. In terms of size ($\sim 200\, {\rm pc}$ in diameter) and stellar density \citep[between $\rho_0\ge 1.5\times 10^4 - 10^7\, {\rm M_\odot/pc^3,}$][]{selman13}, the Nebula is often equate to regions of extreme star 
formation such as the starburst knots observed in interacting galaxies in the Local Universe and young galaxies at high redshift \citep[$z>5$,][]{meurer97,shapley03,heckman04}.

Radcliffe 136 (R136), the core of 30 Doradus' ionizing cluster NGC~2070,  contains the most massive stars ($\sim 300 M_\odot$) known so far \citep{crowther10,bestenlehner11} and is considered a testbed for understanding the early evolution of massive stars. Because of its mass \citep[$\sim 10^5\, {\rm M_\odot}$,][]{selman99, bosch01, andersen09, cignoni15} NGC~2070 has often been compared to young globular clusters. 

By virtue of its location in the LMC \citep[$\sim 50\, {\rm kpc}$;][]{panagia91, pietr13}, the low inclination angle \citep[$\sim 30\degr$;][]{nikolaev04}, and the low foreground reddening, 30 Doradus is an ideal target to study the process of massive star formation in detail. It is thus not a surprise that the entire  region has been surveyed by all the Great Observatories (Spitzer Space Telescope -- \citet{meixner06}; Herschel Space Observatory -- \citet{meixner10}, and Chandra X-ray Observatory -- \citet{townsley06} and the on-going Chandra Survey ``Tarantula-Revealed by X-rays --T-ReX--, PI L. Townsley), and extensively observed at all the wavelengths accessible from the ground, including the VLT-Flames Tarantula Survey \citep{evans11} and the VMC survey in the NIR \citep{cioni11}. 

\begin{figure*}
       \centering
       \includegraphics[trim=2cm 3cm 0cm 2cm, clip,width=18cm]{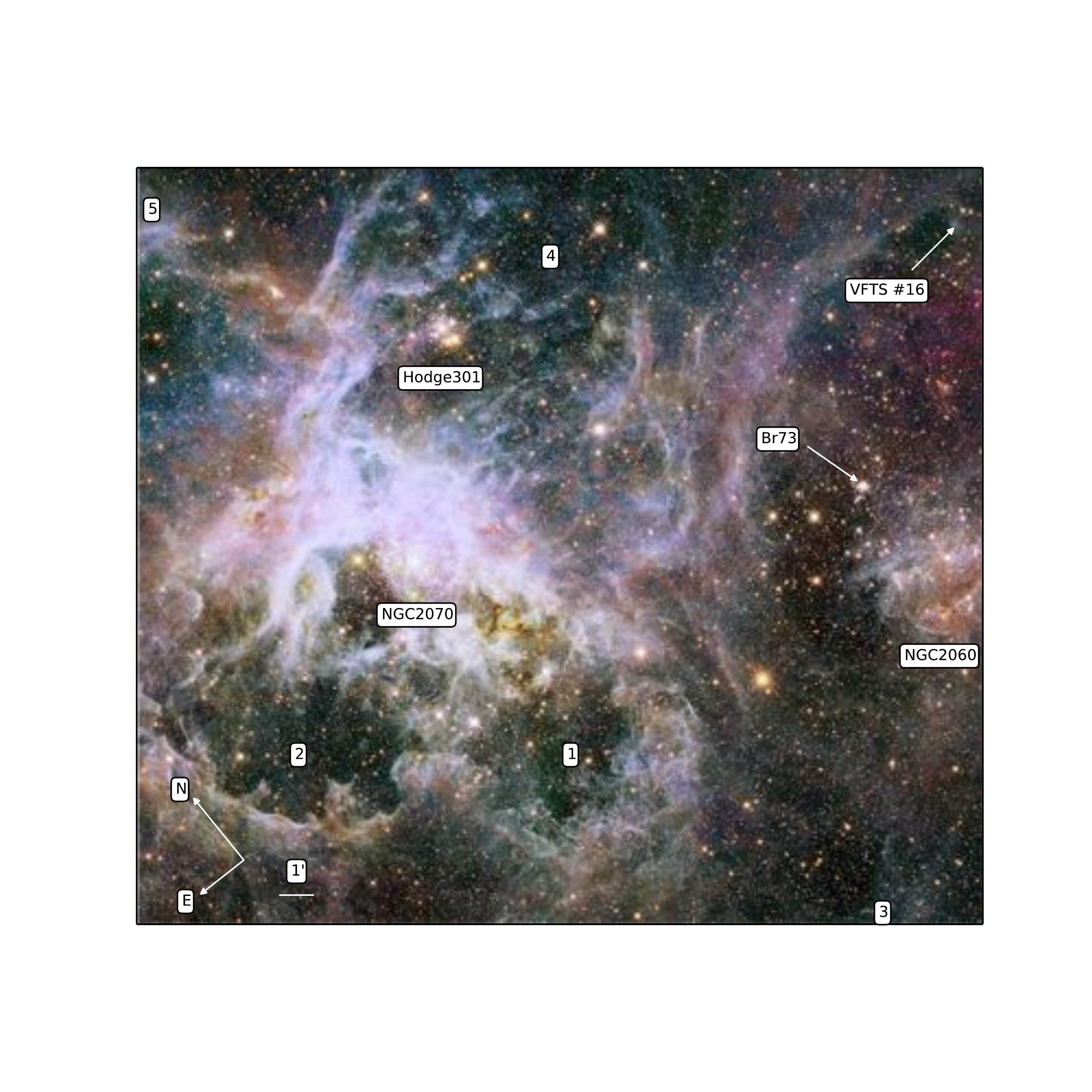}
       \caption{Color composite image of 30 Doradus in the light of F110W 
       (in blue) and F160W (in red). The position of the four larger clusters and associations 
       is highlighted, as well as the runaway candidate VFTS~\#16 and the location of the 5 X-ray hot super-bubbles, identified
       by \citet{wang91}. }
       \label{V-band}
\end{figure*}

High-mass stars in 30 Doradus are so bright that they have been studied for centuries \citep[i.e.][]{caille61, herschel47}. The high angular resolution and sensitivity of the {\it Hubble} Space Telescope ({\it HST}) has allowed in-depth studies of dense clusters and rich associations, such as  R136, Hodge~301, NGC~2060, and Br73 \citep{hunter95, hunter96, walborn99, walborn02, selman99, brandner01, andersen09, demarchi11, sabbi12, selman13, grebel00, mignani05}, whose position relative to the Nebula is shown in Figure~\ref{V-band}.

The Hubble Tarantula Treasury Project (HTTP; Cycle 20, HST GO-12939, PI: E. Sabbi), is a photometric survey of the entire nebula from the near UV to the near IR at the high sensitivity and angular resolution of {\it HST}. HTTP covers approximately $14\arcmin \times 12\arcmin$, that at the distance of the LMC corresponds to $\sim210\times 180\, {\rm pc}$. Preliminary results from the analysis of half of the observations acquired in the IR filters, as well as a discussion of the main goals of the survey have been presented in \citet{sabbi13}.
 
In this paper we present the photometric measurements in all eight filters covered by HTTP. The observations (including filter choice, exposure times and orientation of the various pointings), and the construction of the reference frame are discussed in section 2. In section 3 we introduce the photometric package KS2, used to analyze the data. Details of the photometric catalog and artificial star tests are given in section 4. Color-Magnitude Diagrams (CMDs) and reddening distributions are presented in section 5, while in section 6 we discuss the ages and spatial distribution of the stellar populations found in the Tarantula Nebula. A summary and conclusions are presented in section 7.

\begin{figure}
       \centering
       \includegraphics[trim=0cm 5cm 0 3cm, clip,width=8cm]{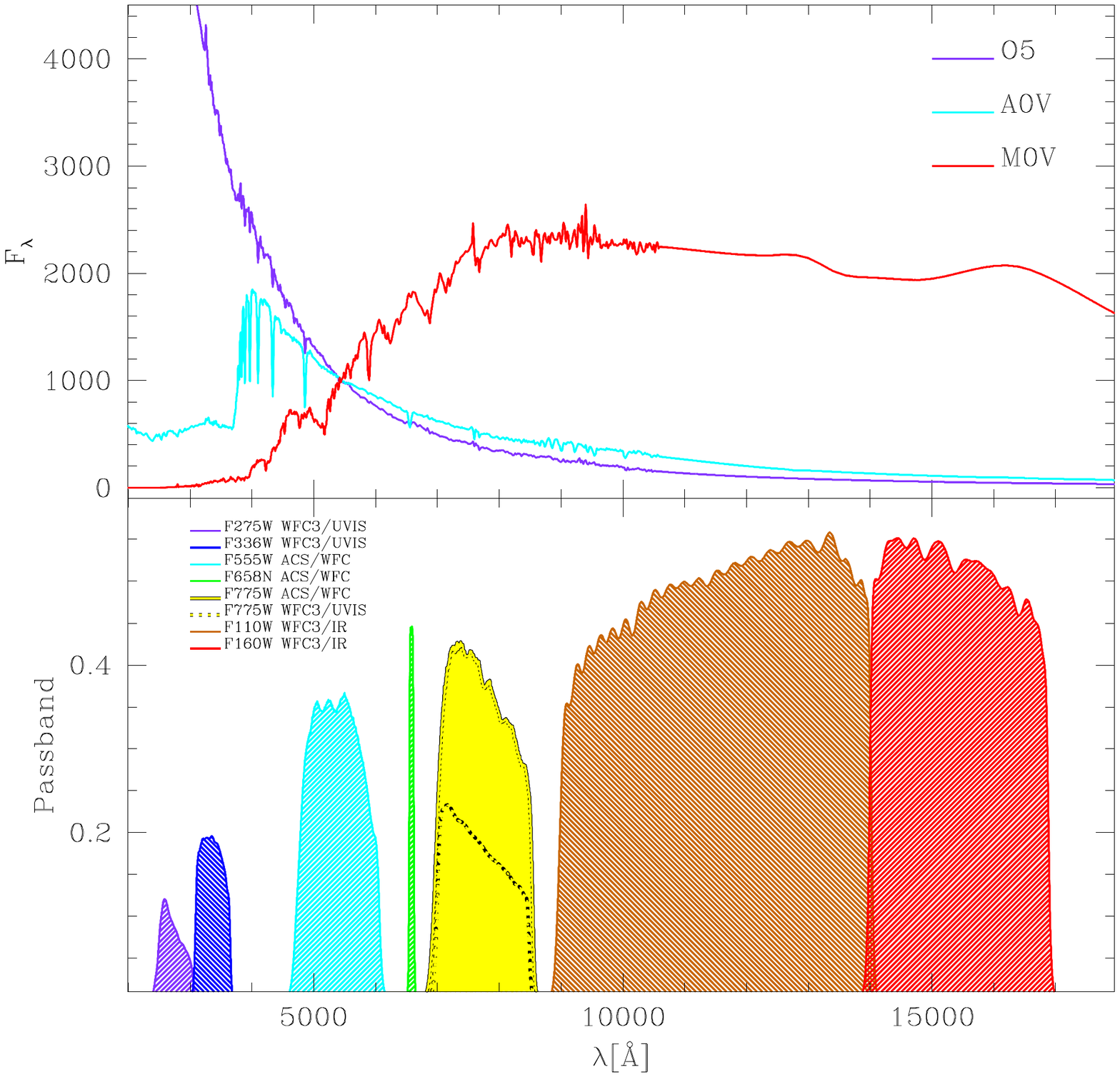}
       \caption{The pass-bands of the filters used in HTTP as a function of wavelength is shown in the {\it Lower Panel}. The {\it Upper Panel} shows Kurucz 1993 models for an O5 (shown in purple), an A0V (in cyan) and a M0V (in red) stars in the range of wavelengths covered by HTTP.}
       \label{f:pass_bands}
\end{figure}

\section{OBSERVATIONS}

\subsection{The data}
\label{data}

HTTP was awarded 60 orbits of HST time in Cycle 20 to survey the entire Tarantula Nebula, using both the UVIS and the IR channels of the Wide Field Camera 3 (WFC3), and, in parallel, the Wide Filed Channel (WFC) of the Advanced Camera for Surveys (ACS). Each of the UVIS orbits consists of one short ($14\, {\rm s}$) and two long ($697\, {\rm s}$ each) exposures in the filter F336W followed by two long exposures ($697$ and $467\, {\rm s}$ respectively) in the F275W filter. Each couple of UVIS long exposures were obtained using a two point dither pattern to enable the rejection of cosmic-rays (CRs) and cover the chip gap. 

HTTP was built on top of the first epoch of the {\it HST} monochromatic survey of 30 Doradus (PI: Lennon, GO-12499) designed to measure proper motions of runaway candidates. That survey used the filter F775W (${\sim}$ Sloan Digital Sky Survey {\it i}-band)  and covered a projected area of $\sim14\arcmin \times 12\arcmin$. The program consisted of a 15 pointings mosaic using WFC3 and ACS in parallel. The orientation angle of the mosaic ($\sim 140\degr$ to the North) was chosen to include the very massive runaway candidate VFTS\#16 \citep[][see Figure~\ref{V-band}]{evans10}. This dataset is described in \citet{sabbi13}. Preliminary proper-motion results, obtained by combining the first epoch dataset to archival WFPC2 images are presented in \citet{platais15}.

The spectral coverage of HTTP+Lennon's programs is shown in Figure~\ref{f:pass_bands}, and the observations are summarized in Table~\ref{log}, including data set names, pointing coordinates, date and time when observations were collected, exposure time in seconds, number of pointings, instrument used, filter, and central wavelength of the filter. Once combined with Lennon's data, HTTP consists of eight filters\footnote{Although they share the same name, the ACS and WFC3 F775W filters differ in terms color-response,  PSF, geometric distortion, and pixel scale. Therefore the two filters have been analyzed independently.} from the near UV (NUV) to the near IR (NIR). 

\begin{figure}
      \centering
      \includegraphics[trim=0 0cm 0 0cm, clip, width=8cm]{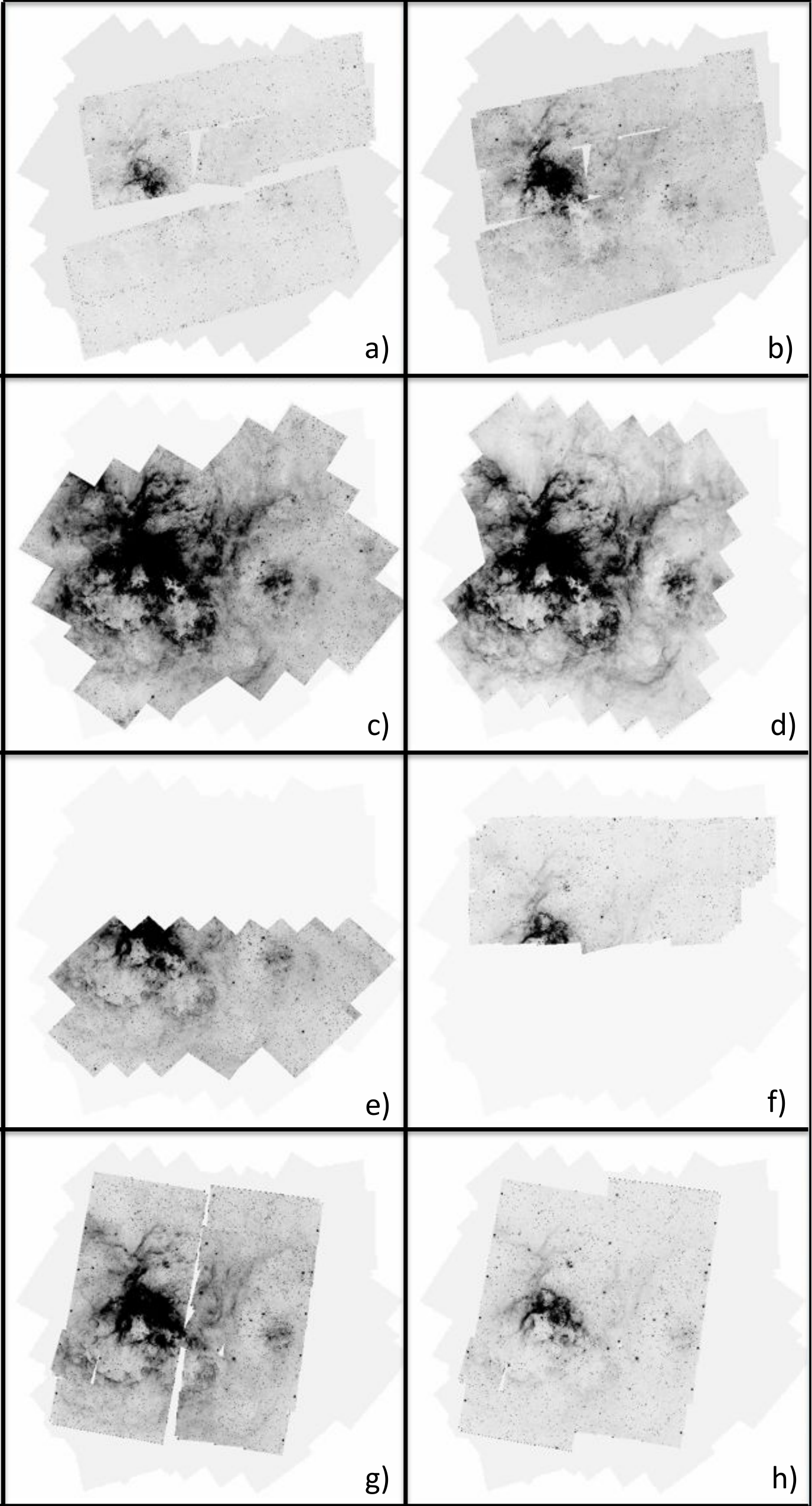}
      \caption{Eight mosaics of the region covered by the HTTP survey in the filters {\it a)} F275W; {\it b)} F336W; {\it c)} F555W; {\it d)} F658N; {\it e)} F775$_{ACS}$; {\it f)} F775W$_{UVIS}$; {\it g)} F110W; and {\it h)} F160W. The faint light-gray area around the mosaics shows total area covered by the survey, when all the filters are considered.}
      \label{mosaics}
\end{figure}

\begin{figure}
      \centering
     \includegraphics[trim=0 0cm 0 0cm, clip, width=8cm]{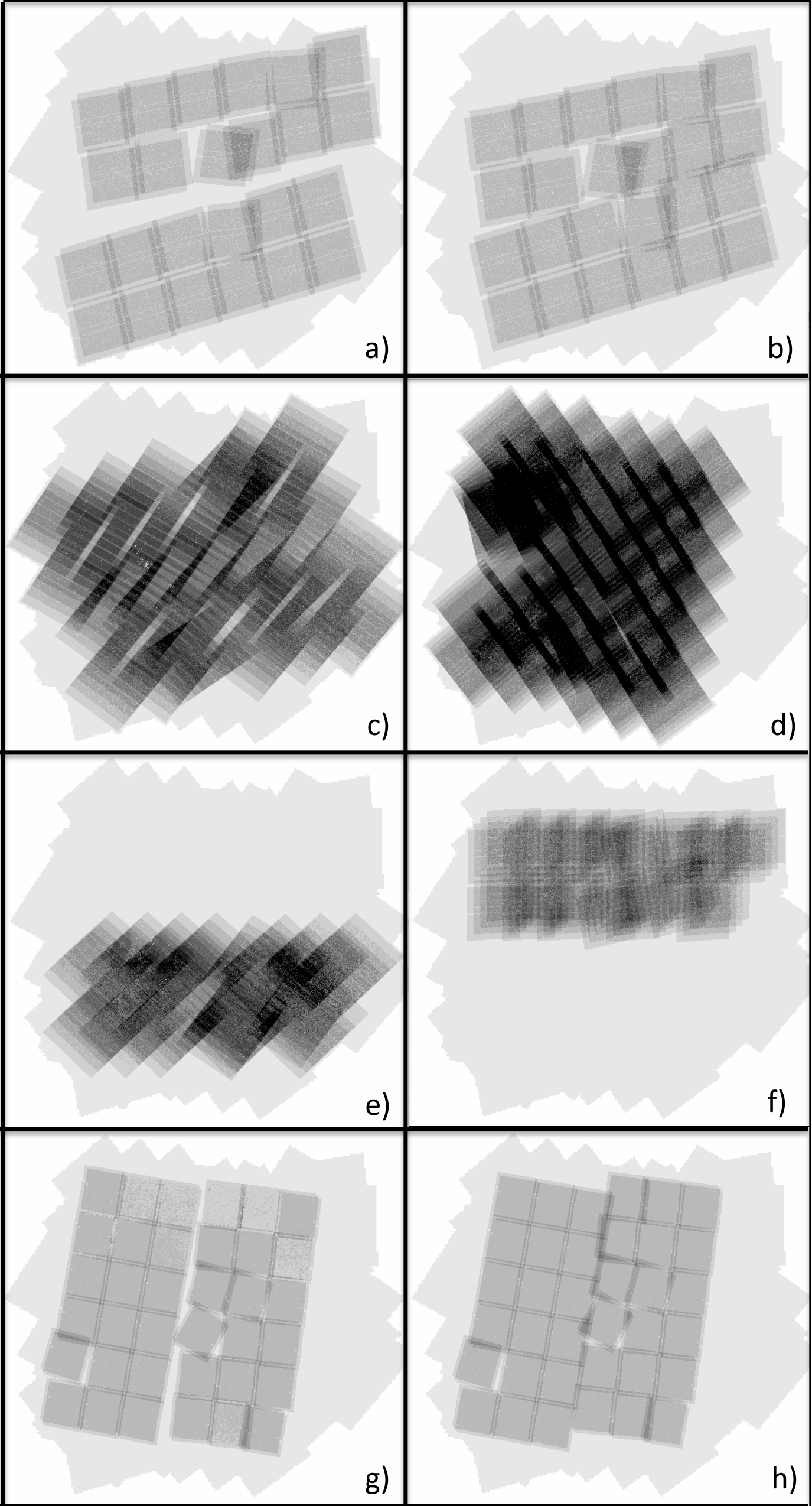}
      \caption{Coverage maps of the eight mosaics shown in Figure~\ref{mosaics}. The darker the image, the greater the number of overlapping images.}
      \label{depth_maps}
\end{figure}

Because of the different camera size and orientations, the entire field of view is not uniformly covered in all filters. Figure~\ref{mosaics} shows for each filter the final mosaic, while the number of exposures contributing to each pixel of each mosaic is shown in Figure~\ref{depth_maps}. The inspection of these two figures shows for example that, because the F775W survey was obtained using the two cameras in parallels, the F775W ACS filter covers only $\sim 57$\% of the total area ({\it Panel e)} of Figure~\ref{mosaics}), while the UVIS coverage in this filter is $\sim 49$\%. Because of the small size of the WFC3 detectors, compared to ACS, the WFC3 mosaics are affected by gaps, the two most noticeable affecting the F275W ({\it Panel a)}) and the F110W ({\it Panel g)}) filters. Table~\ref{mappa_ratio} reports the fraction of area covered by each filter with respect to the region surveyed by Lennon and collaborators by combining the two ACS and WFC3 F775W images.

Although at the moment of collecting HTTP data WCF3 had been in space for less than four years, charge transfer efficiency (CTE) losses were already a concern \citep{bourque13}, especially when the background is low, as is the case for the UV filters \citep{baggett12}. To mitigate the effect of the degrading CTE, we increased the background level of the UV exposures by $12\, {\rm e^-}$  using post-flash\footnote{http$//$www.stsci.edu$/$hst$/$wfc3$/$ins\_performance$/$CTE $/$ANDERSON\_UVIS\_POSTFLASH\_EFFICACY.pdf}. 

While WFC3 was acquiring the UVIS images, ACS was used to collect one short ($13\, {\rm s}$), and four long ($3 \times 640+1\times 337\, {\rm s}$) exposures in the filter F555W. For the deep F555W exposures we used a four point dither pattern, improving the sampling of the point-spread function (PSF). In addition, by smoothing over the spatial variations in the detector response, averaging the flat-field errors, and removing contamination from hot pixels and cosmic rays, the dither pattern allowed us to improve the photometric accuracy of our observations. Because of the important contribution from diffuse emission, the sky background in all ACS images is always higher then $\sim 100e^-$ and therefore there was no need to mitigate CTE losses with post-flash.

The IR orbits consist of two long ($2\times 799\, {\rm s}$) exposures in the F160W filter followed by two long ($799$ and $499\, {\rm s}$ respectively) exposures in the F110W filter. At the same time, ACS was used to obtain four long ($3\times 640 + 1\times 399\, {\rm s}$) exposures in the F658N filter. In the IR the large dither-steps adopted allowed us to maximize the survey area and to mitigate the effects of persistence \citep{long13a, long13b}.

All ACS exposures were acquired using a gain of $2\, {\rm e^-/ADU}$, WFC3/UVIS with gain $1.5\, {\rm e^-/ADU}$, and WFC3/IR with gain $2.5\, {\rm e^-/ADU}$. The entire data set was processed with the standard Space Telescope Science Institute calibration pipelines CALWF3 version 3.1.2 or CALACS version 8.1.2 to subtract bias level, superbias, and superdark, to apply the flat-field correction and, for the IR observations, the non-linearity correction and the up-the-ramp fitting. In addition, CALACS applies a pixel based CTE correction to the flat fielded images.
We applied an analogous correction to the WFC3/UVIS data using the stand-alone software routine {\tt wfc3uv\_ctereverse\_parallel.F} available at the WFC3 webpage\footnote{http://www.stsci.edu/hst/wfc3/tools/cte\_tools}. 

\subsection{The Reference Frame}
\label{reference_frame}

The analysis of the entire HTTP dataset has been carried out directly on the pipeline processed, CTE corrected {\tt \_flc} images. Because IR detectors are not affected by CTE losses, the IR analysis was carried on the {\tt \_flt} images. 

Compared to drizzled (either {\tt \_drz} or {\tt \_drc}) images,  {\tt \_flt} and {\tt \_flc} images have the advantage of not being resampled and therefore they are the most direct representation of the astronomical scene. On the other hand, these images are still affected by geometric distortion. Thus the first step in reducing the HTTP dataset was to create a distortion-free reference frame and relate the astrometry and photometry of each exposure to this frame. We chose the F775W filters to build the reference frame because, compared for example to F555W or F658N, the contribution from the ionized gas in these images is limited, they are among the deepest exposures, and  have much higher spatial resolution than the NIR data.  

To create the reference frame we began by running the publicly available one-pass photometry routine {\tt img2xym} \citep{anderson06} on each exposure.  We used this routine and a library of empirical PSFs to find and measure all the sources that were brighter than 100 counts in a 3 pixel aperture and had no brighter neighbors within a 5-pixel radius.  The 2MASS catalog \citep{skrutskie06} was used to construct a single reference frame that has a pixel scale of 40 mas/pixel (matching that of the WFC3/UVIS camera). The final reference frame measures 32,000 pixels along the X axis and 28,000 pixels along the Y axis to allow the mapping of all pixels covered by the various filters and at the same time minimize the total number of empty pixels in each mosaic.  

In doing so we were able to identify ~110,000 bright, isolated unsaturated stars that could be measured in three or more deep exposures.  Since the 2MASS positions are good to only $\sim 50$ mas, we improved the internal quality of the reference frame by iterating between solving for an average position of each star in the frame and using these average positions to improve the transformation from each exposure into the frame.  Within a few iterations, the average positions converged with RMS residuals of less than 0.01 pixel.

Once the final reference list was constructed from the bright stars in the F775W exposures, we ran the one-pass routine on all the exposures  (short and deep for all filters) and cross-identified stars to determine  the transformation from each image into the reference frame.  The WFC3/IR
images were also mapped to this frame with the WFC3/UVIS pixel scale.

\section{KS2: A Multi-Purpose Finding and Photometry Routine}
\label{star_finder}

The one-pass routine used above finds almost all of the bright stars in a field, but it is not designed to find faint objects, since these often require multiple detections in multiple exposures to be found and well measured. To recover these sources, we used KS2, an evolution
of the program used to measure the ACS Globular Cluster Treasury Survey  \citep[GO-10775, see][]{anderson08}.  A detailed description of KS2  will be presented in Anderson et al. (in prep). In this section we describe the aspects of KS2 that were used for the reduction of the HTTP dataset  and to run the artificial-star tests.

KS2 requires the transformations from each exposure into the reference frame, the  library of PSFs for the relevant filters, and a list of the bright stars that are likely to be saturated in many of the frames.  KS2 can analyze multiple HST instruments, and can process  up to 15 filters and hundreds of exposures at the same time.

KS2 constructs a master catalog of all the sources by going through the reference frame in tiles that are 125$\times$125 pixels in size. For each tile it makes a list of all exposures that cover it  and extracts the relevant raster from each exposure.  It executes several finding passes to identify stars that satisfy various criteria such as isolation within 5 pixels, S/N, quality-of-fit, or number of coincident peaks in multiple frames.  After each finding pass, the stars are measured and the PSF is used to subtract them from each individual exposure. In addition to the subtraction, the routine also makes a mask for each exposure that allows us to verify what kind of PSF artifacts might be present, either related to the diffraction spikes or poor subtraction. Sources detected in following passes have to stand out above this mask to be included in the catalog. 

KS2 is run on the individual {\tt \_flt} and {\tt \_flc} exposures, eliminating the need to  alter the information present in the images, by resampling them with e.g. drizzle. This approach  allows us to determine an independent estimate of each star's flux in each exposure, to give us an estimate of measurement errors, time variability, etc. In performing the photometric analysis KS2 applies a pixel area map correction, thus taking into account the fact that in an image the area of sky covered by different pixels varies as a function of their position. 

\begin{figure}
      \centering
      \includegraphics[trim=0 0cm 0 0cm, clip, width=8cm]{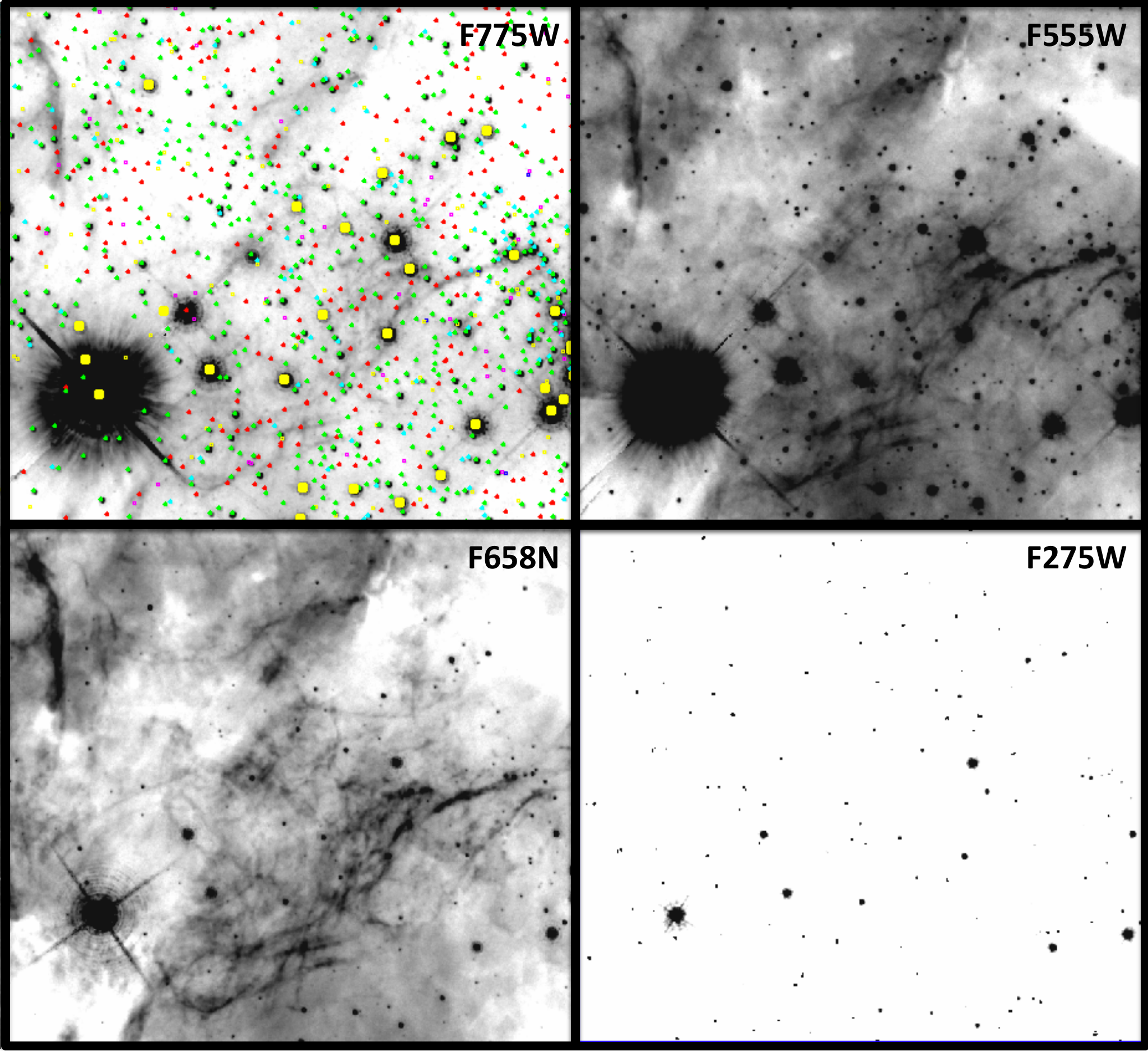}
      \caption{Detail of 30 Doradus in the filters: ACS-F775W ({\it Upper-Left Panel}), F555W ({\it Upper-Right Panel}), F658N ({\it       Lower-Left Panel}), and F275W ({\it Lower-Right Panel}). Stars detected by the finding algorithm are marked in the F775W image. Different colors correspond to different detection passes.}
      \label{detection}
\end{figure}

We chose to use only the F775W, F110W and F160W exposures to find stars, because in these filters both the bluest and reddest stars have the highest signal-to-noise and the contribution of the ionized gas is not as relevant as in the F555W and F658N filters, reducing the possibility to detect spurious sources, for example along sharp filaments of gas (Figure~\ref{detection}). 

Different approaches are needed to measure different stars. Bright stars are best measured by fitting the PSF to the {\tt \_flt}/{\tt \_flc} pixels to solve for position and flux. For simplicity we will call the simultaneous fitting of position and flux ``method \#1''. Faint stars often do not have enough signal to allow their positions to be measured well in individual exposures, so it is best to determine an average position from all the exposures, and then fit each exposure's pixels with the PSF solving only for the flux. From now on we will refer to this method as \#2. 

\begin{figure}
      \centering
      \includegraphics[trim=0.5cm 6cm 0.5cm
       3cm, clip, width=8cm]{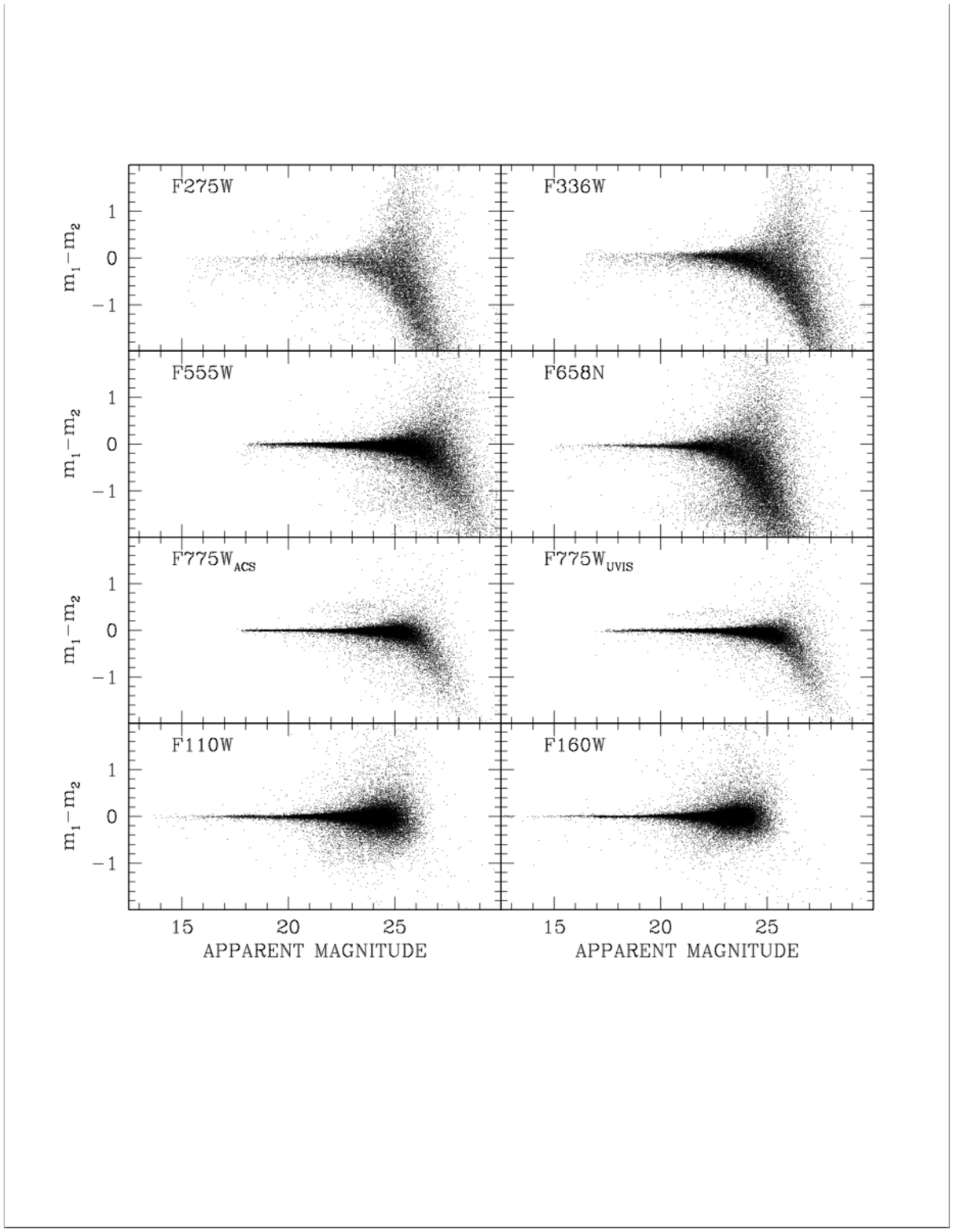}
      \caption{Comparison between the two photometry methods as a function of the apparent magnitude for the filters F275W, F336W, F555W, F658N, ACS/F775W, UVIS/F775W, F110W, and F160W. In each plot only 5\% of the sources are shown.}
      \label{phot_methods}
\end{figure}

We used both ``methods'' on all stars and reported in the final catalog the photometry that is most appropriate for each star's signal to noise. Figure~\ref{phot_methods} shows the comparison between the two photometric measurements. The two methods are in very good agreement in the brighter 5-6 magnitudes, but at fainter magnitudes method \#1 tends to underestimate the magnitude of the measured sources. 

For the ACS/WFC and WFC3/UVIS exposures, our aperture is the 5$\times$5 pixel around the star's central pixel, while for WFC3/IR it is the 3$\times$3 pixel, on account of the detector's severely under-sampled nature. In method~\#1, a robust sky was determined in the source-subtracted image using an annulus between 4 and 8 pixels radii, while in method~\#2, we used a smaller annulus (between 3 and 7 pixels).  In all cases, the PSF was fit to the selected neighbor-subtracted pixels using a least-squares approach that took into account all relevant sources of noise, such as Poisson noise, a $\sim 1\%$ error in the PSF model, error in the sky determination, etc.  

When possible we recovered the information related to saturated stars from the short exposures.  However, many of the known OB stars in the region are so bright to be saturated even in those images.  In these cases we measured the fluxes by taking advantage of the fact that he CCD blooming process does not destroy electrons but simply displaces them along the columns \citep{gilliland04, gilliland10}. In particular we started measuring the flux within a $5\times5$ pixel aperture and additionally identify every pixel that is part of the contiguous distribution centered on the target star, as well as any pixel that is next to any saturated pixel in the contiguous distribution.  We then added the total amount of flux over sky in these pixels and determined the fraction of the PSF that this aperture corresponds to.  The total flux of the star was then determined by dividing the observed amount of light by the fraction of the star's flux that should have landed in these pixels.  

\begin{figure}
      \centering
      \includegraphics[trim=0 1cm 0 1cm, clip, width=8cm]{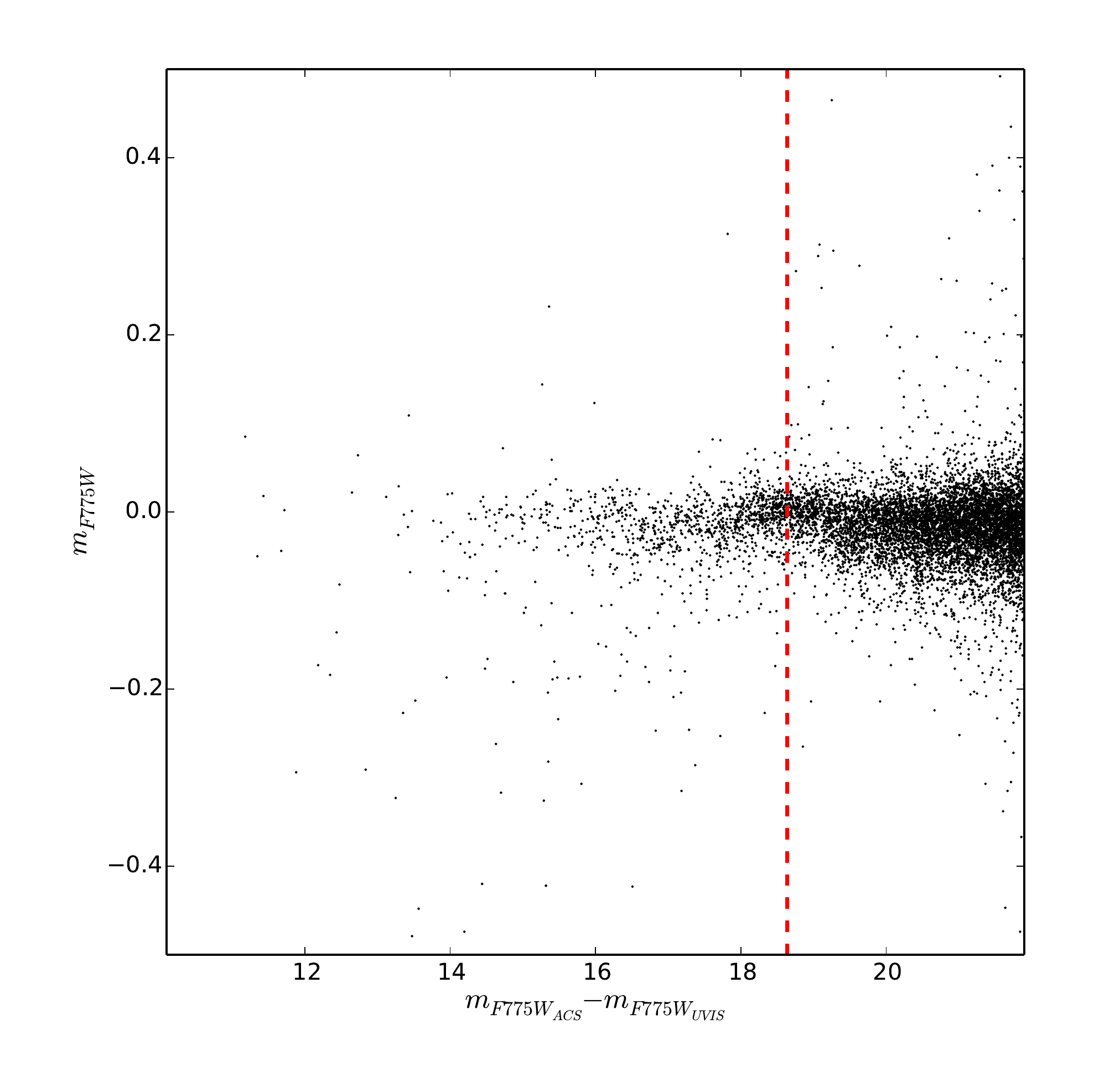}
      \caption{Magnitude difference for the stars measured in both the F775W filters as a function of the $m_{\rm F775W_{ACS}}$ magnitude. The vertical red dashed line marks the saturation point for the ${\rm F775W_{ACS}}$ data.}
      \label{deltaM}
\end{figure}

This process suffers from some flat-fielding errors, since it is impossible to know which pixel each electron landed in, however these are small errors and comparison of fluxes measured for stars that were saturated in the deep exposures and those unsaturated in the short exposures indicates this procedure is generally accurate to better than 5\%. Figure~\ref{deltaM} shows the difference in magnitude for the stars that have been observed with both the F775W filters as a function of magnitude. Above saturation the average difference between the two filters is $m_{\rm F775W_{ACS}}-m-{\rm F775W_{UVIS}}=-0.016\pm 0.028$.

\begin{figure}
      \centering
      \includegraphics[trim=0 6cm 0 3cm, clip, width=8cm]{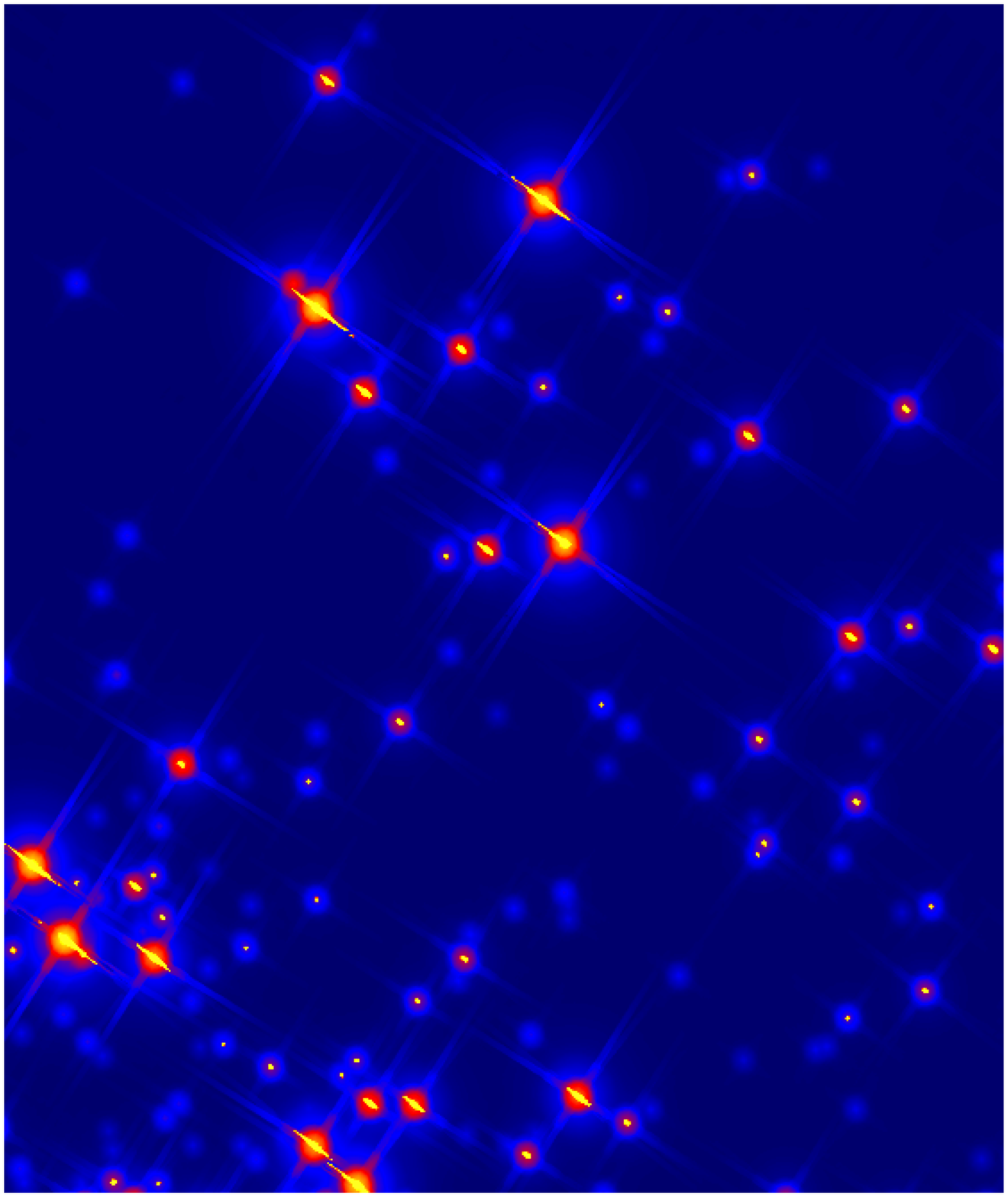}
      \caption{Detail of the photometry mask created by KS2 for the filter F555W. Because of the overlap of exposures acquired in different visits, with slightly different rotation angles, some stars show multiple diffraction spikes.}
      \label{phot_mask}
\end{figure}

The output of KS2 provides a wealth of information to help us evaluate the quality of the photometry. The program creates a stack image for each analyzed filter, allowing us to  verify that each frame was properly matched. The stacked images for all the HTTP filters are shown in Figure~\ref{mosaics}. In addition KS2 provides a coverage map for each filter (shown in Figure~\ref{depth_maps}), useful to verify how many images contributed to each pixel in a mosaic. In addition the code creates stellar-subtracted images to validate that all the point sources have been identified, as well as saturation masks, useful to verify where are the saturated pixels and the diffraction spikes. A portion of the saturation mask derived for the filter F555W is shown in Figure~\ref{phot_mask} as an example. 

\begin{figure}
      \centering
     \includegraphics[trim=0.5cmd 6cm 0.5cmd 3cm, clip,width=8cm]{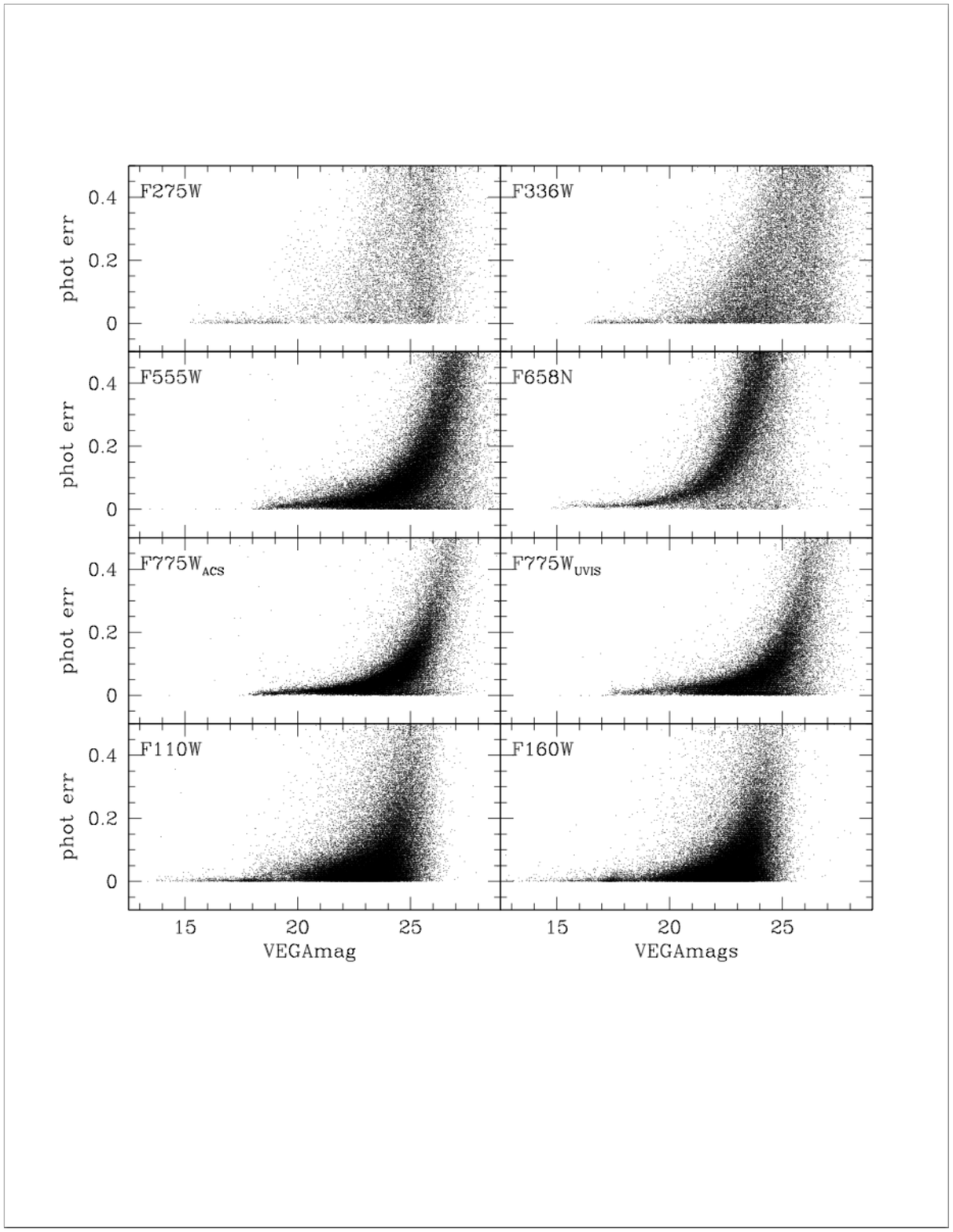}
      \caption{Standard deviation of the independent magnitude measurements. From top to bottom, left  to right we show the filters F275W, F336W, F555W, F658N, ACS/F775W, UVIS/F775W, F110W, and F160W. In each plot only 5\% of the sources are shown.}
      \label{phot_err}
\end{figure}

\begin{figure}
      \centering
     \includegraphics[trim=0 6cm 0 3cm, clip,width=8cm]{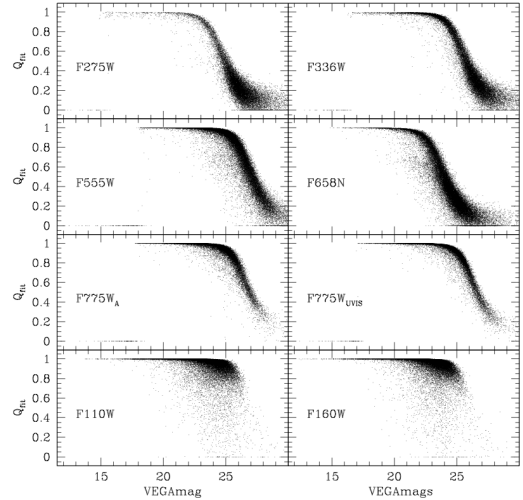}
      \caption{Quality of the PSF-fitting as a function of the apparent magnitude for the filters F275W, F336W, F555W, F658N, ACS/F775W, UVIS/F775W, F110W, and F160W. In each plot only 5\% of the sources are shown.}
      \label{Q_fit}
\end{figure}

For both the photometric methods described above, KS2 creates a separate photometric catalog. Each catalog include: each source's average flux, the RMS of the independent estimates from the contributing exposures, the number of images where the star could have been found, the number in which it was measured well, the finding-pass it was found in, a metric of the quality of the fit of the PSF to its pixels ($Q_{fit}$), etc. Figures~\ref{phot_err} and~\ref{Q_fit} show the distributions of the photometric errors and quality of the fit as a function of magnitudes for each of the 8 filters in the HTTP data set. Both plots show that KS2 is very efficient in detecting very faint sources ($m>28$), however when the S/N is low the quality of the  fit becomes very poor.

\section{The Photometric Catalog}
\label{catalog}

As mentioned before, KS2 measures the flux of all the stars found in different ways. In the final catalog we used the flux measured with  method \#1 if a star was found within the first detection pass. In all other cases we used the flux derived from method \#2.

Fluxes were converted into instrumental magnitudes and then calibrated to the VEGAmag photometric system using the  zero points listed on the STScI Web site.\footnote{http://www.stsci.edu/hst/wfc3/phot/\_zp/\_lbn and http://www.stsci.edu/hst/acs/analysis/zeropoints} The zero points are derived for images combined using ``astrodrizzle'' and cannot be directly applied to {\tt \_flt/\_flc}-based catalogs. 

To calibrate our data to VEGAmag we selected several isolated bright stars in the drizzle images and measured their magnitudes using a $0.5\arcsec$ aperture photometry for ACS data and $0.4\arcsec$ for WFC3. We used the average difference between the KS2 instrumental magnitudes and the calibrated magnitudes obtained from the drizzled images to put our catalog into the VEGAmag photometric system. 

Figure~\ref{deltaM} shows the difference in magnitude between sources that have been observed in both the F775W filters as a function of magnitude in the ACS/F775W filter. After the calibration in VEGAmag, the average difference between the two photometric catalogs is 0.014. Such a difference can be ascribed to a combination of color terms (for red stars the difference drops to 0.006) and uncertainties in the aperture correction. 

Stars too faint to detected in one filter (negative flux) or that fall outside the filter field of view have been assigned magnitude 99.999. We assigned to each source in each filter a quality flag {\tt f}$_{filter}$ ranging from 1 to 7. Stars that cannot be detected in one filter because outside the filter coverage have been assigned {\tt f}$_{filter}=7$, while stars that are too faint to be measured have {\tt f}$_{filter}=6$. 

We tested several combinations of diagnostics to remove as many spurious detections as possible. At the end we found that eliminating the sources with $Q_{fit}<0.75$ removes the majority of the outliers, giving the cleanest CMDs, without severely compromising the completeness of the catalog. Figure~\ref{Q_fit} shows that a consequence of the selection in $Q_{fit}$ is the cut at the fainter magnitudes. As shown in Figure~\ref{phot_err} these are also the sources with the larger photometric errors, therefore it is not a surprise that these objects were not measured as well as those characterized by higher  $Q_{fit}$ values. Sources that have been well fitted by our PSF model are likely stars. If their luminosity have been measured in more than one {\tt \_flt/\_flc} image, they have been assigned {\tt f}$_{filter}=1$.

The photometric errors (Figure~\ref{phot_err}) are determined using the formula $\sigma_{mag}=1.1 \sigma_{flux}/flux$, where $\sigma_{flux}$ is the standard deviation of the independent measurements. For the sources detected in just one exposure it was not possible to measure a photometric error. We assigned to these stars the most probable error for their magnitude, and marked them using the flag {\tt f}$_{filter}=2$ if they were characterized by a good quality of the PSF fitting ($Q_{fit}>0.75$), otherwise we used {\tt f}$_{filter}=5$. Sources with photometric error smaller than 0.25, but with a poor fit of the PSF have been assigned flag {\tt f}$_{filter}=3$; while object with photometric error larger than 0.25 and poor quality of the PSF fitting have been flagged with {\tt f}$_{filter}=4$.

In total, we detected more than 820,000 sources. Of these sources $\sim 620,000$ have been flagged with {\tt f}$_{filter}=1$ in at least one of the F775W filters ( $\sim 19,000$ stars are in common between ACS and WFC3, $\sim 330,000$ stars are in the area covered only with ACS and $\sim 270,000$ in WFC3 only).  More than 520,000 stars  have  {\tt f}$_{filter}=1$ in the F110W  filter and $\sim 570,000$ in F160W. Finally there are more than 30,000 stars with {\tt f}$_{filter}=1$ in the F275W filter,  more than 100,000 in F336W, $\sim 400,000$ in F555W, and $\sim 130,000$ in F658N. The absolute astrometry of the catalog was derived using 2MASS. J2000.0 Right Ascension and Declination in sexagesimal units were also used to define the source IDs in the catalog. An extract of the catalog is shown in Table~\ref{the_catalog}. The Table reports for each source listed the identification number {\tt ID}, the magnitude {\tt m}, the photometric error {\tt err}, the quality of the PSF-fitting parameter {\tt $Q_{fit}$}, and the quality flag {\tt f} for the filters F555W and F775W$_{\rm ACS}$. Coordinates in pixels {\tt x} and {\tt y}, and celestial coordinates in degrees and {\tt Ra} and {\tt Dec} are also listed. 

The final catalog is available for download from the Astrophysical Journal Supplement (ApJS), SAO/NASA Astrophysics DataSystem (ADS) websites. The astro-photometric catalog can be dowloaded also from he Mikulsky Archive for Space Telescope (MAST), at the url https://archive.stsci.edu/prepds/30Dor/. For each of the HTTP filter we also created a larfge mosaic using astrodrizzle\footnote{http://drizzlepac.stsci.edu}. Each mosaic can be downloaded from the url https://archive.stsci.edu/prepds/30Dor/ as well. 

\subsection{Artificial Star Tests}
\label{AS}

Artificial-star tests are a standard procedure to assess the level of completeness and accuracy of a photometric analysis. The tests are performed by inserting stars with known position and flux into the data set, and then repeating the photometric analysis as was used for the real stars. The selection criteria applied to the observed catalog to discard spurious detections are applied to the recovered artificial stars as well.
 
The difference between the input and output magnitudes of the recovered artificial stars provides an estimate of the photometric error. The fraction of recovered artificial stars per bin of magnitude gives an estimate of the photometric completeness. 

In total we simulated almost 20 million artificial stars in each filter. Input magnitudes of the artificial stars in the various filters have been set up to reproduce the same range of colors covered by the observed data. KS2 allows us to add artificial stars on a grid. To avoid the concern that artificial stars may interact with each other we imposed a 20 pixel minimum distance between one artificial star and the nearest simulated source. 

As in the case of the photometry, all the  filters were analyzed at the same time. The input coordinates of the artificial stars were defined in the reference frame, and then KS2 simulated each artificial star on all the single {\tt \_flt}/{\tt \_flc} images corresponding to that position. 

Once the artificial stars were added to the various frames we repeated the photometry as described in section~\ref{star_finder}.  
In our analysis, we considered an artificial star as recovered in a filter if its input and output fluxes agreed to within 0.75 mag. Larger magnitude differences in fact would imply that the flux of the detected object is dominated by a brighter real source. As in the real-stars photometric analysis, we also required that each star is found with $Q_{fit}>0.75$. 

\begin{figure}
      \centering
     \includegraphics[trim=0 5cm 0 3cm, clip, width=8cm]{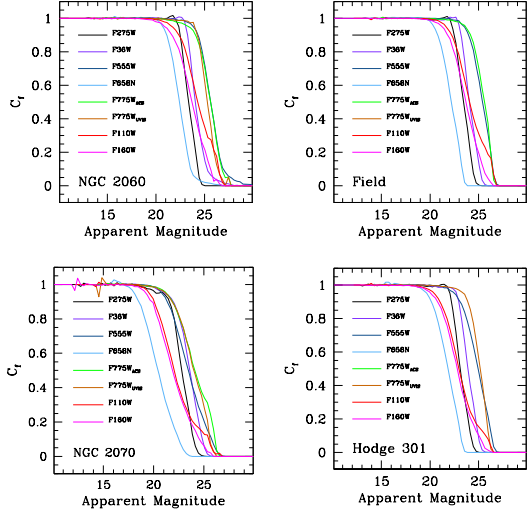}
      \caption{Ratio between the numbers of recovered and simulated artificial stars as a function of magnitude for all the HTTP filters around NGC~2070 ({\it Lower Left Panel}), NGC~2060 ({\it Upper Left Panel}), the field ({\it Upper Right Panel}) and Hodge~301 ({\it Lower Left Panel}). Different filters have been plotted using different colors.}
      \label{compl}
\end{figure}

Figure~\ref{compl} shows the completeness levels for different regions across the Tarantula Nebula for each of the HTTP filters. As expected crowding is the main source of incompleteness, as shown by the fact that around NGC~2070 even the deepest exposures (F55WW, F775W$_{\rm ACS}$ and F775W$_{\rm WFC3}$) are already less then 90\% complete at mag=22, while in less crowded regions (i.e. NGC~2060 and in the field) in these filters the completeness remains above 90\% down to mag=24.5-25.  

Not surprisingly the photometry in the narrow band filter F658N is less complete in all regions, dropping below 90\% between 19 and 21 depending on the crowding. It is interesting to note that the photometry in the NUV (F275W and F336W) remains almost 100\% down to mag=23, mimicking very well the completeness level of the broadband optical filters. However while for the latter the completeness decays more smoothly, in the NUV it rapidly drops to zero. This can probably be ascribed to the fact that in the NUV we forced the detection. 

Finally the completeness declines in the NIR is more gentle than in in the broadband optical filters. As a result, although even in regions of moderate crowding the NIR photometry is already less then 90\% complete at mag=22.5, its detection threshold is as deep as in the optical. 

\section{Color-Magnitude Diagrams}
\label{cmds}

\begin{figure}
      \centering
     \includegraphics[width=8cm]{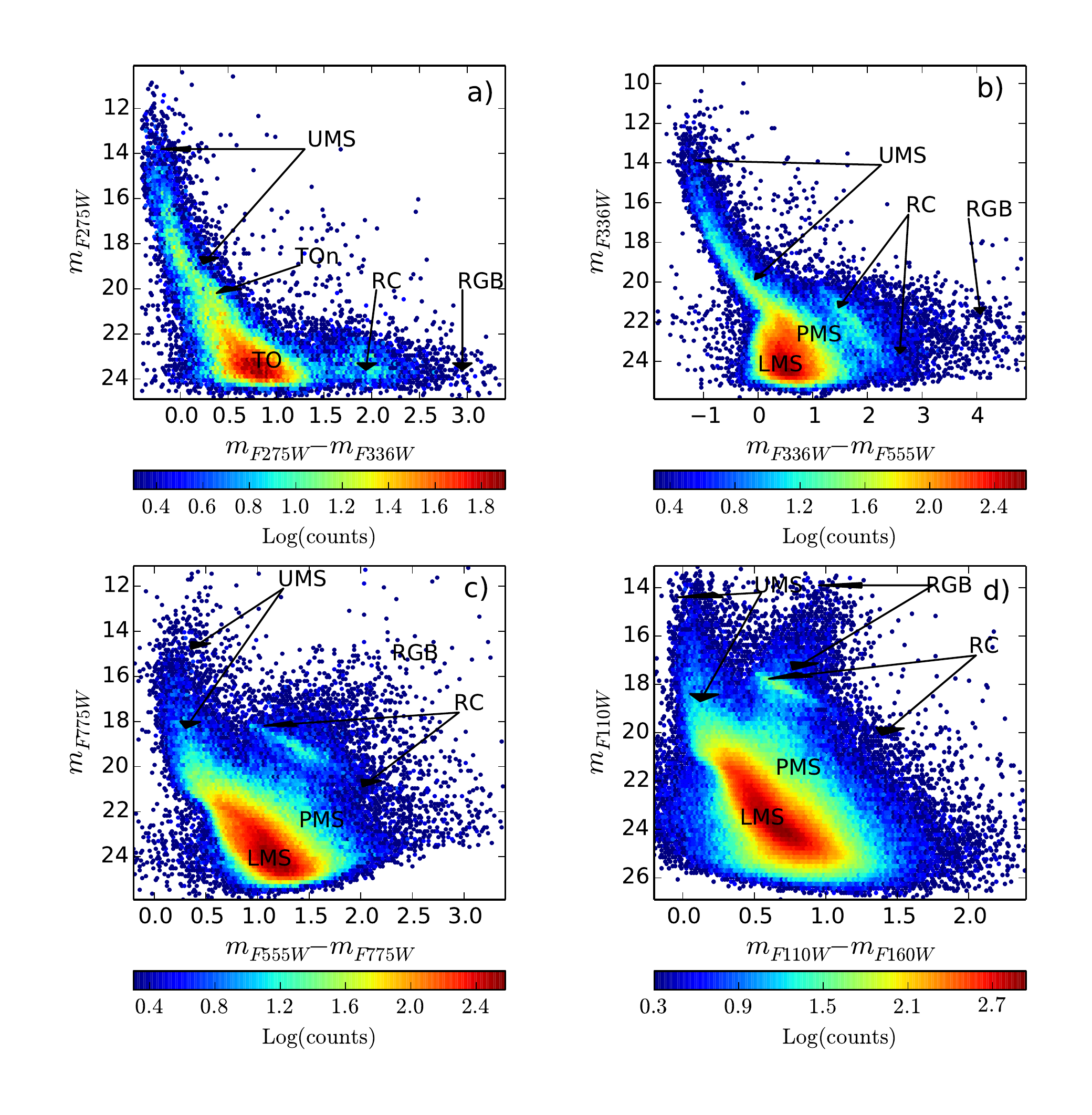}
      \caption{CMDs derived from the HTTP photometric catalog for different combinations of filters. The NUV CMD $m_{\rm F275W}$ vs $m_{\rm F275W} -m_{\rm F336W}$ is shown in {\it Panel a)}. The $m_{\rm F336W}$ vs $m_{\rm F336W}-m_{\rm F555W}$ CMD is shown in {\it Panel b)}, {\it Panel c)} shows the $m_{\rm F555W}$ vs $m_{\rm F555W} -m_{\rm F775W} $, while the NIR CMD $m_{\rm F110W}$ vs $m_{\rm F110W} -m_{\rm F160W}$ is shown in {\it Panel d)}. The position of evolutionary phases typical of stars of different ages and masses are indicated in all 4 CMDs.}
      \label{4hessi}
\end{figure}

CMDs realized from the HTTP photometric catalog can be used to interpret the stellar populations found in 30 Doradus. CMDs at near ultraviolet (NUV, $m_{\rm F275W}$ vs $m_{\rm F275W} -m_{\rm F336W} $, {\it Panel a)}), optical ($m_{\rm F336W}$ vs $m_{\rm F336W}-m_{\rm F555W}$, and$m_{\rm F555W}$ vs $m_{\rm F555W} -m_{\rm F775W} $, {\it Panels b)} and {\it c)} respectively), and near-infrared (NIR, $m_{\rm F110W}$ vs $m_{\rm F110W} -m_{\rm F160W}$, {\it Panel d)}) wavelengths are shown in Figure~\ref{4hessi}. In plotting the optical CMD, we combined the ACS and UVIS F775W magnitudes, and when both magnitudes were available we used the average value. Evolutionary phases for stars of different ages and masses are highlighted in each CMD to aid with the interpretation.

An inspection of the four CMDs shows that a variety of stellar populations are present in the region of the Tarantula Nebula. Because of the wide wavelength coverage, the shown CMDs can be used to highlight the properties of stars of different ages and in different evolutionary phases. 

\subsection{The NUV CMD}
\label{the nuv}

The most prominent feature in the NUV CMD (Figure~\ref{4hessi} - {\it Panel a)}) is the narrow and well-defined upper main sequence (UMS) characterized by stars brighter than $m_{\rm F275W}\le19.5$, and bluer than $m_{\rm F275W} -m_{\rm F336W} \le0.25$. The UMS is typical of a young stellar population, and comprises intermediate- ($M\ga 4M_\odot$) and high-mass  stars.

\begin{figure}
      \centering
     \includegraphics[width=8cm]{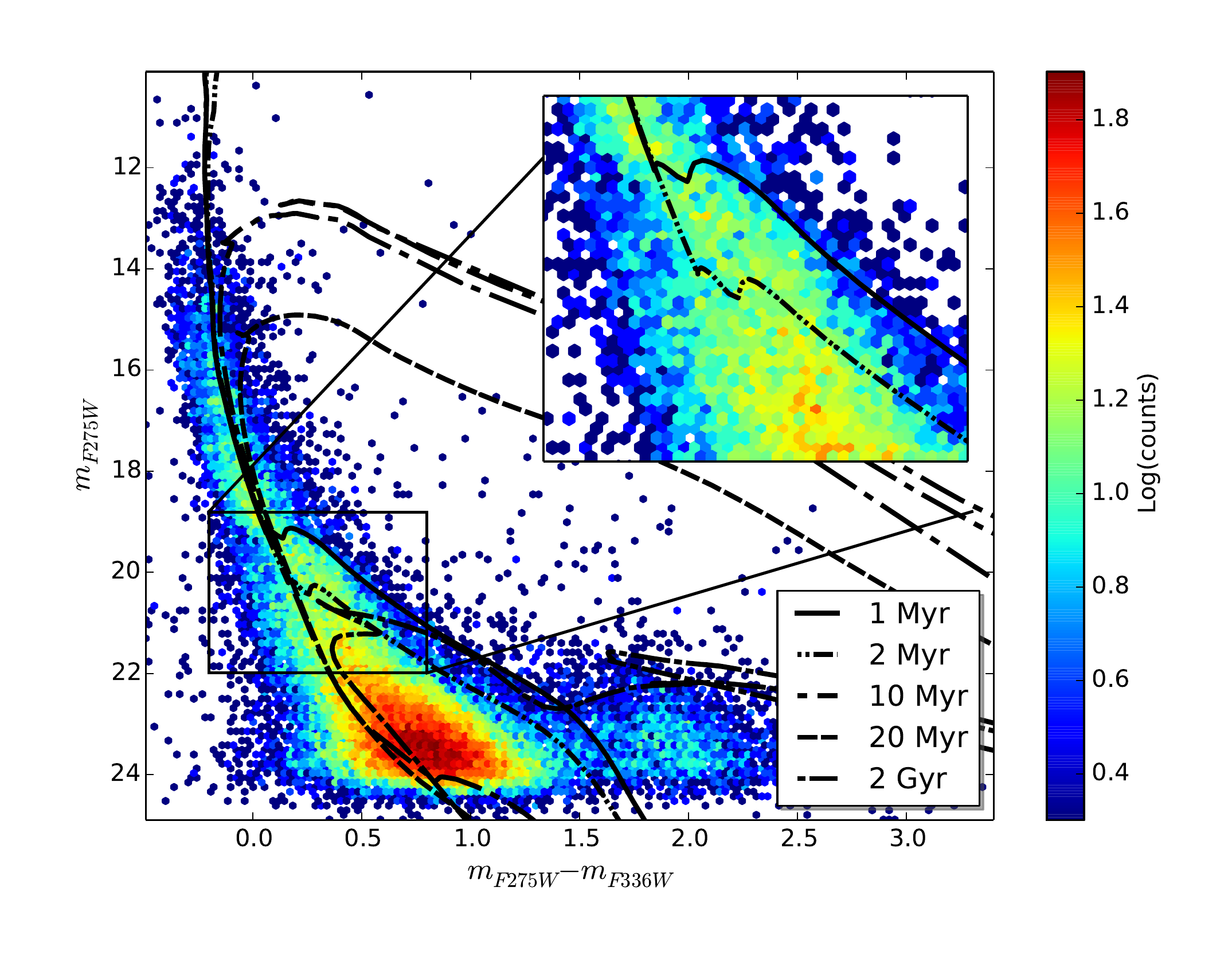}
      \caption{NUV CMD $m_{\rm F275W}$ vs $m_{\rm F275W} -m_{\rm F336W}$ CMD for the entire region covered by HTTP. Padova isochrones for metallicity for 1 (dashed-dotted-dotted line), 2 (long-short dashed line), 5 (dotted line) Myr,  and ${\rm age}=2.0\, {\rm Gyr}$ (continuous line) are superimposed assuming a distance modulus of 18.5, and a reddening E(B-V)=0.3 in addition to the Galactic foreground extinction. Extinction coefficients $A_{F275W}=7.9$ and $A_{F336W}=6.4$ are from De Marci et al (2015). For the younger populations (age$\le5$ Myr) we assumed metallicity Z=0.008, while for the older populations we assumed Z=0.004. The locus of the PMS-TOn is shown in the zoomed-in inset.}
      \label{cmd_uv}
\end{figure}

Figure~\ref{cmd_uv} shows the NUV CMD with superimposed Padova isochrones \citep{bressan12, chen14, tang14} for different ages. We assumed the distance modulus of 18.5, consistent with the literature \citep[i.e.][]{panagia91, pietr13}, and the extinction law of de Marchi et. (2015). For the younger stellar populations we assumed a metallicity $Z=0.008$, \citep[as proposed by][]{dufour82, bernard08}. The comparison with theoretical models indicates that the brightest stars in the CMD are likely O-type dwarfs as massive as $\sim 60-80\, {\rm M_{\odot}}$. 

At magnitude  $m_{\rm F275W}\simeq 19.4$ the UMS bends toward redder colors (see the inset in Figure~\ref{cmd_uv}). This is likely where young intermediate-mass pre-main sequence (PMS) stars join the UMS. This point is often called PMS turn-on (TOn). The brighter TOn can be reproduced with a 1 Myr old isochrone  and $E(B-V)=0.3$. Fainter TOns can be found down to $\sim 5-10$ Myr, suggesting that the region has been forming stars for a prolonged time, in agreement with the fact that the stellar population of the 30~Doradus region covers multiple ages, from regions of ongoing star formation \citep{rubio92, rubio99, walborn99, brandner01, walborn12} to the 15-25 Myr old Hodge~301 \citep{grebel00, evans15}. 

\begin{figure}
      \centering
     \includegraphics[width=8cm]{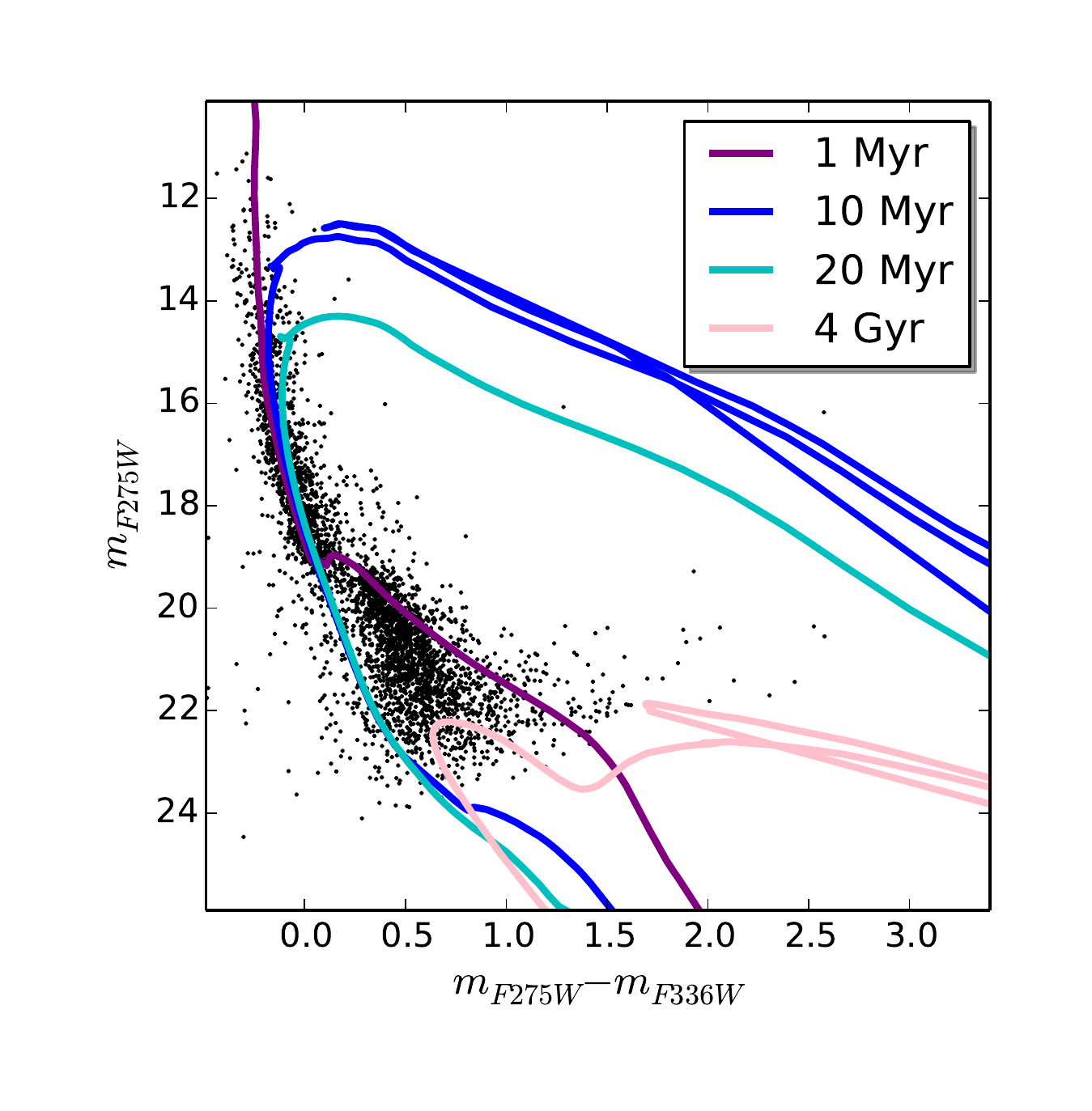}
      \caption{NUV CMD $m_{\rm F275W}$ vs $m_{\rm F275W} -m_{\rm F336W}$ CMD for the stars within 15 pc from the center of R136. Padova isochrones for metallicity $Z=0.008$ and ages=1, 10, and 20 Myr are shown in purple, blue and cyan respectively. The assumed distance modulus is 18.5 and we added E(B-V)=0.28 to the Galactic foreground reddening. A 4 Gyr old isochrone for metallicity  $Z=0.004$ is shown in pink assuming a distance modulus of 18.5.}
      \label{n2070_uv}
\end{figure}

The majority of the brighter and younger PMS stars are within 10-15 pc from the center of R136 \citep[$R.A._{J2000}=05^h 38^m 42.3^s Dec_{J2000}=-69\degr 06\arcmin 03.3\arcsec$][]{hog00}, consistent with the fact that about one fourth of the O stars of the LMC are found in NGC~2070 \citep{kennicutt89}. The NUV CMD for the NGC~2070 region is shown in Figure~\ref{n2070_uv}. Unfortunately the southern part of NGC~2070 falls within the gap that affects the mosaic in the F275W filter (Figure~\ref{mosaics}), therefore this CMD underestimates the total number of UV sources in this region. 

While Padova ischrones reproduce very well the properties of the UMS and of the RC, therefore supporting the validity of the assumed extinction law, they fail in reproducing the colors of the PMS candidates. PMS in particular are constantly bluer than what is predicted by the models. This discrepancy is likely due to problems with the models of atmosphere, and in particular to the characteristic excess radiation shortward of the Balmer discontinuity due to accretion \citep{kuhi74, gullbring98, robberto04}. 

Moving away from NGC~2070, the contamination from older main sequence and turn-off (TO) stars may become important. For example the TO of a $\sim 1$ Gyr old stellar population affected only by the Galactic foreground extinction would be as bright as $m_{\rm F275W}\simeq 20$. To better understand the origin of the bending in the UMS observed in Figure~\ref{cmd_uv}), we compared the spatial distribution of the stars found at luminosity of the TOn with that of fainter and bluer sources, that likely belong to the older lower main-sequence (LMS), and to the red clump stars (RC, corresponding to a stellar population older than 1-2 Gyr.). 

\begin{figure}
      \centering
     \includegraphics[width=8cm]{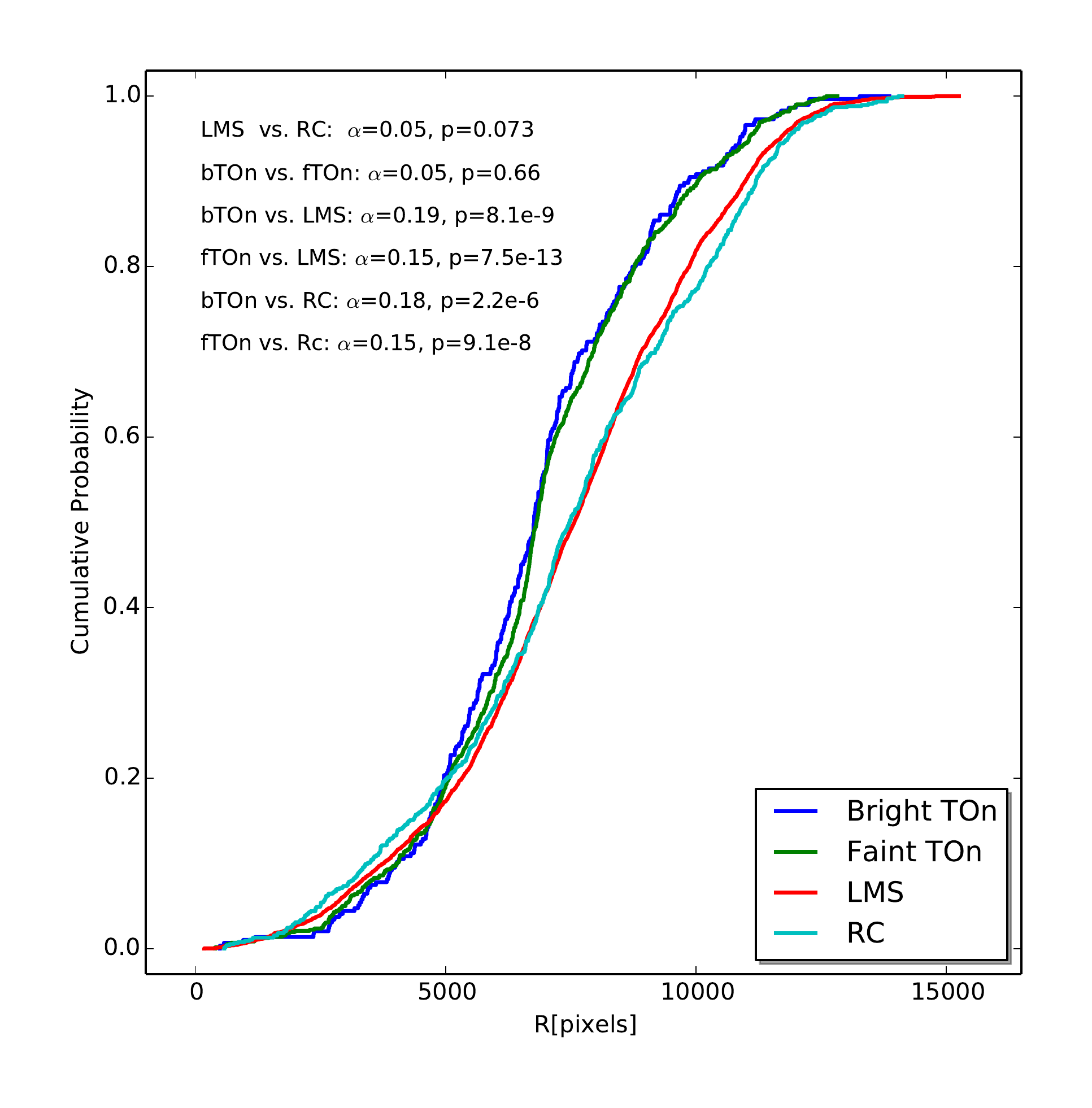}
      \caption{Cumulative radial distributions for the bright (blue line) and the faint (green line) TOn candidates, the LMS (red line) and the RC (cyan line) stars as a function of their projected distance (R) from the center of the mosaic (R.A.=05:37:52.285; Dec=-69:07:21.57; J=2000). Results from KS tests comparing the spatial distribution of TOn candidates, LMS and RCs are listed.}
      \label{ks_test}
\end{figure}

We used a Kolmogorov-Smirnov test to estimate the probability that TOn candidates have the same  spatial distribution of RC or LMS stars (Figure~\ref{ks_test}). In our analysis we divided the TOns candidates in two groups. The bright TOn (bTOn) stars are in the magnitude range $19.5< m_{\rm F275W}<20$, and are bluer than $m_{\rm F275W} -m_{\rm F336W} \le0.5$. The faint TOn (fTOn) are in the magnitude range $21< m_{\rm F275W}<21.5$ and are bluer than $m_{\rm F275W} -m_{\rm F336W} \le0.5$. We consider LMS stars those sources fainter than $m_{\rm F275W}\ge22$ and bluer than $m_{\rm F275W} -m_{\rm F336W} \le1.5$, while sources redder than $m_{\rm F275W} -m_{\rm F336W}>1.5$ are considered RC stars.

While we cannot reject the hypothesis that bTOn and fTOn candidates, or LMS and RC stars, are extracted from the same distribution, we can exclude that TOns candidates are compatible with LMS (the probability is $p\le10^{-8}$ for bTOn and $p\le 10^{-12}$ for fTOn candidates) or RC stars ( $p\le10^{-6}$ for bTOn and $p\le 10^{-7}$ for fTOn candidates), supporting the hypothesis that the change in the slope of the UMS at $m_{\rm F275W}=19.5$ is mainly due to the MS-TOn. 

\subsection{Optical and NIR CMDs}
\label{the vis_ir}

The optical and NIR CMDs (Figure~\ref{4hessi} - {\it Panels c)} and {\it d)}) can be used to study the properties of the stellar populations that formed in the region during the LMC's lifetime. The older stars in the CMDs belong to the LMC field. Evolutionary sequences typical of these older stellar populations are the LMS, the subgiant branch (SGB), the red giant branch (RGB), and the RC. 

The LMS are low mass ($<2\, {\rm M_\odot}$) stars with ages spanning from a few tens of million of years to several billion years. In the optical ({\it Panel c)} of Figure~\ref{4hessi}) most of LMS stars are fainter than $m_{\rm F775W}\ge 19$ and bluer than $m_{\rm F555W}-m_{\rm F775W}<0.17m_{\rm F775W}-2.6$, while in the NIR ({\it Panel d)}are fainter than $m_{\rm F110W}\ge18$ and bluer than  $m_{\rm F110W}-m_{\rm F160W}<0.16m_{\rm F110W}-3$.

\begin{figure}
      \centering
     \includegraphics[width=8cm]{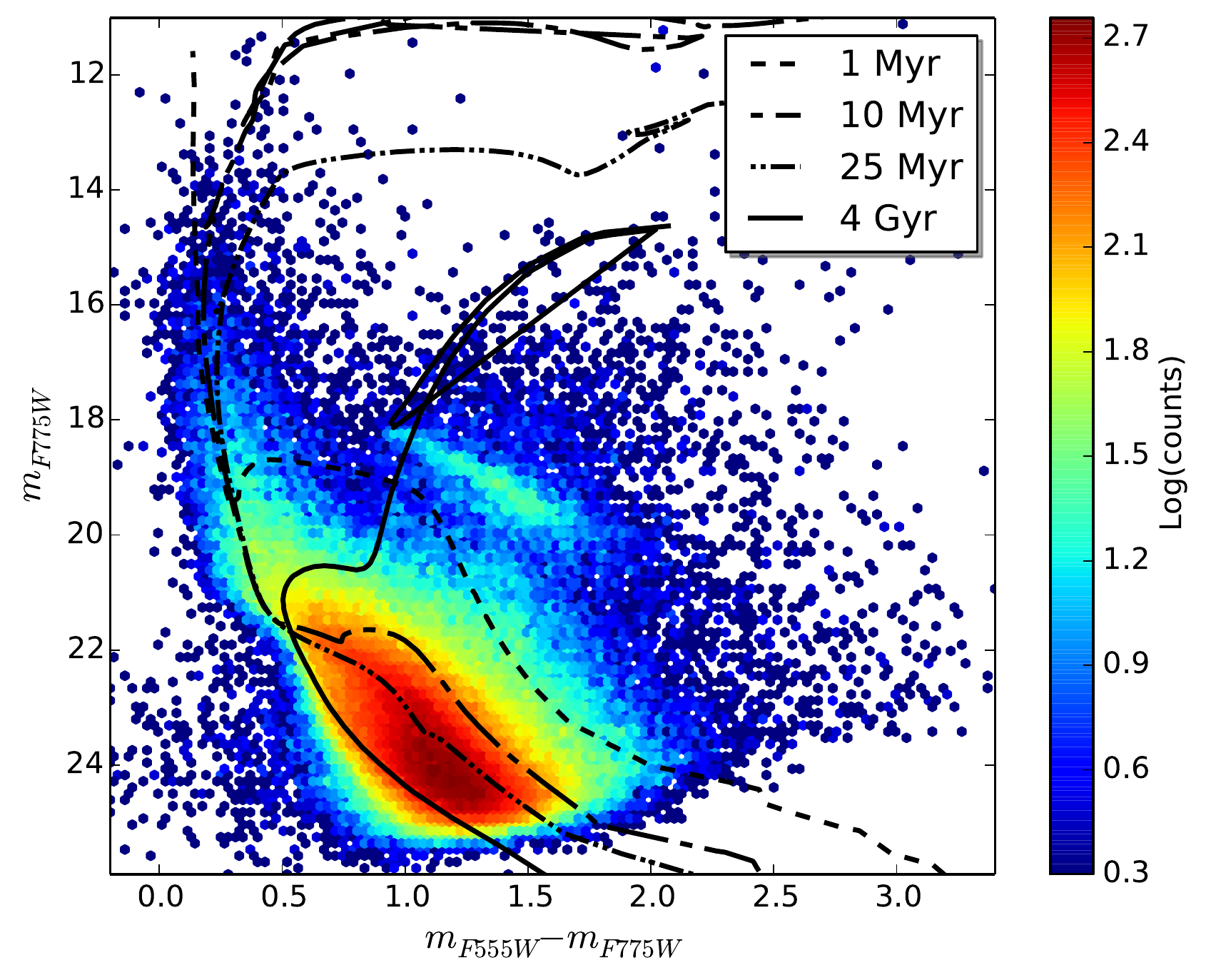}
      \caption{Optical  $m_{\rm F775W}$ vs $m_{\rm F555W} -m_{\rm F775W} $ CMD for the entire region covered by HTTP. Padova isochrones for metallicity $Z=0.008$ and ${\rm ages}=1, 10$, and $25\, {\rm Myr}$ are superimposed using  black dashed, short-dashed-long-dashed, and  dashed-dotted-dotted lines respectively. The continuous black line corresponds to a 4 Gyr old isochrone with metallicity Z=0.004. For all isochrones we assumed a distance modulus of 18.5. To fit the bluer edge of the younger ($\le 500\, {\rm M yr}$ old) stellar populations, in addition to the Galactic extinction ($R_{V}=3.1$ and $A_{V}=0.06$), a minimum E(B-V)=0.3 was needed. Extinction coefficients $A_{F555W}=4.6$ and $A_{F775W}=3.1$ are from De Marci et al (2015). The Galactic foreground extinction was sufficient to fit the bluer edge of the RC.}
      \label{cmd_vi}
\end{figure}

\begin{figure}
      \centering
     \includegraphics[width=8cm]{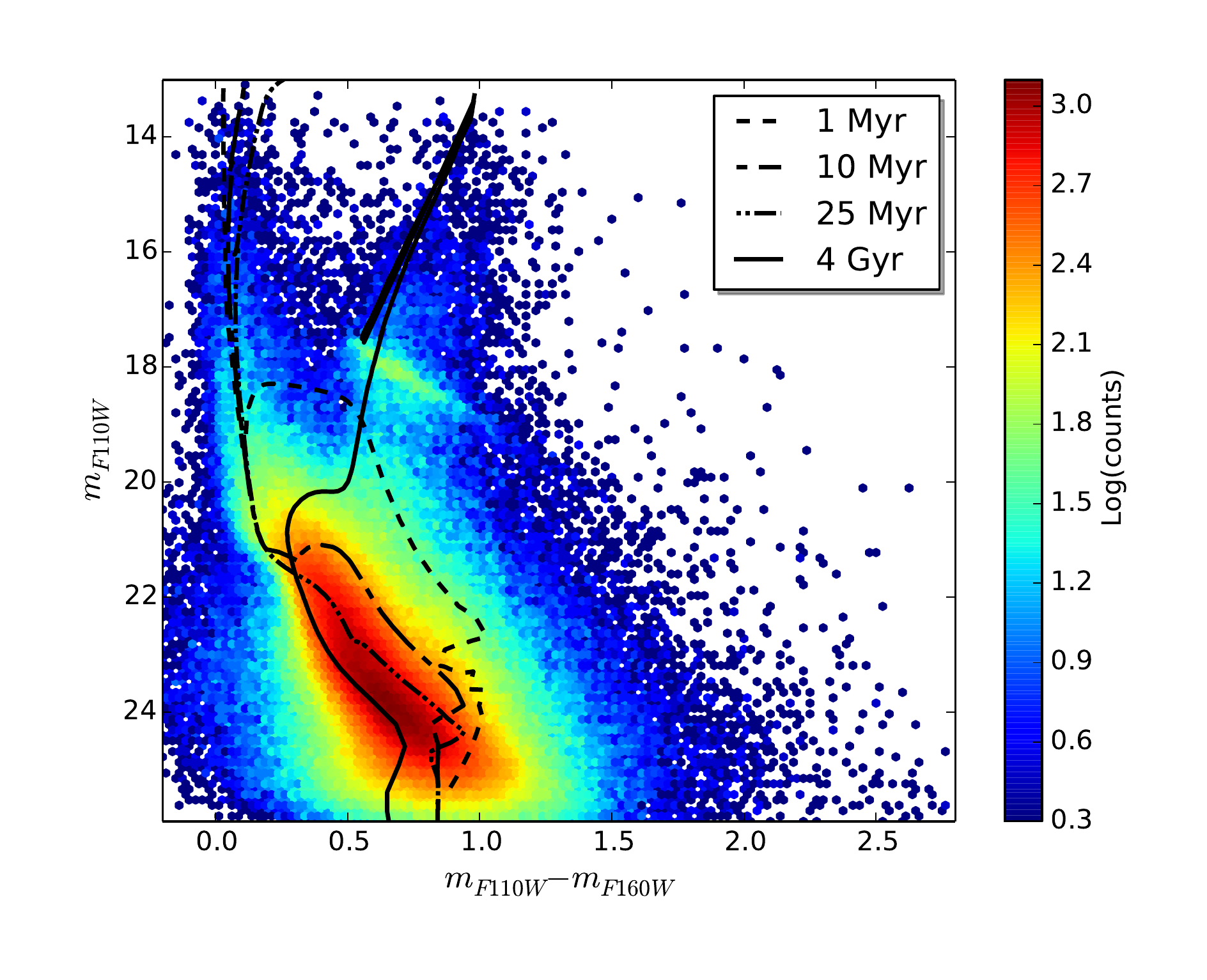}
      \caption{Same as Figure~\ref{cmd_vi}, but for the CMD in the NIR $m_{\rm F110W}$ vs $m_{\rm F110W} -m_{\rm F160W}$.  Extinction coefficients $A_{F110W}=1.9$ and $A_{F160W}=1.3$ are from De Marci et al (2015). }
      \label{cmd_ir}
\end{figure}

Once stars evolve off the main sequence, they become brighter and redder ($m_{\rm F775W}< 21.5$,  $m_{\rm F555W} -m_{\rm F775W}>1$ in {\it Panel c)}, and $m_{\rm F110W}<20.5$,  $m_{\rm F110W} -m_{\rm F160W} >0.7$ in  {\it Panel d)}  respectively) and migrate into the SGB and then RGB evolutionary phases. Padova isochrones for different ages and metallicities have been superimposed on the optical CMD in Figure~\ref{cmd_vi} and on the NIR CMD in Figure~\ref{cmd_ir} to help with the interpretation. 

As in the NUV,  in both optical and NIR CMDs the brighter and bluer objects are the UMS stars. In the optical CMD they are brighter than $m_{\rm F775W}<19$ and bluer than $m_{\rm F555W} -m_{\rm F775W} <1$, while in the NIR they occupy the portion of CMD above $m_{\rm F110W}<18$, with colors bluer than $m_{\rm F110W} -m_{\rm F160W} <0.4$.

The majority of the faint ($m_{\rm F775W}< 19$) red ($m_{\rm F555W}-m_{\rm F775W}>0.17m_{\rm F775W}-2.6$) objects in {\it Panel c)} of Figure~\ref{4hessi} (corresponding in {\it Panel d)} to the stars fainter than $m_{\rm F110W}<18$, and redder than $m_{\rm F110W}-m_{\rm F160W}>0.16m_{\rm F110W}-3$) are likely low-mass PMS stars. When looking at the CMDs for the entire region a combination of differential reddening (that pushes LMS stars toward redder colors) and age spread (older PMS stars are cooler and lie closer to the LMS than younger objects) blurs the transition between LMS and PMS starts. Rotational variability, accretion excess, unresolved binaries, and/or dusty disks \citep[see e.g.,][]{gouliermis12} can further broaden the portion of the CMD populated by PMS stars, making it even harder to separate PMS stars from the LMS.  

\subsection{Reddening and Dust Distribution}
\label{the dust}

 \begin{figure}
     \centering
     \includegraphics[trim=0cm 0cm 0 0cm, clip,width=8cm]{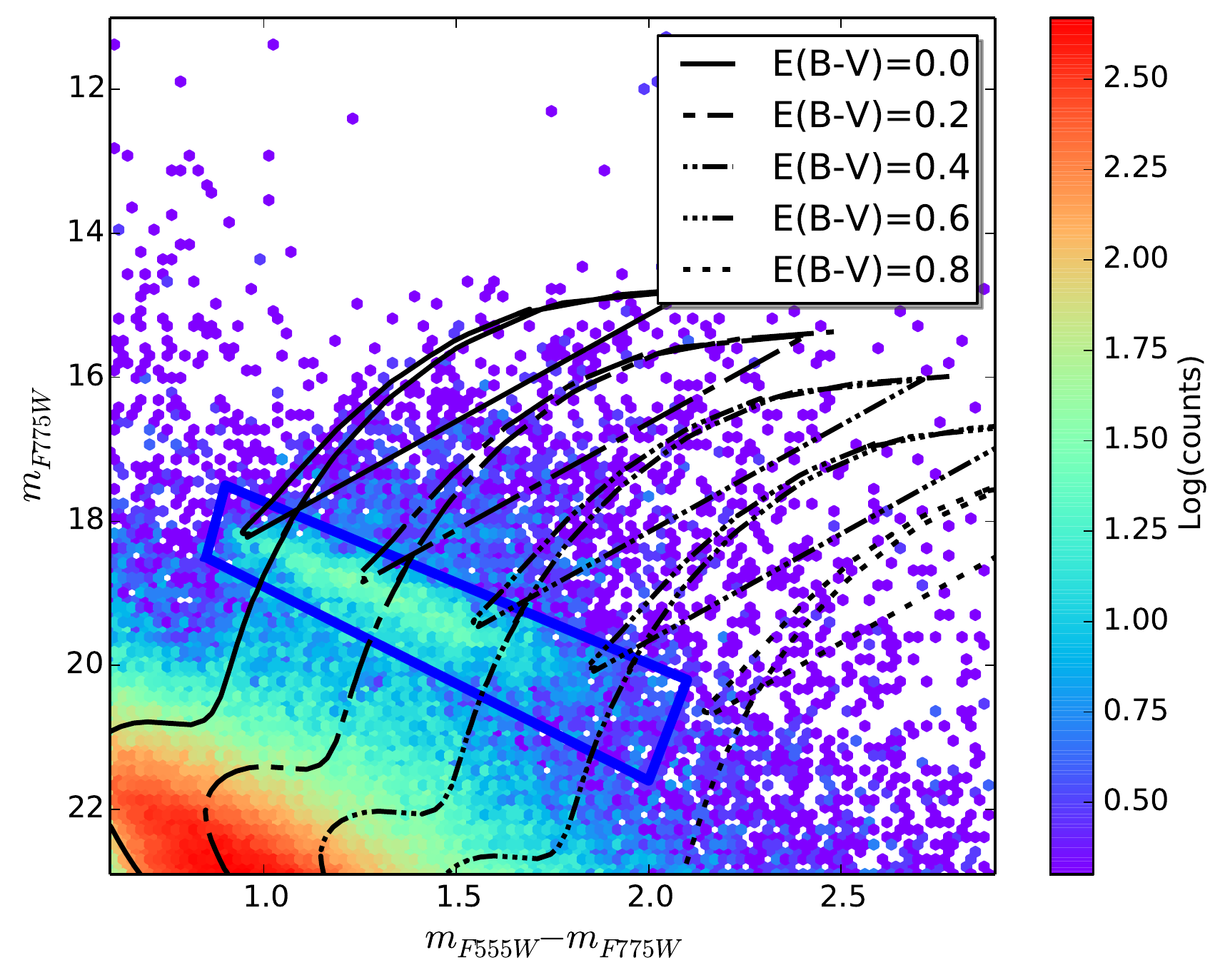}
     \caption{Detail of the $m_{\rm F775W} $ vs $m_{\rm F555W} -m_{\rm F775W} $ CMD centered around the RC. Padova isochrones for a 4 Gyr old stellar population of metallicity Z=0.004 and different values of E(B-V) are superimposed. In each isochrone the contribution of the Galactic foreground reddening is taken into account. The sources in the black box  have been used to study the spatial distribution of RC stars. The color scale is the same as in Figure~\ref{cmd_vi}.}
      \label{rc}
\end{figure}

 \begin{figure*}
     \centering
     \includegraphics[trim=0cm 0cm 0 0cm, clip,width=16cm]{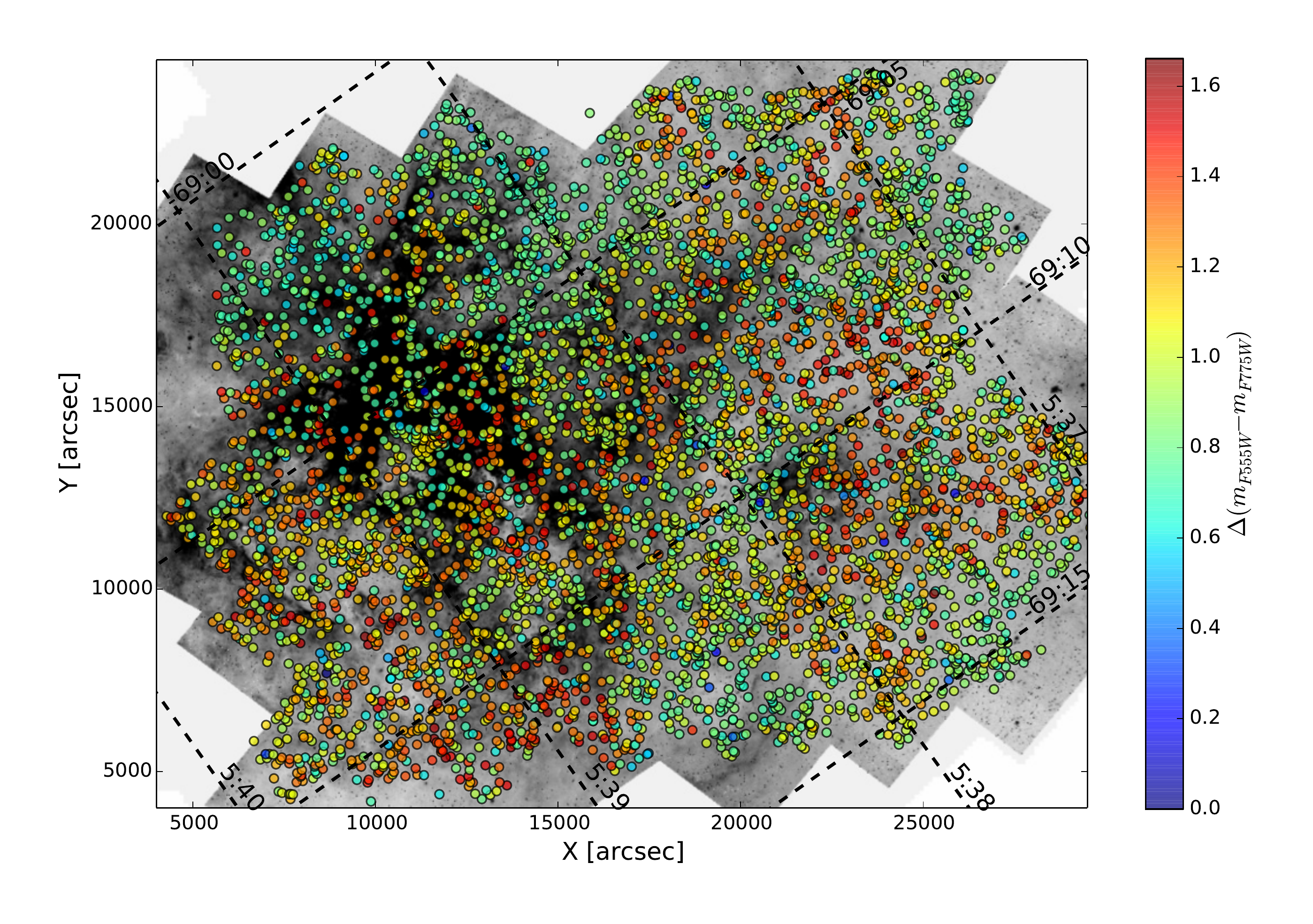}
     \caption{Spatial distribution of the RC stars selected from the $m_{\rm F555W}$ vs $m_{\rm F555W} -m_{\rm F775W} $ CMD. Sources have been color-coded according to their of their color excess in the $m_{\rm F555W} -m_{\rm F775W}$ color, with the zero point of the distribution being the color of the RC affected only by the Galactic reddening.}
      \label{red old stars}
\end{figure*}

RC stars are evolved core helium burning sources, whose intrinsic luminosity is relatively insensitive to age. Once their metallicity is known, RC stars can be used as standard candles to determine distances \citep[e.g.][]{stanek98, girardi01} and reddening \citep{udalski99a, udalski99b, zaritsky04,haschke11, demarchi14a}. 

Figure~\ref{rc} shows a detail of the entire optical CMD centered on the RC. RC stars are distributed along a narrow and extended strip, whose stretch in magnitude and color indicates that the reddening is varying considerably across the region. While the Galactic extinction ($R_{V}=3.1$ and $A_{V}=0.06$) is sufficient to fit the bluer edge of the RC, reddening values larger than ${\rm E(B-V)}> 1.0$ are necessary to reproduce the color of the most extinguished stars. 

Figure~\ref{red old stars} shows the spatial distribution of the RC stars that in Figure~\ref{rc} fall within the blue rectangle. The stars have been superimposed on the mosaic in the F555W filter to help with the identification of main structures. Stars have been color-coded from dark-blue to red as a function of their color excess in the $m_{\rm F555W} -m_{\rm F775W}$ color, with the color of the RC affected only by the Galactic extinction being the zero point of the distribution.

Although the spectral energy distribution of few RC stars indicate that they are affected only by foreground reddening, and are likely in front of 30 Dor, the majority of the stars suffer from a much higher reddening. It is interesting to note that the color of RC stars correlates with the distribution of the ionized gas on the eastern side of the Nebula, but not on the western side. 

To the east (lower left corner of Figure~\ref{red old stars}) the reddest RC stars coincide with regions of high dust and gas concentrations. These are among the most extinguished RC stars with E(B-V)$>0.8$ and coincides with the edges of  hot super-bubbles \#1 and \#2 (see also Figure~\ref{V-band} for the location of the super bubbles). Inside the bubbles the interstellar medium (ISM) becomes more transparent (E(B-V)$\sim0.3$), suggesting a filamentary distribution for the gas with a low volume-filling factor.

On average the region north to NGC~2070 ($x\le18000$ and $y\ge 15000$), corresponding to hot super bubble \#5, Hodge~301 and hot super bubble \#4, is characterized by the lowest (E(B-V)$<0.2$) extinction.  Another region where the ISM is relatively transparent is hot super-bubble \#3. Supernovae explosions and stellar winds are probably responsible for having removed the majority of the dust from these regions. Chandra X-rays images \citep{townsley06}, for example, show that hot super-bubble \#5 is a chimney and that hot plasma is escaping toward the north. 

The west and south portions of 30 Doradus do not show high concentrations of ionized gas (see for example Figure~\ref{mosaics}, {\it Panel d)}). Although on the majority of the region the ISM is relatively transparent (E(B-V)$<0.3$), in some areas (i.e. to the west and the south of NGC~2060) the background stars are completely blocked. 

Several of the dark clouds in Figure~\ref{red old stars} have basically no blue RC stars. Since RC stars are too old to be associated with 30 Doradus, it is possible that at least these parts of the 30~Doradus system are above the majority of the LMC's stellar disk, suggesting a possible small offset of the Tarantula Nebula from the disk toward us. Similar offsets have been seen for example in the H{\sc ii} regions on NGC 55 \citep{ferguson96}. The possibility that 30 Doradus could be at the nearer side of the LMC was already proposed by \citet{pellegrini11}, and \citet{sabbi13}.

In De Marchi et al. (2015) we used the $\sim 2500$ RC stars present in the HTTP field to further investigate the extinction properties of the ISM. Following the same approach used by  \citet{demarchi14a} and  \citet{demarchi14b}  we discussed local variations in the extinction law,  and their relation to the presence/absence of massive hot stars. In another paper (Arab et al. 2015, in preparation) we will use a Bayesian fit of the spectral energy distributions from UMS and RGB stars to infer properties of the dust, such as the size of the grains and their three-dimensional distribution. 

\section{Spatial distribution and Ages of the Stars in 30 Doradus}
\label{the age}

\begin{figure*}
     \centering
     \includegraphics[trim=0cm 0cm 0 0cm, clip,width=17cm]{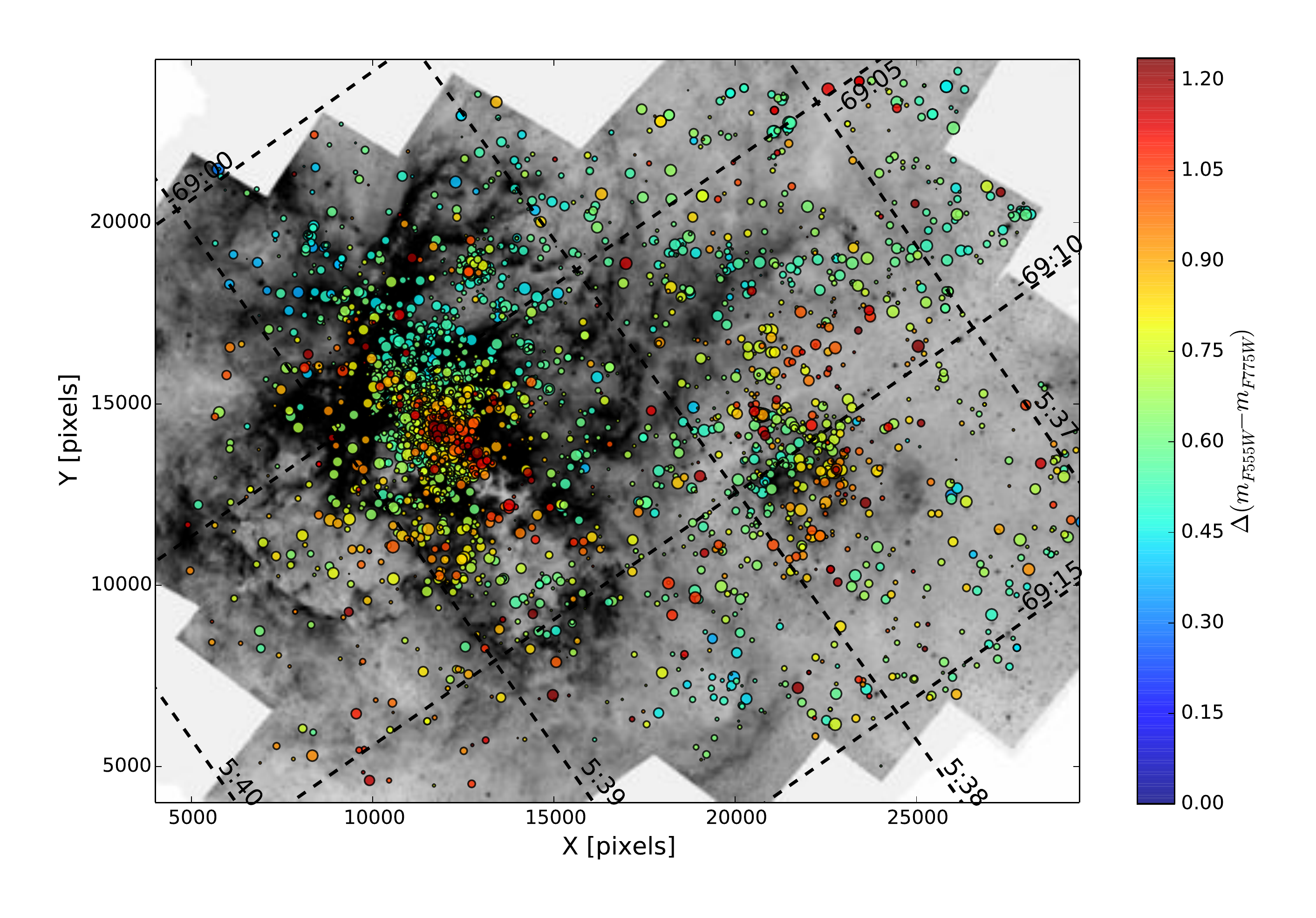}
      \caption{Spatial distribution of the UMS stars in the magnitude range between $15< m_{F775W}< 19$ and colors  $-0.1<m_{\rm F555W} -m_{\rm F775W} <0.6$, overlaid on the image of 30 Doradus obtained in the ACS filter F555W. The size of the circles relates to the luminosity of the source in the F775W filter. Stars are color-coded according to their  color excess relative to the color of UMS stars affected only by the galactic extinction.}
      \label{red young stars}
\end{figure*}

UMS stars are short-lived objects that can be used to highlight sites of recent star formation. Figure~\ref{red young stars} shows the spatial distribution of UMS candidates, selected in the magnitude range between $15< m_{F775W}< 19$ and with colors bluer than $m_{\rm F555W} -m_{\rm F775W} <0.6$. Similarly to what we have done in Figure~\ref{red old stars}, UMS stars have been superimposed on the 30 Doradus mosaic obtained in the filter F555W. The size of the circles is proportional to the apparent magnitude of the stars, with the larger symbols corresponding to the brighter sources. 

Although rotation, binarity, and age spread can contribute to the broadening of the UMS, for the majority of the source in the HTTP catalog the excess in the $m_{\rm F555W} -m_{\rm F775W}$ color is likely proportional to the amount of extinction. To have a sense of how much the reddening is changing across the Tarantula Nebula, we color-coded the UMS stars from dark blue to red on the basis of their color excess with the respect of $m_{\rm F555W} -m_{\rm F775W}$ color of UMS stars affected only by the galactic extinction. To do so we had to rectify the UMS by deriving its fiducial line and then subtracting the color of the fiducial line at the corresponding F775W magnitude from the color of each selected UMS star. 

As mentioned in the previous section, RC stars (Figure~\ref{red old stars}) trace the distribution of the old ($> 1- 2$ Gyr) field of the LMC, and  these stars are uniformly distributed over the entire region. The distribution of the UMS stars in Figure~\ref{red young stars} instead is variable and complex, preserving at least partial memory of where the stars formed. The majority of the UMS stars are found in the three most prominent clusters and associations Hodge~301, NGC~2060, and the mini-starburst NGC~2070 \citep{leitherer98}. In addition, several small associations of UMS stars can be found along the extended filaments of gas and dust between NGC~2070 and Hodge~301, around the edges of hot super-bubble \#5, and inside the two hot super-bubbles \#1 and \#3. Finally chains of massive blue stars can be found along the giant ($\sim 140$ pc long)  arc of dust and gas that divides the region in two parts and clearly separates NGC~2070 from NGC~2060. 

\begin{figure}
      \centering
     \includegraphics[width=8cm]{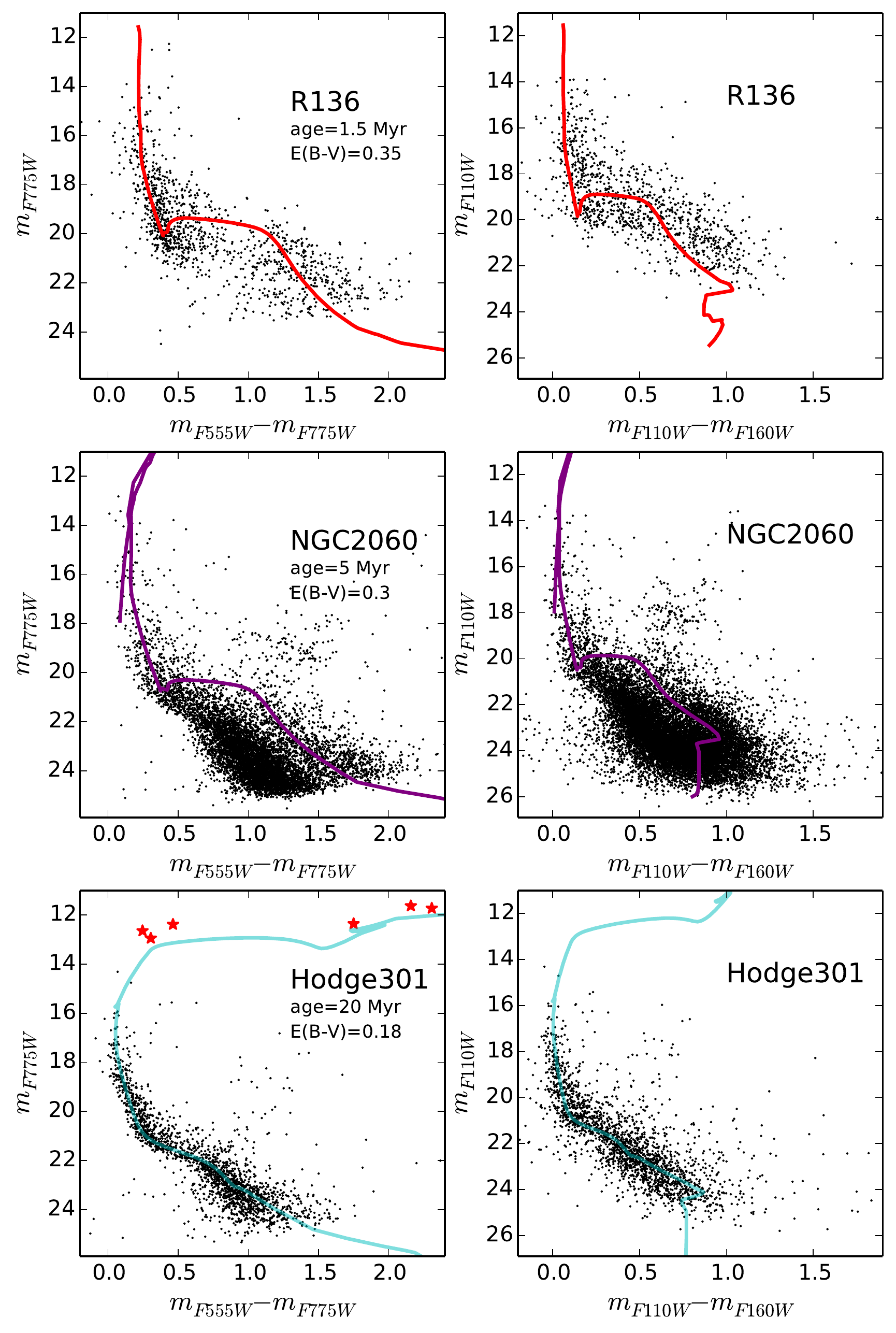}
      \caption{Optical and NIR CMDs for the three main clusters in the HTTP field: the starburst region NGC~2070 is shown in the two {\it Upper Panels} , NGC~2060 is shown in the two {\it Central Panels}, and Hodge~301 is shown in the two {\it Bottom Panels}. Padova isochrones for 1.5, 4 and 20 Myr old stellar populations are shown in red, purple and cyan respectively. Red stellar symbols in the optical CMD of Hodge~301 highlight the position of the blue and red supergiants.}
      \label{ammassi}
\end{figure}

Optical and NIR CMDs for the three main clusters are shown in Figure~\ref{ammassi}. For each cluster we computed the ridge line of the UMS in the optical CMD and then fitted it with Padova isochrones to derive an estimate of the reddening. We then constrain the age of the cluster using the luminosity of the TOn. We then used the derived reddening and age to superimpose the isochrone on the NIR CMD. For all the clusters we assumed a distance modulus of $(m-M)_0=18.5$, as proposed by \citet{panagia91} and \citet{pietr13}.

{\bf R136} 

R136 ($R.A._{J2000}=05^h 38^m 42.3^s Dec_{J2000}=-69\degr 06\arcmin 03.3\arcsec$) is the youngest cluster in the 30~Doradus region. It is at the center of the mini-starburst NGC~2070 and contains the most massive stars known so far \citep{crowther10}. R136's 10 brightest stars  are responsible for almost 30\% of the entire ionizing flux in 30~Doradus \citep{doran13}. The structure of NGC~2070 is complex and includes multiple clusters and associations of different ages and sizes \citep{walborn97}. \citet{selman99} found that NGC~2070 likely formed stars over a relatively prolonged interval of time, with star formation progressing from the outside in. 

Recently \citet{sabbi12} noted that NGC~2070 can be separated into two components: a slightly older elongated and more diffuse system extending toward the north-east (called the NE-clump), and the very compact and younger R136. The two top panels of Figure~\ref{ammassi} show the optical and NIR CMDs for stars within 5 pc from R136 (thus excluding the contribution of the NE-clump). 

From isochrone fitting we found that R136 is affected by a relatively high amount of reddening, with $E(B-V)=0.35$ and formed most of its stars between 1 and 4 Myr ago, although some star formation could be still ongoing, in agreement with the findings of \citet{rubio92, walborn97, massey98, rubio99, selman99}. In addition, a multitude of embedded objects at the top of extended pillars, as well as dark Bok globules, surrounding R136 could be the result of a very recent episode of star formation triggered by the interaction of the stellar winds from R136 and the supernova remnants associated to Hodge~301 as suggested by  \citet{walborn97}. In \citet{cignoni15} we compared HTTP data with synthetic CMDs to reconstruct the SFH of NGC~2070 over the last 20 Myr. In doing so we found that the peak of star formation occurred in the last 3 Myr, even if within the 20 pc from the center of R136 the star formation rate was significantly above the average value of the LMC already $\sim 7$ Myr ago. 

To reproduce the colors of the stars in the UMS in Figure~\ref{ammassi} we had to assume a higher reddening value than what we used in the UV (Figure~\ref{n2070_uv}). This difference is likely an artifact due to the gap in the F275W mosaic (Figures~\ref{mosaics} and \ref{depth_maps}, {\it Panel A)}). Because of the gap in fact the NUV CMD covers only the northern part of NGC~2070, characterized by a lower extinction.   

{\bf NGC~2060} 

NGC~2060 ($R.A._{J2000}=5^h:37^m:46.9^s$; $Dec_{J2000}=-69\degr:10\arcmin:18\arcsec$) is an extended ($r\simeq 21$ pc) star forming region that hosts the young ($\sim 5000$ yr), fast-rotating X-ray pulsars PSR J0537$-$6910 \citep{marshall98, cusumano98}, and many Wolf-Reyet (WR) stars \citep{breysacher81}. Part of the region is obscured by a bird-like dark cloud of dust, that makes it extremely hard to characterize the spatial distribution and stellar density of the system. 

The reddening varies considerably across the system, with the ISM becoming more transparent in the peripheral areas, particularly toward the North-East. To estimate the age of NGC~2060 we selected the stars within $\sim 5$ pc from $R.A.=5^h:37^m:51.6^s$; $Dec=-69\degr:10\arcmin:22.3\arcsec$ and around the compact cluster that hosts the WR stars Br73 (Figure~\ref{ammassi}, two central panels). In this region the average $E(B-V)=0.3$ and the cluster is likely $5\pm 1$ Myr old, confirming the young ages proposed by \citet{schild92} and \citet{walborn95}, who spectroscopically identified several early O-type stars across the region. 

{\bf Hodge~301} 

Hodge~301 ($R.A._{J2000}=5^h:38^m:17^s$; $Dec_{J2000}=-69\degr:04\arcmin:00\arcsec$) is the older cluster in the Tarantula Nebula. Compared to the previous systems the reddening across the region is more uniform and relatively low (E(B-V)$=0.18$), probably because stellar winds first and supernova explosions later \citep[based on the present day stellar mass function, ][ estimated that the cluster may have survived up to 40 supernova explosions]{grebel00} have cleaned the region from most of the dust. 

The Hodge 301 optical and NIR CMDs are shown in the two lower panels of Figure~\ref{ammassi}. Colors and luminosities of the bright red supergiants ($1.6<m_{\rm F555W} -m_{\rm F775W}<2.2$, $13.8<m_{\rm F775W}<14.3$) and the blue supergiants ($0.15<m_{\rm F555W} -m_{\rm F775W}<0.3$, $12.9<m_{\rm F775W}<13.3$) stars can be reproduced with A Padova isochrones for metallicity Z=0.008, and ages between 16 and 20 Myr, in excellent agreement with the age derived by \citet{evans15} from the spectroscopic analysis of 15 B-type stars. It is interesting to note that this age estimate is in good agreement with the values obtained by \citet{grebel00} using the isochrones from the Geneva group \citep{schaerer93}, but younger than the age derived by the same team using an older set of Padova isochrones \citep{bertelli94}.  

It is also worth noting that the new set of Padova isochrones fails in reproducing the characteristics of the low mass stars, whose colors and magnitudes in both the optical and IR CMDs are more consistent with those of a $\sim 25$ Myr old stellar population. This discrepancy is not surprising, because the evolutionary time for the PMS phase are still very uncertain. 

Furthermore recent studies in both the Milky Way \citep{sana12} and 30 Dor \citep{sana13, dunstall15} show that most of the massive O and B stars are close binaries. In Hodge 301, however the fraction of OB-type binaries is significantly lower than in the rest of the Tarantula Nebula \citep{dunstall15}, possibly as a results the binary dynamical evolution \citep{demink14}. If this is the case, the direct comparison of UMS stars in Hodge 301 with single star evolutionary models would be hazardous. Indeed mergers and mass transfer can artificially rejuvenate the most massive stars, creating an apparent discrepancy between these objects and the sources at lower masses already 5 Myr after the cluster formation \citep{schneider14}. 

\subsection{PMS stars and evidence for triggered star formation}
\label{the babies}

As mentioned in section~\ref{the vis_ir}, most of the faint and red objects in both the optical and in the NIR CMDs (Figures~\ref{cmd_vi} and \ref{cmd_ir}) are likely PMS stars. Because of their young age, PMS stars have not had time to migrate far away from their birth place, and therefore their spatial distribution can be used to trace how star formation occurred across a region. For this purpose, in the optical CMD we selected all the sources to the right of the Padova Z=0.008 5 Myr old isochrone (assuming the distance modulus 18.5 and E(B-V)=0.3), in the magnitude range $21.5<m_{\rm m_{F775W}}<23.0$ and bluer than ${\rm m_{F555W}-m_{F775W}}<2.0$. 

\begin{figure}
     \centering
     \includegraphics[trim=2.5cm 1cm 0 0cm, clip,width=10cm]{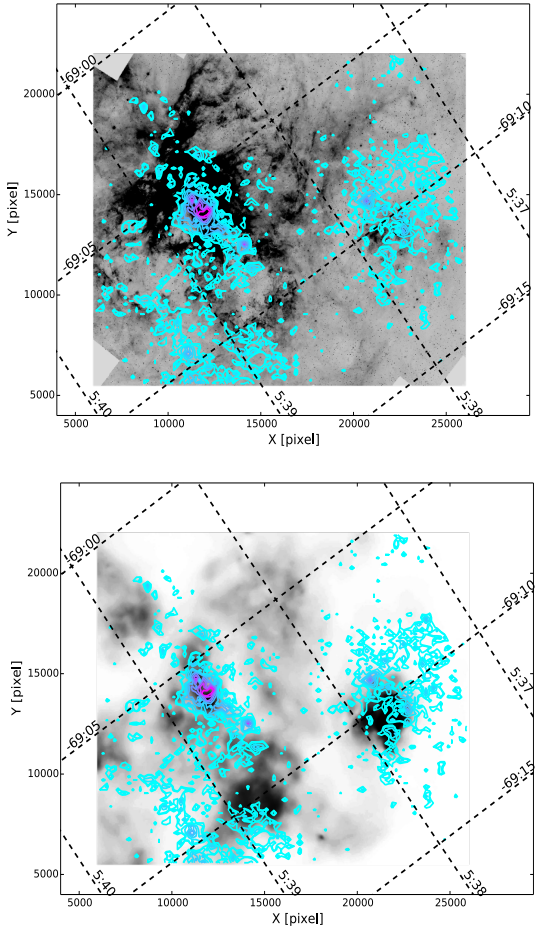}
      \caption{Spatial density distribution of PMS candidates. From purple to cyan, contours show 99\%, 97.5\%, 96.5\%, 95\%, 93\%, 90\% 85\%, 80\%, 65\% and 50\% of the PMS candidates. Contours are overlaid on: a portion of the HTTP mosaic in the F555W filter ({\it Upper Panel}); a portion of the Chandra/ACIS image of 30 Doradus ({\it Lower Panel}).}
\label{pms}
\end{figure}

PMS candidates can be found almost everywhere in the region (see Figure 4 in Cignoni et al. 2015), although clear over-densities can be noted, for example, in correspondence of the two young clusters NGC~2060 and NGC~2070. Because highly extinguished MS and SGB stars can contaminate the distribution of PMS candidates, in Figure~\ref{pms} we show the density contours only for regions whose number of PMS candidates is above the 50 percentile. This selection allows us also to highlight the clustering properties of these sources \citep{lada03}. To help with the interpretation in the {\it Upper Panel} we superimposed the density contours on the mosaic acquired in the F555W filter. 

The Figure shows an excellent correlation between the distribution of the PMS candidates and the ionized filaments of ionized gas and warm dust that envelope and confine several extended hot ($10^6-10^7$ K)  bubbles of plasmas. The {\it Lower Panel} of Figure~\ref{pms} shows that PMS candidates also correlate well with soft X-rays detected by Chandra ACIS  at 700-1120eV \citep{townsley06}. 
In particular we find that:
 \begin{itemize}
\item To the North, chains of small clusters clearly  depart from NGC~2070 and envelop the giant hot super bubble \#5. The densest point in this system is only partially resolved into stars by HST and is still heavily embedded in its dusty cocoon. This system, nicknamed the ``Skull Nebula'', is also one of the brightest sources in mid-IR \citep{walborn12}.
\item The boundary between hot super-bubbles \#1 and \#2, as well as the east side of hot super-bubble \#2 are marked by several clumps of PMS candidates.
\item PMS stars surround NGC~2060. The two richer systems coincide with diffuse soft X-rays to the North and the East of the SNR, the larger being associated with the compact bright cluster hosting Br73. 
\item The large arc of gas and dust that divides NGC~2060 and NGC~2070 in the X-ray is emitting at 700-1120eV and is the birth-site of several compact associations. The X-Ray peak of hot super-bubble \# 4, on the contrary, occurs at 350-700 eV and the region appears devoid of PMS stars.
\end{itemize}

The clumps of PMS candidates identified with HTTP also correlate with the population of embedded young stellar objects (YSOs) identified by \citet{seale14} combining Herschel and Spitzer data. The strong correlation between cool gas, warm dust and PMS candidates, as well as the anti-correlation between hot gas and very recent star formation, once again shows how, by carving cavities in the interstellar medium, strong stellar winds from massive stars, WR stars, and supernovae explosions, shut down star formation (disruptive feedback) in some regions, and ignite new generations of stars several parsecs away (constructive feedback). 

\section{Conclusions}

HTTP (HST GO-12939, PI: E. Sabbi) is a multi-wavelength study of 30 Doradus, the closest extragalactic giant H{\sc ii} region and an excellent example of large scale star formation in an environment that, in many ways (such as metallicity, dust content and SF rate) resembles the extreme conditions of the early universe. HTTP covers the 30 Dor region in the NUV (F275W, F336W), optical (F555W, F658N, F775W), and NIR (F110W, F160W) wavelengths using simultaneously both the imagers (WFC3 and ACS) currently operating on HST.

In this paper we have presented the observing strategy, the data analysis, and the photometric catalog for more than 600,000 stars. The stellar photometry was measured by PSF-fitting all the filters simultaneously. The astrometric RMS are less then 0.4 mas. The completeness of the catalog varies across the field depending both on crowding and back ground values. 

Since HTTP is a Treasury project, all the data collected  have been immediately available to the astronomical community. To enhance the scientific return of HTTP, we are now releasing the astro-photometric catalog, results from the artificial star tests, and the mosaicked images for all the HTTP's filters for download from the ApJS , ADS and the MAST websites. 

Our study probes the stellar content of the Tarantula Nebula down to $\sim 0.5$ M$_\odot$ and provides a snapshot of the history of star formation of the entire region, confirming that 30 Dor is a complex region that has built up its stellar content over several million years. The oldest stellar population observed at all wavelengths belongs to the field of the LMC. This component is uniformly distributed over the entire region. The presence of RGB and RC stars confirms the finding of \citet{zaritsky04} that the field has been forming stars for several billion years, and that it has been relatively active even in the  recent past. 

The spread in color of RC stars indicates that extinction is variable, with some areas totally blocking the background and others being transparent. This suggests the nebula is a partially filled structure offset towards the near side of the local LMC stellar disk.

The younger stellar population is organized in several clusters and associations. The larger systems in order of increasing age are NGC~2070, NGC~2060 , and Hodge~301. In using Padova isochrones to obtain a first estimate of the clusters ages we find that NGC~2070 has likely formed the majority of its stars in the last 3 Myr and that the majority of the stars in NGC~2060 are likely as young as $\sim 5 {\rm Myr}$ in agreement with previous photometric and spectroscopic studies. 

The comparison of the UV CMDs with theoretical models shows that at these wavelengths the models of atmosphere fail to reproduce the colors of the PMS candidates. The models and observations are in very good agreement in the optical and in the NIR. It is similarly possible that theoretical models fail to properly reproduce the characteristics of low mass stars at older ages. For example in the case of Hodge~301 Padova isochrones were not able to reproduce at the same time the properties of high and low-mass stars in the optical, implying a difference in age between the brightest stars and the low mass stars as large as 8-10 Myr. On the other hand if the models are correct, then the most massive stars could be the by-product of mergers and mass transfer in binary systems and appear younger than their real age, like in the case of the blue stragglers stars \citep{sandage53} found older clusters. 

While several studies have focused on Tarantula's massive stars, HTTP has provided for the first time a rich and statistically significant census of the low mass PMS stars. Although the complicated kinematics \citep{chu94} and the highly variable reddening make it hard to infer the tridimensional structure of 30 Dor, the high spatial resolution and sensitivity of HST have allowed us to trace how star formation has been developing in the region. 

The distribution of UMS stars and PMS candidates in clumps and filaments mimics the predictions of recent hydrodynamical simulation \citep[i.e.][]{schneider12, krumholz12, bate12}  supporting the hypothesis that stars may preferentially form in those filaments of gas that built up more mass  during the turbulent formation of the cloud. In addition the fact that PMS candidates surround many of the super-bubbles carved by supernova explosions suggests that fast supernova explosions had a major role in triggering the most recent episodes of star formation and that stellar feedback is still shaping the region. The overlap between PMS candidates and embedded YSOs indicates that, once initiated, the process of star formation continues for at least a few million years, even in the presence of fast stellar winds and high UV radiation, and that the SF episode in 30 Dor is not complete, continuing to evolve. 



\acknowledgments
We thank the anonymous referee for the thorough review and highly appreciate the comments and suggestions, which significantly contributed to improving the quality of this paper.
Based on observations with the NASA/ESA Hubble Space Telescope, obtained at the Space Telescope Science Institute, which is operated by AURA Inc., under NASA contract NAS 5-26555. 
These observations were associated with Programs 12499 and 12939. Support for both Programs 12499 and 12939 was provided by NASA through grants from the Space Telescope Science Institute. 
D.A.G. kindly acknowledges financial support by the German Research Foundation (DFG) through grant GO\,1659/3-2. 
SdM acknowledges support by the European Commission, grant H2020-MSCA-IF-2014, project id 661502.
M.T. was partially funded by the Italian MIUR through the grant PRIN-MIUR 2010LY5N2T.
EKG acknowledges support by Sonderforschungsbereich SFB 881 ``The Milky Way System'' of the German Research Foundation (DFG), particularly subproject B5.



Facilities: \facility{HST(WFC3)}, \facility{HST(ACS)}.

\clearpage

\begin{deluxetable}{cccccccccc}
\tablecolumns{10}
\tabletypesize{\scriptsize}
\tablecaption{Log of the Observations\label{log}} 
\tablewidth{0pt}
\tablehead{
\colhead{Data set} & \colhead{Ra}         & \colhead{Dec}        & \colhead{Obs. Date} & \colhead{Start Time} & \colhead{Exp. Time} & \colhead{$\#$ of}     & \colhead{Instrument} & \colhead{Filter} & \colhead{Central}
\\
\colhead{}              & \colhead{(2000.0)} & \colhead{(2000.0)} & \colhead{}                 & \colhead{}                 & \colhead{s}                & \colhead{Pointings} & \colhead{}                  & \colhead{}         & \colhead{ Wavelength}
}
\startdata
\multicolumn{10}{}{} \\
\multicolumn{4}{}{}& \multicolumn{1}{c}{GO-12939} & \multicolumn{5}{}{} \\
\multicolumn{10}{}{} \\
\hline

JBY001010 & 05 39 35.626  & -69 10 04.27 & 09$/$17$/$2013 & 16:29:40 & 2270.000 & 4 & ACS$/$WFC & F555W & 5360.755 \\  
JBY07B010 & 05 39 10.427  & -69 11 04.56 & 09$/$23$/$2013 & 21:29:47 & 2270.000 & 4 & ACS$/$WFC & F555W & 5360.755 \\  
JBY07C010 & 05 38 52.450  & -69 11 17.86 & 09$/$10$/$2013 & 14:01:57 & 2270.000 & 4 & ACS$/$WFC & F555W & 5360.755 \\  
JBY07D010 & 05 38 19.969  & -69 12 57.32 & 09$/$25$/$2013 & 00:34:19 & 2270.000 & 4 & ACS$/$WFC & F555W & 5360.755 \\  
JBY07E010 & 05 37 53.035  & -69 13 38.70 & 09$/$21$/$2013 & 21:53:23 & 2270.000 & 4 & ACS$/$WFC & F555W & 5360.755 \\  
JBY07F010 & 05 37 27.388  & -69 14 31.71 & 09$/$24$/$2013 & 22:28:23 & 2270.000 & 4 & ACS$/$WFC & F555W & 5360.755 \\  
JBY07G010 & 05 39 21.634  & -69 07 26.59 & 09$/$21$/$2013 & 04:22:46 & 2270.000 & 4 & ACS$/$WFC & F555W & 5360.755 \\  
JBY07H010 & 05 38 55.986  & -69 08 20.43 & 09$/$25$/$2013 & 22:23:46 & 2270.000 & 4 & ACS$/$WFC & F555W & 5360.755 \\  
JBY07I010 & 05 38 30.339  & -69 09 14.03 & 09$/$21$/$2013 & 16:06:30 & 2270.000 & 4 & ACS$/$WFC & F555W & 5360.755 \\  
JBY07J010 & 05 38 04.691  & -69 10 07.39 & 09$/$26$/$2013 & 00:12:20 & 2270.000 & 4 & ACS$/$WFC & F555W & 5360.755 \\  
JBY07K010 & 05 37 45.476  & -69 10 56.15 & 09$/$26$/$2013 & 14:44:21 & 2270.000 & 4 & ACS$/$WFC & F555W & 5360.755 \\  
JBY07L010 & 05 37 13.395  & -69 11 53.39 & 09$/$26$/$2013 & 19:17:40 & 2270.000 & 4 & ACS$/$WFC & F555W & 5360.755 \\  
IBY013020 & 05 37 13.569  & -69 11 02.28 & 04$/$06$/$2013 & 03:23:50 & 1408.000 & 2 & WFC3$/$UVIS & F336W & 3354.874 \\  
IBY013030 & 05 37 23.334  & -69 11 24.44 & 04$/$06$/$2013 & 03:55:41 & 1164.000 & 2 & WFC3$/$UVIS & F275W & 2706.949 \\  
IBY07M020 & 05 37 39.876  & -69 10 16.78 & 04$/$09$/$2013 & 20:13:33 & 1408.000 & 2 & WFC3$/$UVIS & F336W & 3354.874 \\  
IBY07M030 & 05 37 49.644  & -69 10 38.84 & 04$/$09$/$2013 & 20:45:24 & 1164.000 & 2 & WFC3$/$UVIS & F275W & 2706.949 \\  
IBY07N020 & 05 38 03.992  & -69 09 59.17 & 03$/$28$/$2013 & 09:22:59 & 1408.000 & 2 & WFC3$/$UVIS & F336W & 3354.874 \\  
IBY07N030 & 05 38 14.331  & -69 10 11.76 & 03$/$28$/$2013 & 10:50:10 & 1164.000 & 2 & WFC3$/$UVIS & F275W & 2706.949 \\  
IBY07O020 & 05 38 32.399  & -69 08 45.02 & 04$/$09$/$2013 & 21:41:28 & 1408.000 & 2 & WFC3$/$UVIS & F336W & 3354.874 \\  
IBY07O030 & 05 38 42.170  & -69 09 06.89 & 04$/$09$/$2013 & 22:13:19 & 1164.000 & 2 & WFC3$/$UVIS & F275W & 2706.949 \\  
IBY07P020 & 05 38 58.614  & -69 07 58.76 & 04$/$09$/$2013 & 06:00:34 & 1408.000 & 2 & WFC3$/$UVIS & F336W & 3354.874 \\  
IBY07P030 & 05 39 08.386  & -69 08 20.54 & 04$/$09$/$2013 & 07:19:59 & 1164.000 & 2 & WFC3$/$UVIS & F275W & 2706.949 \\  
IBY07Q020 & 05 39 24.798  & -69 07 12.25 & 04$/$05$/$2013 & 08:43:55 & 1408.000 & 2 & WFC3$/$UVIS & F336W & 3354.874 \\  
IBY07Q030 & 05 39 34.572  & -69 07 33.95 & 04$/$05$/$2013 & 09:29:59 & 1164.000 & 2 & WFC3$/$UVIS & F275W & 2706.949 \\  
IBY07R020 & 05 37 25.828  & -69 13 44.30 & 04$/$10$/$2013 & 10:33:36 & 1408.000 & 2 & WFC3$/$UVIS & F336W & 3354.874 \\  
IBY07R030 & 05 37 35.618  & -69 14 06.41 & 04$/$10$/$2013 & 12:02:32 & 1164.000 & 2 & WFC3$/$UVIS & F275W & 2706.949 \\  
IBY07S020 & 05 37 52.183  & -69 12 58.68 & 04$/$10$/$2013 & 21:56:22 & 1408.000 & 2 & WFC3$/$UVIS & F336W & 3354.874 \\   
IBY07S030 & 05 38 01.974  & -69 13 20.70 & 04$/$07$/$2013 & 22:28:13 & 1164.000 & 2 & WFC3$/$UVIS & F275W & 2706.949 \\  
IBY07T020 & 05 38 18.507  & -69 12 12.80 & 04$/$11$/$2013 & 01:20:34 & 1408.000 & 2 & WFC3$/$UVIS & F336W & 3354.874 \\  
IBY07T030 & 05 38 28.299  & -69 12 34.73 & 04$/$11$/$2013 & 02:59:17 & 1164.000 & 2 & WFC3$/$UVIS & F275W & 2706.949 \\  
IBY07U020 & 05 38 44.799  & -69 11 26.68 & 04$/$11$/$2013 & 20:38:02 & 1408.000 & 2 & WFC3$/$UVIS & F336W & 3354.874 \\  
IBY07U030 & 05 38 54.594  & -69 11 48.51 & 04$/$11$/$2013 & 21:45:19 & 1164.000 & 2 & WFC3$/$UVIS & F275W & 2706.949 \\  
IBY07V020 & 05 39 11.061  & -69 10 40.31 & 04$/$12$/$2013 & 00:47:54 & 1408.000 & 2 & WFC3$/$UVIS & F336W & 3354.874 \\  
IBY07V030 & 05 39 20.857  & -69 11 02.05 & 04$/$12$/$2013 & 02:17:42 & 1164.000 & 2 & WFC3$/$UVIS & F275W & 2706.949 \\   
IBY07W020 & 05 39 37.292  & -69 09 53.68 & 04$/$11$/$2013 & 23:07:06 & 1408.000 & 2 & WFC3$/$UVIS & F336W & 3354.874 \\  
IBY07W030 & 05 39 47.089  & -69 10 15.33 & 04$/$11$/$2013 & 23:38:57 & 1164.000 & 2 & WFC3$/$UVIS & F275W & 2706.949 \\  
JBY013010 & 05 36 51.308  & -69 05 25.54 & 04$/$06$/$2013 & 03:25:14 & 2270.000 & 4 & ACS$/$WFC & F555W & 5360.755 \\  
JBY07M010 & 05 37 17.516  & -69 04 40.25 & 04$/$09$/$2013 & 20:14:57 & 2270.000 & 4 & ACS$/$WFC & F555W & 5360.755 \\  
JBY07N010 & 05 37 30.966  & -69 04 48.73 & 03$/$28$/$2013 & 09:24:23 & 2270.000 & 4 & ACS$/$WFC & F555W & 5360.755 \\  
JBY07O010 & 05 38 09.841  & -69 03 08.91 & 04$/$09$/$2013 & 21:42:52 & 2270.000 & 4 & ACS$/$WFC & F555W & 5360.755 \\  
JBY07P010 & 05 38 35.957  & -69 02 22.87 & 04$/$09$/$2013 & 06:01:58 & 2270.000 & 4 & ACS$/$WFC & F555W & 5360.755 \\  
JBY07Q010 & 05 39 02.044  & -69 01 36.58 & 04$/$05$/$2013 & 08:45:19 & 2270.000 & 4 & ACS$/$WFC & F555W & 5360.755 \\  
JBY07R010 & 05 37 03.469  & -69 08 07.65 & 04$/$10$/$2013 & 10:35:00 & 2270.000 & 4 & ACS$/$WFC & F555W & 5360.755 \\  
JBY07S010 & 05 37 29.724  & -69 07 22.24 & 04$/$07$/$2013 & 21:57:46 & 2270.000 & 4 & ACS$/$WFC & F555W & 5360.755 \\  
JBY07T010 & 05 37 55.948  & -69 06 36.59 & 04$/$11$/$2013 & 01:21:58 & 2270.000 & 4 & ACS$/$WFC & F555W & 5360.755 \\  
JBY07U010 & 05 38 22.142  & -69 05 50.68 & 04$/$11$/$2013 & 20:39:26 & 2270.000 & 4 & ACS$/$WFC & F555W & 5360.755 \\  
JBY07V010 & 05 38 48.305  & -69 05 04.52 & 04$/$11$/$2013 & 00:49:18 & 2270.000 & 4 & ACS$/$WFC & F555W & 5360.755 \\  
JBY07W010 & 05 39 14.438  & -69 04 18.11 & 04$/$11$/$2013 & 23:08:30 & 2270.000 & 4 & ACS$/$WFC & F555W & 5360.755 \\   
JBY01N010 & 05 38 37.689  & -69 01 38.30 & 05$/$29$/$2013 & 17:22:56 & 2220.000 & 4 & ACS$/$WFC & F658N & 6583.956 \\   
JBY01O010 & 05 38 46.440  & -69 02 31.80 & 05$/$22$/$2013 & 10:20:48 & 2220.000 & 4 & ACS$/$WFC & F658N & 6583.956 \\  
JBY01P010 & 05 38 51.411  & -69 03 06.40 & 05$/$29$/$2013 & 23:06:25 & 2220.000 & 4 & ACS$/$WFC & F658N & 6583.956 \\   
JBY01Q010 & 05 39 28.323  & -69 06 28.59 & 05$/$30$/$2013 & 16:41:54 & 2220.000 & 4 & ACS$/$WFC & F658N & 6583.956 \\   
JBY01R010 & 05 39 45.201  & -69 08 05.15 & 05$/$31$/$2013 & 12:12:18 & 2220.000 & 4 & ACS$/$WFC & F658N & 6583.956 \\  
JBY01S010 & 05 38 04.811  & -69 01 20.60 & 06$/$01$/$2013 & 12:31:47 & 2220.000 & 4 & ACS$/$WFC & F658N & 6583.956 \\  
JBY01T010 & 05 38 21.689  & -69 02 57.67 & 06$/$02$/$2013 & 18:01:35 & 2220.000 & 4 & ACS$/$WFC & F658N & 6583.956 \\  
JBY01U010 & 05 38 31.372  & -69 03 49.02 & 06$/$10$/$2013 & 14:18:59 & 2220.000 & 4 & ACS$/$WFC & F658N & 6583.956 \\  
JBY01V010 & 05 38 55.445  & -69 06 11.50 & 06$/$03$/$2013 & 13:00:52 & 2220.000 & 4 & ACS$/$WFC & F658N & 6583.956 \\  
JBY01W010 & 05 39 12.323  & -69 07 48.26 & 06$/$11$/$2013 & 04:02:08 & 2220.000 & 4 & ACS$/$WFC & F658N & 6583.956 \\  
JBY01X010 & 05 39 29.201  & -69 09 24.92 & 06$/$04$/$2013 & 11:14:42 & 2220.000 & 4 & ACS$/$WFC & F658N & 6583.956 \\  
JBY01Y010 & 05 37 48.811  & -69 02 39.78 & 06$/$04$/$2013 & 16:22:00 & 2220.000 & 4 & ACS$/$WFC & F658N & 6583.956 \\  
JBY01Z010 & 05 38 05.689  & -69 04 16.95 & 06$/$06$/$2013 & 06:08:32 & 2220.000 & 4 & ACS$/$WFC & F658N & 6583.956 \\  
JBY02A010 & 05 38 22.567  & -69 05 54.01 & 06$/$07$/$2013 & 14:33:07 & 2220.000 & 4 & ACS$/$WFC & F658N & 6583.956 \\  
JBY02B010 & 05 38 39.445  & -69 07 30.97 & 06$/$07$/$2013 & 20:55:29 & 2220.000 & 4 & ACS$/$WFC & F658N & 6583.956 \\  
JBY02C010 & 05 38 56.323  & -69 09 07.82 & 06$/$11$/$2013 & 18:59:39 & 2220.000 & 4 & ACS$/$WFC & F658N & 6583.956 \\  
JBY02D010 & 05 39 13.147  & -69 10 19.49 & 06$/$12$/$2013 & 17:16:09 & 2220.000 & 4 & ACS$/$WFC & F658N & 6583.956 \\ 
JBY069010 & 05 38 19.716  & -69 00 08.84 & 05$/$28$/$2013 & 11:21:23 & 2220.000 & 4 & ACS$/$WFC & F658N & 6583.956 \\  
IBY01N020 & 05 37 46.039  & -69 05 19.30 & 05$/$29$/$2013 & 17:21:51 & 1298.465 & 2 & WFC3$/$IR & F110W & 11534.459 \\  
IBY01N030 & 05 37 46.377  & -69 04 51.07 & 05$/$29$/$2013 & 18:14:36 & 1598.466 & 2 & WFC3$/$IR & F160W & 15369.176 \\ 
IBY01O020 & 05 38 01.824  & -69 06 53.35 & 05$/$22$/$2013 & 10:19:43 & 1298.465 & 2 & WFC3$/$IR & F110W & 11534.459 \\  
IBY01O030 & 05 38 01.332  & -69 06 25.19 & 05$/$22$/$2013 & 12:45:56 & 1598.466 & 2 & WFC3$/$IR & F160W & 15369.176 \\  
IBY01P020 & 05 38 21.522  & -69 08 22.52 & 05$/$22$/$2013 & 23:05:21 & 1298.465 & 2 & WFC3$/$IR & F110W & 11534.459 \\  
IBY01P030 & 05 38 19.630  & -69 07 56.10 & 05$/$29$/$2013 & 23:29:29 & 1598.466 & 2 & WFC3$/$IR & F160W & 15369.176 \\  
IBY01Q020 & 05 38 36.482  & -69 10 09.59 & 05$/$30$/$2013 & 16:40:49 & 1298.465 & 2 & WFC3$/$IR & F110W & 11534.459 \\  
IBY01Q030 & 05 38 36.821  & -69 09 41.36 & 05$/$30$/$2013 & 17:04:57 & 1598.466 & 2 & WFC3$/$IR & F160W & 15369.176 \\  
IBY01R020 & 05 38 53.296  & -69 11 46.15 & 05$/$31$/$2013 & 12:11:13 & 1298.465 & 2 & WFC3$/$IR & F110W & 11534.459 \\  
IBY01R030 & 05 38 53.636  & -69 11 17.92 & 05$/$31$/$2013 & 13:17:24 & 1598.466 & 2 & WFC3$/$IR & F160W & 15369.176 \\  
IBY01S020 & 05 37 13.173  & -69 05 01.60 & 06$/$01$/$2013 & 12:30:42 & 1298.465 & 2 & WFC3$/$IR & F110W & 11534.459 \\  
IBY01S030 & 05 37 13.511  & -69 04 33.37 & 06$/$01$/$2013 & 13:26:50 & 1598.466 & 2 & WFC3$/$IR & F160W & 15369.176 \\ 
IBY01T020 & 05 37 29.987  & -69 06 38.67 & 06$/$02$/$2013 & 18:00:30 & 1298.465 & 2 & WFC3$/$IR & F110W & 11534.459 \\ 
IBY01T030 & 05 37 30.326  & -69 06 10.44 & 06$/$02$/$2013 & 18:24:38 & 1598.466 & 2 & WFC3$/$IR & F160W & 15369.176 \\  
IBY01U020 & 05 37 47.743  & -69 08 15.51 & 06$/$10$/$2013 & 14:17:55 & 1298.465 & 2 & WFC3$/$IR & F110W & 11534.459 \\  
IBY01U030 & 05 37 47.140  & -69 07 47.41 & 06$/$10$/$2013 & 15:12:03 & 1598.466 & 2 & WFC3$/$IR & F160W & 15369.176 \\  
IBY01V020 & 05 38 03.616  & -69 09 52.50 & 06$/$03$/$2013 & 12:59:47 & 1298.465 & 2 & WFC3$/$IR & F110W & 11534.459 \\ 
IBY01V030 & 05 38 03.955  & -69 09 24.27 & 06$/$03$/$2013 & 13:23:55 & 1598.466 & 2 & WFC3$/$IR & F160W & 15369.176 \\ 
IBY01W020 & 05 38 20.430  & -69 11 29.26 & 06$/$11$/$2013 & 04:01:03 & 1298.465 & 2 & WFC3$/$IR & F110W & 11534.459 \\  
IBY01W030 & 05 38 20.770  & -69 11 01.03 & 06$/$11$/$2013 & 04:25:11 & 1598.466 & 2 & WFC3$/$IR & F160W & 15369.176 \\  
IBY01X020 & 05 38 37.244  & -69 13 05.92 & 06$/$04$/$2013 & 11:13:37 & 1298.465 & 2 & WFC3$/$IR & F110W & 11534.459 \\  
IBY01X030 & 05 38 37.584  & -69 12 37.69 & 06$/$04$/$2013 & 11:37:45 & 1598.466 & 2 & WFC3$/$IR & F160W & 15369.176 \\  
IBY01Y020 & 05 36 57.121  & -69 06 20.78 & 06$/$04$/$2013 & 16:20:55 & 1298.465 & 2 & WFC3$/$IR & F110W & 11534.459 \\ 
IBY01Y030 & 05 36 57.459  & -69 05 52.55 & 06$/$04$/$2013 & 16:45:03 & 1598.466 & 2 & WFC3$/$IR & F160W & 15369.176 \\  
IBY01Z020 & 05 37 13.935  & -69 07 57.95 & 06$/$06$/$2013 & 06:07:27 & 1298.465 & 2 & WFC3$/$IR & F110W & 11534.459 \\ 
IBY01Z030 & 05 37 14.274  & -69 07 29.72 & 06$/$06$/$2013 & 06:31:35 & 1598.466 & 2 & WFC3$/$IR & F160W & 15369.176 \\  
IBY02A020 & 05 37 30.749  & -69 09 35.01 & 06$/$07$/$2013 & 14:32:02 & 1298.465 & 2 & WFC3$/$IR & F110W & 11534.459 \\  
IBY02A030 & 05 37 31.088  & -69 09 06.78 & 06$/$07$/$2013 & 15:32:07 & 1598.466 & 2 & WFC3$/$IR & F160W & 15369.176 \\  
IBY02B020 & 05 37 47.564  & -69 11 11.97 & 06$/$07$/$2013 & 20:54:24 & 1298.465 & 2 & WFC3$/$IR & F110W & 11534.459 \\  
IBY02B030 & 05 37 47.903  & -69 10 43.74 & 06$/$07$/$2013 & 21:54:11 & 1598.466 & 2 & WFC3$/$IR & F160W & 15369.176 \\   
IBY02C020 & 05 38 04.378  & -69 12 48.83 & 06$/$11$/$2013 & 18:58:34 & 1298.465 & 2 & WFC3$/$IR & F110W & 11534.459 \\  
IBY02C030 & 05 38 04.718  & -69 12 20.60 & 06$/$11$/$2013 & 19:52:43 & 1598.466 & 2 & WFC3$/$IR & F160W & 15369.176 \\  
IBY02D020 & 05 38 24.165  & -69 14 19.25 & 06$/$12$/$2013 & 17:15:04 & 1298.465 & 2 & WFC3$/$IR & F110W & 11534.459 \\ 
IBY02D030 & 05 38 24.135  & -69 13 50.97 & 06$/$12$/$2013 & 18:11:09 & 1598.466 & 2 & WFC3$/$IR & F160W & 15369.176 \\  
IBY069020 & 05 37 28.125  & -69 03 49.84 & 05$/$28$/$2013 & 11:20:18 & 1298.465 & 2 & WFC3$/$IR & F110W & 11534.459 \\  
IBY069030 & 05 37 28.462  & -69 03 21.61 & 05$/$28$/$2013 & 12:19:11 & 1598.466 & 2 & WFC3$/$IR & F160W & 15369.176 \\  
IBY03A020 & 05 39 00.765  & -69 08 22.47 & 12$/$12$/$2012 & 03:33:41 & 1298.465 & 2 & WFC3$/$IR & F110W & 11534.459 \\  
IBY03A030 & 05 39 00.799  & -69 08 50.76 & 12$/$12$/$2012 & 03:57:49 & 1598.466 & 2 & WFC3$/$IR & F160W & 15369.176 \\   
IBY03B020 & 05 38 43.186  & -69 06 49.63 & 12$/$12$/$2012 & 08:56:26 & 1298.465 & 2 & WFC3$/$IR & F110W & 11534.459 \\  
IBY03B030 & 05 38 43.214  & -69 07 17.91 & 12$/$12$/$2012 & 09:54:40 & 1598.466 & 2 & WFC3$/$IR & F160W & 15369.176 \\  
IBY03C020 & 05 38 25.648  & -69 05 16.68 & 12$/$12$/$2012 & 20:06:19 & 1298.465 & 2 & WFC3$/$IR & F110W & 11534.459 \\   
IBY03C030 & 05 38 25.670  & -69 05 44.96 & 12$/$12$/$2012 & 21:37:28 & 1598.466 & 2 & WFC3$/$IR & F160W & 15369.176 \\  
IBY03D020 & 05 38 08.152  & -69 03 43.61 & 12$/$13$/$2012 & 21:00:36 & 1298.465 & 2 & WFC3$/$IR & F110W & 11534.459 \\ 
IBY03D030 & 05 38 08.168  & -69 04 11.90 & 12$/$13$/$2012 & 21:47:08 & 1598.466 & 2 & WFC3$/$IR & F160W & 15369.176 \\  
IBY03E020 & 05 37 50.697  & -69 02 10.44 & 12$/$13$/$2012 & 04:19:38 & 1298.465 & 2 & WFC3$/$IR & F110W & 11534.459 \\ 
IBY03E030 & 05 37 50.706  & -69 02 38.72 & 12$/$13$/$2012 & 05:16:41 & 1598.466 & 2 & WFC3$/$IR & F160W & 15369.176 \\  
IBY03F020 & 05 39 33.742  & -69 08 31.86 & 12$/$14$/$2012 & 14:57:02 & 1298.465 & 2 & WFC3$/$IR & F110W & 11534.459 \\  
IBY03F030 & 05 39 33.788  & -69 09 00.15 & 12$/$14$/$2012 & 15:21:10 & 1598.466 & 2 & WFC3$/$IR & F160W & 15369.176 \\  
IBY03G020 & 05 39 16.122  & -69 06 59.23 & 12$/$13$/$2012 & 13:07:40 & 1298.465 & 2 & WFC3$/$IR & F110W & 11534.459 \\  
IBY03G030 & 05 39 16.162  & -69 07 27.52 & 12$/$13$/$2012 & 13:31:48 & 1598.466 & 2 & WFC3$/$IR & F160W & 15369.176 \\  
IBY03H020 & 05 38 58.543  & -69 05 26.49 & 12$/$17$/$2012 & 15:00:52 & 1298.465 & 2 & WFC3$/$IR & F110W & 11534.459 \\  
IBY03H030 & 05 38 58.577  & -69 05 54.77 & 12$/$17$/$2012 & 16:13:51 & 1598.466 & 2 & WFC3$/$IR & F160W & 15369.176 \\  
IBY03I020 & 05 38 41.006  & -69 03 53.63 & 12$/$14$/$2012 & 09:43:38 & 1298.465 & 2 & WFC3$/$IR & F110W & 11534.459 \\   
IBY03I030 & 05 38 41.033  & -69 04 21.92 & 12$/$14$/$2012 & 10:07:46 & 1598.466 & 2 & WFC3$/$IR & F160W & 15369.176 \\   
IBY03J020 & 05 38 23.510  & -69 02 20.67 & 12$/$15$/$2012 & 07:09:47 & 1298.465 & 2 & WFC3$/$IR & F110W & 11534.459 \\   
IBY03J030 & 05 38 23.531  & -69 02 48.95 & 12$/$15$/$2012 & 08:01:28 & 1598.466 & 2 & WFC3$/$IR & F160W & 15369.176 \\  
IBY03K020 & 05 38 06.055  & -69 00 47.59 & 12$/$16$/$2012 & 07:20:27 & 1298.465 & 2 & WFC3$/$IR & F110W & 11534.459 \\  
IBY03K030 & 05 38 06.070  & -69 01 15.87 & 12$/$16$/$2012 & 08:09:42 & 1598.466 & 2 & WFC3$/$IR & F160W & 15369.176 \\  
IBY03L020 & 05 39 49.066  & -69 07 08.44 & 12$/$16$/$2012 & 18:16:19 & 1298.465 & 2 & WFC3$/$IR & F110W & 11534.459 \\  
IBY03L030 & 05 39 49.117  & -69 07 36.72 & 12$/$16$/$2012 & 19:38:33 & 1598.466 & 2 & WFC3$/$IR & F160W & 15369.176 \\  
IBY03M020 & 05 39 31.753  & -69 05 11.27 & 12$/$09$/$2012 & 03:50:39 & 1298.465 & 2 & WFC3$/$IR & F110W & 11534.459 \\   
IBY03M030 & 05 39 32.378  & -69 05 39.36 & 12$/$09$/$2012 & 04:14:47 & 1598.466 & 2 & WFC3$/$IR & F160W & 15369.176 \\  
IBY03N020 & 05 39 13.868  & -69 04 03.26 & 12$/$18$/$2012 & 21:19:16 & 1298.465 & 2 & WFC3$/$IR & F110W & 11534.459 \\  
IBY03N030 & 05 39 13.907  & -69 04 31.55 & 12$/$18$/$2012 & 21:59:59 & 1598.466 & 2 & WFC3$/$IR & F160W & 15369.176 \\   
IBY03O020 & 05 38 56.331  & -69 02 30.51 & 12$/$17$/$2012 & 08:37:55 & 1298.465 & 2 & WFC3$/$IR & F110W & 11534.459 \\  
IBY03O030 & 05 38 56.364  & -69 02 58.79 & 12$/$17$/$2012 & 09:23:49 & 1598.466 & 2 & WFC3$/$IR & F160W & 15369.176 \\  
IBY03P020 & 05 38 38.836  & -69 00 57.64 & 12$/$17$/$2012 & 23:43:25 & 1298.465 & 2 & WFC3$/$IR & F110W & 11534.459 \\  
IBY03P030 & 05 38 38.863  & -69 01 25.92 & 12$/$18$/$2012 & 00:07:33 & 1598.466 & 2 & WFC3$/$IR & F160W & 15369.176 \\  
IBY03Q020 & 05 38 21.381  & -68 59 24.66 & 12$/$18$/$2012 & 06:58:18 & 1298.465 & 2 & WFC3$/$IR & F110W & 11534.459 \\  
IBY03Q030 & 05 38 21.402  & -68 59 52.94 & 12$/$18$/$2012 & 07:40:52 & 1598.466 & 2 & WFC3$/$IR & F160W & 15369.176 \\  
IBY087020 & 05 39 18.385  & -69 09 55.20 & 12$/$19$/$2012 & 03:38:09 & 1298.465 & 2 & WFC3$/$IR & F110W & 11534.459 \\  
IBY087030 & 05 39 18.426  & -69 10 23.48 & 12$/$19$/$2012 & 04:21:42 & 1598.466 & 2 & WFC3$/$IR & F160W & 15369.176 \\  
JBY03A010 & 05 38 11.741  & -69 12 21.56 & 12$/$12$/$2012 & 03:34:45 & 2220.000 & 4 & ACS$/$WFC & F658N & 6583.956 \\  
JBY03B010 & 05 37 54.166  & -69 10 48.40 & 12$/$12$/$2012 & 08:57:30 & 2220.000 & 4 & ACS$/$WFC & F658N & 6583.956 \\ 
JBY03C010 & 05 37 36.634  & -69 09 15.14 & 12$/$12$/$2012 & 20:07:23 & 2220.000 & 4 & ACS$/$WFC & F658N & 6583.956 \\  
JBY03D010 & 05 37 19.142  & -69 07 41.76 & 12$/$13$/$2012 & 21:01:40 & 2220.000 & 4 & ACS$/$WFC & F658N & 6583.956 \\  
JBY03E010 & 05 37 01.693  & -69 06 08.28 & 12$/$13$/$2012 & 04:20:42 & 2220.000 & 4 & ACS$/$WFC & F658N & 6583.956 \\  
JBY03F010 & 05 38 44.812  & -69 12 31.53 & 12$/$14$/$2012 & 14:58:06 & 2220.000 & 4 & ACS$/$WFC & F658N & 6583.956 \\   
JBY03G010 & 05 38 27.196  & -69 10 58.59 & 12$/$13$/$2012 & 13:08:44 & 2220.000 & 4 & ACS$/$WFC & F658N & 6583.956 \\  
JBY03H010 & 05 38 09.622  & -69 09 25.54 & 12$/$17$/$2012 & 15:01:56 & 2220.000 & 4 & ACS$/$WFC & F658N & 6583.956 \\  
JBY03I010 & 05 37 52.090  & -69 07 52.37 & 12$/$14$/$2012 & 09:44:42 & 2220.000 & 4 & ACS$/$WFC & F658N & 6583.956 \\  
JBY03J010 & 05 37 34.599  & -69 06 19.09 & 12$/$15$/$2012 & 07:10:51 & 2220.000 & 4 & ACS$/$WFC & F658N & 6583.956 \\  
JBY03K010 & 05 37 17.149  & -69 04 45.70 & 12$/$16$/$2012 & 07:21:31 & 2220.000 & 4 & ACS$/$WFC & F658N & 6583.956 \\  
JBY03L010 & 05 39 00.235  & -69 11 08.38 & 12$/$16$/$2012 & 18:17:23 & 2220.000 & 4 & ACS$/$WFC & F658N & 6583.956 \\  
JBY03M010 & 05 38 48.157  & -69 09 38.14 & 12$/$09$/$2012 & 03:51:43 & 2220.000 & 4 & ACS$/$WFC & F658N & 6583.956 \\  
JBY03N010 & 05 38 25.046  & -69 08 02.58 & 12$/$18$/$2012 & 21:20:20 & 2220.000 & 4 & ACS$/$WFC & F658N & 6583.956 \\  
JBY03O010 & 05 38 07.513  & -69 06 29.51 & 12$/$17$/$2012 & 08:38:59 & 2220.000 & 4 & ACS$/$WFC & F658N & 6583.956 \\  
JBY03P010 & 05 37 50.022  & -69 04 56.33 & 12$/$17$/$2012 & 23:44:29 & 2220.000 & 4 & ACS$/$WFC & F658N & 6583.956 \\  
JBY03Q010 & 05 37 32.573  & -69 03 23.04 & 12$/$18$/$2012 & 06:59:22 & 2220.000 & 4 & ACS$/$WFC & F658N & 6583.956 \\ 
JBY087010 & 05 38 29.357  & -69 13 54.60 & 12$/$19$/$2012 & 03:39:13 & 2220.000 & 4 & ACS$/$WFC & F658N & 6583.956 \\  
IBY001020 & 05 39 06.435  & -69 04 43.03 & 09$/$17$/$2013 & 16:28:16 & 1408.000 & 2 & WFC3$/$UVIS & F336W & 3354.874 \\  
IBY001030 & 05 38 56.323  & -69 04 26.74 & 09$/$17$/$2013 & 17:52:45 & 1164.000 & 2 & WFC3$/$UVIS & F275W & 2706.949 \\  
IBY07B020 & 05 38 41.214  & -69 05 43.32 & 09$/$23$/$2013 & 21:28:23 & 1408.000 & 2 & WFC3$/$UVIS & F336W & 3354.874 \\  
IBY07B030 & 05 38 31.094  & -69 05 27.03 & 09$/$23$/$2013 & 22:23:09 & 1164.000 & 2 & WFC3$/$UVIS & F275W & 2706.949 \\  
IBY07C020 & 05 38 07.740  & -69 06 52.21 & 09$/$10$/$2013 & 14:00:33 & 1408.000 & 2 & WFC3$/$UVIS & F336W & 3354.874 \\ 
IBY07C030 & 05 37 57.163  & -69 06 51.48 & 09$/$10$/$2013 & 15:16:40 & 1164.000 & 2 & WFC3$/$UVIS & F275W & 2706.949 \\  
IBY07D020 & 05 37 50.714  & -69 07 36.08 & 09$/$25$/$2013 & 00:32:55 & 1408.000 & 2 & WFC3$/$UVIS & F336W & 3354.874 \\  
IBY07D030 & 05 37 40.580  & -69 07 19.79 & 09$/$25$/$2013 & 01:04:46 & 1164.000 & 2 & WFC3$/$UVIS & F275W & 2706.949 \\  
IBY07E020 & 05 37 23.765  & -69 08 17.46 & 09$/$21$/$2013 & 21:51:59 & 1408.000 & 2 & WFC3$/$UVIS & F336W & 3354.874 \\  
IBY07E030 & 05 37 13.625  & -69 08 01.17 & 09$/$21$/$2013 & 23:24:21 & 1164.000 & 2 & WFC3$/$UVIS & F275W & 2706.949 \\  
IBY07F020 & 05 36 58.097  & -69 09 10.47 & 09$/$24$/$2013 & 22:26:59 & 1408.000 & 2 & WFC3$/$UVIS & F336W & 3354.874 \\  
IBY07F030 & 05 36 47.950  & -69 08 54.18 & 09$/$24$/$2013 & 23:08:11 & 1164.000 & 2 & WFC3$/$UVIS & F275W & 2706.949 \\  
IBY07G020 & 05 38 52.502  & -69 02 05.35 & 09$/$21$/$2013 & 04:21:22 & 1408.000 & 2 & WFC3$/$UVIS & F336W & 3354.874 \\  
IBY07G030 & 05 38 42.410  & -69 01 49.06 & 09$/$21$/$2013 & 04:53:13 & 1164.000 & 2 & WFC3$/$UVIS & F275W & 2706.949 \\ 
IBY07H020 & 05 38 26.834  & -69 02 59.19 & 09$/$25$/$2013 & 22:22:22 & 1408.000 & 2 & WFC3$/$UVIS & F336W & 3354.874 \\  
IBY07H030 & 05 38 16.735  & -69 02 42.90 & 09$/$25$/$2013 & 23:00:17 & 1164.000 & 2 & WFC3$/$UVIS & F275W & 2706.949 \\  
IBY07I020 & 05 38 01.167  & -69 03 52.79 & 09$/$21$/$2013 & 16:05:06 & 1408.000 & 2 & WFC3$/$UVIS & F336W & 3354.874 \\  
IBY07I030 & 05 37 51.061  & -69 03 36.50 & 09$/$21$/$2013 & 17:32:17 & 1164.000 & 2 & WFC3$/$UVIS & F275W & 2706.949 \\ 
IBY07J020 & 05 37 35.499  & -69 04 46.15 & 09$/$26$/$2013 & 00:10:56 & 1408.000 & 2 & WFC3$/$UVIS & F336W & 3354.874 \\  
IBY07J030 & 05 37 25.386  & -69 04 29.86 & 09$/$26$/$2013 & 00:49:39 & 1164.000 & 2 & WFC3$/$UVIS & F275W & 2706.949 \\  
IBY07K020 & 05 37 10.123  & -69 05 53.03 & 09$/$26$/$2013 & 14:42:57 & 1408.000 & 2 & WFC3$/$UVIS & F336W & 3354.874 \\  
IBY07K030 & 05 36 59.739  & -69 05 42.49 & 09$/$26$/$2013 & 15:38:36 & 1164.000 & 2 & WFC3$/$UVIS & F275W & 2706.949 \\  
IBY07L020 & 05 36 44.164  & -69 06 32.15 & 09$/$21$/$2013 & 19:16:16 & 1408.000 & 2 & WFC3$/$UVIS & F336W & 3354.874 \\  
IBY07L030 & 05 36 34.037  & -69 06 15.86 & 09$/$21$/$2013 & 20:14:00 & 1164.000 & 2 & WFC3$/$UVIS & F275W & 2706.949 \\  
\hline
\hline
\multicolumn{10}{}{} \\
\multicolumn{4}{}{} & \multicolumn{1}{c}{GO-12499} & \multicolumn{5}{}{} \\
\multicolumn{10}{}{} \\
\hline
IBSF01020 & 05 39 01.607 & -69 03 07.51 & 10$/$03$/$2011 & 17:32:38 & 2639.000 & 4 & WFC3$/$UVIS &  F775W & 7647.629 \\ 
IBSF02020 & 05 38 45.473 & -69 04 09.22 & 10$/$03$/$2011 & 23:56:07 & 2639.000 & 4 & WFC3$/$UVIS &  F775W & 7647.629 \\ 
IBSF03020 & 05 38 29.314 & -69 05 10.83 & 10$/$04$/$2011 & 14:18:11 & 2639.000 & 4 & WFC3$/$UVIS &  F775W & 7647.629 \\ 
IBSF04020 & 05 38 11.082 & -69 06 37.57 & 10$/$08$/$2011 & 13:09:23 & 2639.000 & 4 & WFC3$/$UVIS &  F775W & 7647.629 \\ 
IBSF05020 & 05 37 56.920 & -69 07 13.77 & 10$/$04$/$2011 & 17:16:27 & 2639.000 & 4 & WFC3$/$UVIS &  F775W & 7647.629 \\  
IBSF06020 & 05 37 35.134 & -69 08 08.28 & 10$/$05$/$2011 & 09:28:03 & 2639.000 & 4 & WFC3$/$UVIS &  F775W & 7647.629 \\ 
IBSF07020 & 05 37 24.424 & -69 09 16.32 & 10$/$05$/$2011 & 19:03:07 & 2639.000 & 4 & WFC3$/$UVIS &  F775W & 7647.629 \\ 
IBSF09020 & 05 38 42.710 & -69 00 45.31 & 10$/$05$/$2011 & 23:50:49 & 2639.000 & 4 & WFC3$/$UVIS &  F775W & 7647.629 \\ 
IBSF10020 & 05 38 26.590 & -69 01 46.90 & 10$/$06$/$2011 & 03:15:22 & 2639.000 & 4 & WFC3$/$UVIS &  F775W & 7647.629 \\ 
IBSF11020 & 05 38 10.445 & -69 02 48.40 & 10$/$06$/$2011 & 12:36:51 & 2639.000 & 4 & WFC3$/$UVIS &  F775W & 7647.629 \\ 
IBSF12020 & 05 37 54.275 & -69 03 49.81 & 10$/$06$/$2011 & 15:21:17 & 2639.000 & 4 & WFC3$/$UVIS &  F775W & 7647.629 \\ 
IBSF13020$^\ast$ & 05 37 39.196 & -69 04 44.99 & 10$/$06$/$2011 & 18:46:44 &   35.000 & 1 & WFC3$/$UVIS &  F775W & 7647.629 \\ 
IBSF14020 & 05 37 21.961 & -69 05 54.51 & 10$/$07$/$2011 & 18:57:36 & 2639.000 & 4 & WFC3$/$UVIS &  F775W & 7647.629 \\ 
IBSF15020 & 05 37 05.614 & -69 06 53.46 & 10$/$07$/$2011 & 21:18:00 & 2639.000 & 4 & WFC3$/$UVIS &  F775W & 7647.629 \\ 
IBSF16020 & 05 36 49.343 & -69 07 54.48 & 10$/$07$/$2011 & 23:45:17 & 2639.000 & 4 & WFC3$/$UVIS &  F775W & 7647.629 \\ 
IBSF53020$^\star$ & 05 37 39.196 & -69 04 44.99 & 10$/$29$/$2011 & 22:45:17 & 2132.000 & 4 & WFC3$/$UVIS &  F775W & 7647.629 \\ 
JBSF01010 & 05 39 38.121 & -69 08 06.41 & 10$/$03$/$2011 & 17:33:43 & 2329.000 & 4 & ACS$/$WFC & F775W & 7693.671 \\ 
JBSF02010 & 05 39 21.954 & -69 09 08.33 & 10$/$03$/$2011 & 23:57:12 & 2329.000 & 4 & ACS$/$WFC & F775W & 7693.671 \\ 
JBSF03010 & 05 39 05.762 & -69 10 10.16 & 10$/$04$/$2011 & 14:19:16 & 2329.000 & 4 & ACS$/$WFC & F775W & 7693.671 \\ 
JBSF04010 & 05 38 37.186 & -69 12 06.39 & 10$/$08$/$2011 & 13:10:28 & 2329.000 & 4 & ACS$/$WFC & F775W & 7693.671 \\ 
JBSF05010 & 05 38 33.301 & -69 12 13.52 & 10$/$04$/$2011 & 17:17:32 & 2329.000 & 4 & ACS$/$WFC & F775W & 7693.671 \\  
JBSF06010 & 05 38 01.117 & -69 13 37.45 & 10$/$05$/$2011 & 09:29:08 & 2329.000 & 4 & ACS$/$WFC & F775W & 7693.671 \\ 
JBSF07010 & 05 38 00.738 & -69 14 16.51 & 10$/$05$/$2011 & 19:04:12 & 2329.000 & 4 & ACS$/$WFC & F775W & 7693.671 \\ 
JBSF09010 & 05 39 19.085 & -69 05 44.46 & 10$/$05$/$2011 & 23:51:54 & 2329.000 & 4 & ACS$/$WFC & F775W & 7693.671 \\  
JBSF10010 & 05 39 02.933 & -69 06 46.27 & 10$/$06$/$2011 & 03:16:27 & 2329.000 & 4 & ACS$/$WFC & F775W & 7693.671 \\ 
JBSF11010 & 05 38 46.755 & -69 07 47.98 & 10$/$06$/$2011 & 12:37:56 & 2329.000 & 4 & ACS$/$WFC & F775W & 7693.671 \\ 
JBSF12010 & 05 38 30.552 & -69 08 49.60 & 10$/$06$/$2011 & 15:22:22 & 2329.000 & 4 & ACS$/$WFC & F775W & 7693.671 \\ 
JBSF13010$^\ast$ & 05 38 05.129 & -69 10 14.12 & 10$/$06$/$2011 & 18:47:49 &   32.000 & 1 & ACS$/$WFC & F775W & 7693.671 \\ 
JBSF14010 & 05 37 59.148 & -69 10 51.30 & 10$/$07$/$2011 & 18:58:41 & 2329.000 & 4 & ACS$/$WFC & F775W & 7693.671 \\ 
JBSF15010 & 05 37 41.790 & -69 11 53.89 & 10$/$07$/$2011 & 21:19:05 & 2329.000 & 4 & ACS$/$WFC & F775W & 7693.671 \\ 
JBSF16010 & 05 37 25.485 & -69 12 55.12 & 10$/$07$/$2011 & 23:46:22 & 2329.000 & 4 & ACS$/$WFC & F775W & 7693.671 \\ 
JBSF53010$^\star$ & 05 38 05.129 & -69 10 14.12 & 10$/$29$/$2011 & 22:46:22 & 1952.000 & 4 & ACS$/$WFC & F775W & 7693.671 \\ 
\enddata
\end{deluxetable}

\begin{table}[htbp]
  \centering
  \begin{tabular}{@{} cc @{}}
    \hline
    Filter & Percentage \\ 
    \hline
    F775W$_{combined}$ & 100\% \\ 
    F555W & 89\% \\ 
    F658N & 85\% \\ 
    F275W & 82\% \\ 
    F336W & 79\% \\ 
    F110W & 76\% \\ 
    F160W & 75\% \\ 
    F775W$_{ACS}$ & 57\% \\ 
    F775W$_{UVIS}$ &49\%  \\ 
    \hline
  \end{tabular}
  \caption{Field of view covered by each filter with the respect to the two WFC3 and ACS F775W filters combined together.}
  \label{mappa_ratio}
\end{table}

\begin{deluxetable}{lllllllllllllll}
\tablecolumns{15}
\tabletypesize{\scriptsize}
\tablecaption{HTTP Photometric Catalog\label{the_catalog}}
\tablewidth{0pt}
\tablehead{
\multicolumn{2}{}{} & \multicolumn{4}{c}{F555W} & \multicolumn{4}{c}{F775W$_{ACS}$} & \multicolumn{5}{}{} \\
\colhead{ID} & \colhead{} &  \colhead{m} &  \colhead{err} & \colhead{Q} & \colhead{f} & \colhead{m} & \colhead{err} & \colhead{Q} & \colhead{f} &  \colhead{} &  \colhead{X} & \colhead{Y} & \colhead{RA} & \colhead{Dec }\\
\colhead{} & \colhead{} &  \colhead{} &  \colhead{} &  \colhead{} & \colhead{} & \colhead{} & \colhead{} & \colhead{} & \colhead{} &\colhead{} &  \colhead{pixels} & \colhead{pixels} & \colhead{degr} & \colhead{degr} 
}

\startdata
SABBI 053850.819-691404.92	& ... &      24.528  &        0.349       &    0.444   &  1	 &   24.163     &     0.016  &    0.992  &   1   & ... &  17895.980 &  3636.450  &	84.711746  & -69.234700\\
SABBI 053851.156-691400.99	& ... &      24.896  &        0.921       &    0.373   &  1	 &   25.331     &     0.139  &    0.930  &   1   & ... &  17801.800 &  3691.600  &	84.713150  & -69.233608\\
SABBI 053850.777-691403.44	& ... &      19.384  &        0.018       &    0.989   &  1	 &   19.225     &     0.005  &    1.000  &   1   & ... &  17878.980 &  3670.330  &	84.711571  & -69.234289\\
SABBI 053850.977-691405.11	& ... &      23.503  &        0.342       &    0.758   &  1	 &   23.523     &     0.014  &    0.997  &   1   & ... &  17881.240 &  3620.410  &	84.712404  & -69.234753\\
SABBI 053850.848-691404.25	& ... &      24.860  &        0.648       &    0.309   &  1	 &   25.089     &     0.004  &    0.952  &   1   & ... &  17883.060 &  3648.090  &	84.711867  & -69.234514\\
SABBI 053851.098-691400.90	& ... &      25.496  &        1.185       &    0.205   &  1	 &   25.108     &     0.120  &    0.942  &   1   & ... &  17806.840 &  3697.850  &	84.712908  & -69.233583\\
SABBI 053850.553-691403.61	& ... &      22.614  &        0.184       &    0.913   &  1	 &   22.579     &     0.041  &    0.999  &   1   & ... &  17906.060 &  3684.190  &	84.710638  & -69.234336\\
SABBI 053850.382-691402.62	& ... &      24.774  &        0.359       &    0.434   &  1	 &   24.473     &     0.026  &    0.991  &   1   & ... & 17910.450 &  3717.820  &	84.709925  & -69.234061\\
SABBI 053850.525-691404.28	& ... &      25.702  &        0.401       &    0.071   &  1	 &   25.169     &     0.121  &    0.958  &   1   & ... &  17918.860 &  3672.460  &	84.710521  & -69.234522\\
SABBI 053850.700-691406.15	& ...  &      21.985  &        0.071       &    0.967   &  1	 &   21.540     &     0.031  &    0.999  &   1   & ... &   17926.770 &  3620.400  &	84.711250  & -69.235042\\
\enddata
\end{deluxetable}



\begin{thebibliography}{}

\bibitem[Andersen et al.(2009)]{andersen09}
Andersen, M., Zinnecker, H., Moneti, A., et al. 2009, ApJ, 707, 1347 

\bibitem[Anderson \& King(2006)]{anderson06}
Anderson, J., \& King, J.R. 2006, STSCI Institute Science Report ACS 2006-01 (Baltimore, MD: STScI)

\bibitem[Anderson \& King(2000)]{anderson00}
Anderson, J., \& King, I.R. 2000, PASP, 112, 1360

\bibitem[Anderson et al.(2008)]{anderson08}
Anderson, J., Sarajedini, A., Bedin, L.R., et al. 2008, AJ, 135, 2055

\bibitem[Baggett \& Anderson(2012)]{baggett12}
Baggett, S., \& Anderson, J. 2012, STSCI Institute Science Report WFC3 2012-12 (Baltimore, MD: STScI)

\bibitem[Bate(2012)]{bate12}
Bate, M.R. 2012, MNRAS, 419, 3115

\bibitem[Bernard et al.(2008)]{bernard08}
Bernard, J.-P., Reach, W. T., Paradis, D., et al. 2008, AJ, 136, 919

\bibitem[Bertelli et al.(1994)]{bertelli94}
Bertelli, G., Bressan, A., Chiosi, C., Fagotto, F., Nasi, E. 1994, A\&AS, 106, 275

\bibitem[Bestenlehner et al.(2011)]{bestenlehner11}
Bestenlehner, J. M., Vink, J. S., Gr\"{a}fener, G., et al. 2011, A\&A, 530, L14

\bibitem[Bourque \& Kozurina-Platais(2013)]{bourque13}
Bourque, M., \& Kozurina-Platais, V. 2013, STSCI Institute Science Report WFC3 2013-03 (Baltimore, MD: STScI)

\bibitem[Bosch et al.(2001)]{bosch01}
Bosch, G., Selman, F., Melnick, J., Terlevich, R. 2001, A\&A, 380, 137 

\bibitem[Brandner et al.(2001)]{brandner01}
Brandner, W., Grebel, E. K., Barb\'a, R. H., Walborn, N.R., Moneti, A. 2001, AJ, 122, 858

\bibitem[Bressan et al.(2012)]{bressan12}
Bressan, A., Marigo, P., Girardi, L., et al. 2012, MNRAS, 427, 127

\bibitem[Breysacher(1981)]{breysacher81}
Breysacher, J. 1981, A\&AS, 43, 203

\bibitem[Chen et al.(2014)]{chen14}
Chen, Y., Girardi, L., Bressan, A., et al. 2014, MNRAS, 444, 2525

\bibitem[Chu \& Kennicutt(1994)]{chu94}
Chu, You-Hua, \& Kennicutt, R.C., Jr. 1994, AJ, 108, 1696 

\bibitem[Cignoni et al.(2015)]{cignoni15}
Cignoni, M., Sabbi, E., van der Marel, R.P., et al. 2015, ApJ, (accepted), astro-ph/1505.04799

\bibitem[Cioni et al.(2011)]{cioni11}
Cioni, M.-R., Clementini, G., Girardi, L., et al. 2011, A\&A, 527A, 116

\bibitem[Crowther et al.(2010)]{crowther10}
Crowther, P. A., Schnurr, O., Hirschi, R., et al. 2010, MNRAS, 408, 731

\bibitem[Cusumano et al.(1998)]{cusumano98}
Cusumano, G., Maccarone, M. C., Mineo, T., et al. 1998, A\&A, 333, L55

\bibitem[de la Caille(1761)]{caille61}
de la Caille, N. L. 1761, RSPT, 52, 21

\bibitem[De Marchi et al.(2011)]{demarchi11}
De Marchi, G., Paresce, F., Panagia, N., et al. 2011, ApJ, 739, 27

\bibitem[De Marchi et al.(2014)]{demarchi14a}
De Marchi, G., Panagia, N., Girardi, L. 2014, MNRAS, 438, 1

\bibitem[De Marchi \& Panagia(2014)]{demarchi14b}
De Marchi, G., \& Panagia, N. 2014, MNRAS, 445, 1

\bibitem[de Mink, et al.(2014)]{demink14}
de Mink, S.E., Sana, H., Langer, N., Izzard, R.G., Schneider, F.R.N. 2014, ApJ, 782, 7

\bibitem[Doran et al.(2013)]{doran13}
Doran, E.I., Crowther, P.A., de Koter, A., et al. 2013, A\&A, 558, A134

\bibitem[Dufour et al.(1982)]{dufour82}
Dufour, R. J., Shields, G. A., \& Talbot, R. J., Jr. 1982, ApJ, 252, 461

\bibitem[Dunstall et al.(2015)]{dunstall15}
Dunstall, P.R., Dufton, P.L., Sana., H., et al. 2015, A\&A, (accepted), astro.ph/150507121

\bibitem[Evans et al.(2015)]{evans15}
Evans, C.J., Kennedy, M.B., Dufton, P.L., et al. 2015, A\&A, 574, 13

\bibitem[Evans et al.(2011)]{evans11}
Evans, C.J., Taylor, W., H\'enault-Brunet, V., et al. 2011, A\&A, 530A, 108

\bibitem[Evans et al.(2010)]{evans10}
Evans, C.J., Walborn, N.R., Crowther, P.A., et al. 2010, ApJL, 715, L74

\bibitem[Ferguson et al.(1996)]{ferguson96}
Ferguson, A.M.N., Wyse, R.F.G., Gallagher, J.S. 1996, AJ, 112, 256  

\bibitem[Gilliland(2004)]{gilliland04}
Gilliland, R. 2004, STSCI Institute Science Report  ACS 2004-01 (Baltimore, MD: STScI)

\bibitem[Gilliland et al.(2010)]{gilliland10}
Gilliland, R.L., Rajan, A., \& Deustua, S. 2010, STSCI Institute Science Report  WFC3 2010-10 (Baltimore, MD: STScI)

\bibitem[Girardi \& Salaris(2001)]{girardi01}
Girardi, L. \& Salaris, M. 2001, MNRAS, 323, 109

\bibitem[Grebel \& Chu(2000)]{grebel00}
Grebel, E.K., \& Chu, Y.-H. 2000, AJ, 111, 787

\bibitem[Gouliermis(2012)]{gouliermis12}
Gouliermis, D. 2012, SSR, 169, 1

\bibitem[Gullbring et al. (1998)]{gullbring98}
Gullbring, E., Hartmann, L., Briceno, C., Calvet, N. 1998, ApJ, 492, 323

\bibitem[Harris \& Zaritsky(2009)]{harris09}
Harris, J., \& Zaritsky, D. 2004, AJ, 138, 1243

\bibitem[Haschke et al.(2011)]{haschke11}
Haschke, R., Grebel, E.K., Duffau, S. 2011, AJ, 141, 158

\bibitem[Heckman et al.(2004)]{heckman04}
Heckman, T.M., Kauffmann, G., Brinchmann, J., et al. 2004, ApJ, 613, 109

\bibitem[Herschel(1847)]{herschel47}
Herschel, J. F. W. 1847, Results of Astronomical Observations Made during the
Years 1834, 5, 6, 7, 8, at the Cape of Good Hope; Being the Completion of a Telescopic Survey of the Whole Surface of the Visible Heavens, Commenced in 1825 (London: Smith, Elder, \& Co.)

\bibitem[Hog et al.(2000)]{hog00}
Hog, E., Fabricius, C., Makarov, V.V., et al. 2000, A\&A, 355, L27

\bibitem[Hunt \& Hirashita(2009)]{hunt09}
Hunt, L.K., Hirashita, H. 2009, A\&A, 507, 1327

\bibitem[Hunter et al(1996)]{hunter96}
Hunter, D., O'Neil, E. J., Lynds, R., et al. 1996, ApJL, 459, L27

\bibitem[Hunter et al.(1995)]{hunter95}
Hunter, D., Shaya, E. J., Holtzman, J. A., et al. 1995, ApJ, 448, 179

\bibitem[Kennicutt et al.(1989)]{kennicutt89}
Kennicutt, R.C., Edgar, B.K., Hodge, P.W. 1989, ApJ, 337, 761  

\bibitem[Kennicutt \& Hodge(1986)]{kennicutt86}
Kennicutt, R.C., \& Hodge, P.W. 1986, ApJ, 306, 130

\bibitem[Krumholz et al(2012)]{krumholz12}
Krumholz, M.R., Klein, R.I., McKee, C.F. 2012, ApJ, 754, 71

\bibitem[Kuhi(1974)]{kuhi74}
Kuhi, L. V. 1974, A\&AS, 15, 47

\bibitem[Lada \& Lada(2003)]{lada03}
Lada, C.J., \& Lada, E.A. 2003, ARA\&A, 41, 57

\bibitem[Leitherer(1998)]{leitherer98}
Leitherer, C. 1998, in Stellar Astrophysics for the Local Group: VIII Canary
Islands Winter School of Astrophysics, ed. A. Aparicio, A. Herrero, \& F.
Sanchez (Cambridge: Cambridge Univ. Press), 527

\bibitem[Long et al.,(2013a)]{long13a}
Long, K.S., Baggett, S.M., \& MacKenty, J.W. 2013, STSCI Institute Science Report  WFC3 2013-06 (Baltimore, MD:STScI)

\bibitem[Long et al.,(2013b)]{long13b}
Long, K.S., Baggett, S.M., \& MacKenty, J.W. 2013 STSCI Institute Science Report  WFC3 2013-07 (Baltimore, MD:STScI)

\bibitem[Marshall et al.(1998)]{marshall98}
Marshall, F.E., Gotthelf, E.V., Zhang, W., Middleditch, J., Wang, Q.D. 1998, ApJ, 499, L179

\bibitem[Massey \& Hunter(1998)]{massey98}
Massey, P., \& Hunter, D.A. 1998, ApJ, 493, 180

\bibitem[Meixner et al(2006)]{meixner06}
Meixner, M., Gordon, K.D., Indebetouw, R., et al.  2006, AJ, 132, 2268

\bibitem[Meixner et al(2010)]{meixner10}
Meixner, M., Galliano, F., Hony, S., et al. 2010, A\&A, 518, L71

\bibitem[Meurer et al.(1997)]{meurer97}
Meurer, G.R., Heckman, T.M., Lehnert, M.D., Leitherer, C., Lowenthal, J. 1997, AJ, 114, 54

\bibitem[Mignani et al.(2005)]{mignani05}
Mignani, R.P., Pulone, L., Iannicola, G, et al. 2005, 431, 659

\bibitem[Nikolaev et al.(2004)]{nikolaev04}
Nikolaev S., Drake, A.J., Keller, S.C., et al. 2004, ApJ,  601, 260 

\bibitem[Oey et al.(2003)]{oey03}
Oey, M.S., Parker, J.S., Mikles, V.J., Zhang, X. 2003, AJ, 126, 2317

\bibitem[Panagia et al.(1991)]{panagia91}
Panagia, N., Gilmozzi, R., Macchetto,  F., Adorf, H.-M., Kirshner, R. P. 1991, ApJ, 380, L23  

\bibitem[Pellegrini et al.(2011)]{pellegrini11}
Pellegrini, E.W., Baldwin, J.A., \& Ferland, G.J. 2011, ApJ, 738, 34

\bibitem[Pietrzy\'nski et al.(2013)]{pietr13}
Pietrzy\'nski, G., Graczyk, D., Gieren, W., et al. 2013, Nature, 495, 76

\bibitem[Platais et al.(2015)]{platais15}
Platais, I., van der Marel, R.P, Lennon, D.J., et al. 2015, ApJ, (accepted), astro.ph/1507.06653

\bibitem[Robberto et al.(2004)]{robberto04}
Robberto, M., Song, J., Mora Carrillo, G. et al. 2004, ApJ, 606, 952

\bibitem[Rubio et al.(1992)]{rubio92}
Rubio, M., Roth, M., Garcia, J. 1992, A\&A, 261, L29

\bibitem[Rubio et al.(1999)]{rubio99}
Rubio, M., Barb\'{a}, R.H., Walborn, N.R. 1999, A\&A, 347, 532

\bibitem[Sabbi et al.(2013)]{sabbi13}
Sabbi, E., Anderson, J., Lennon, D.J. et al. 2013, AJ, 146, 53

\bibitem[Sabbi et al.(2012)]{sabbi12}
Sabbi, E., Lennon, D. J., Gieles, M., et al. 2012, ApJL, 754, L37

\bibitem[Sana et al.(2013)]{sana13}
Sana, H., de Koter, A., de Mink, S.E., et al. 2013, A\&A, 550, 107

\bibitem[Sana et al.(2012)]{sana12}
Sana, H., de Mink,S.E., de Koter, A. et al. 2012, Sci, 337, 444 

\bibitem[Sandage(1953)]{sandage53}
Sandage, A.R. 1953, AJ, 58, 61 

\bibitem[Seale et al.(2014)]{seale14}
Seale, J.P., Meixner, M., Sewilo, M. et al. 2014, AJ, 148, 124 

\bibitem[Schaerer et al.(1993)]{schaerer93}
Schaerer, D., Meynet, G., Maeder, A., Schaller, G. 1993, A\&AS, 98, 523

\bibitem[Schild \& Testor(1992)]{schild92}
Schild, H., \& Testor, G. 1992, A\&AS, 92, 729

\bibitem[Schneider et al.(2014)]{schneider14}
Schneider, F.R.N., Izzard, R.G., de Mink, S.E., et al. 2014, ApJ, 780, 117

\bibitem[Selman \& Melnick(2013)]{selman13}
Selman, F.J., Melnick, J. 2013, A\&A, 552, A94

\bibitem[Selman et al.(1999)]{selman99}
Selman, F.J., Melnick, J., Bosch, G., \& Terlevich, R. 1999, A\&A, 347, 532

\bibitem[Shapley et al.(2003)]{shapley03}
Shapley, A.E., Steidel, C.C., Pettini, M., Adelberg, K.L. 2003, ApJ, 588, 65

\bibitem[Schneider et al.(2012)]{schneider12}
Schneider, N., Csengeri, T., Hennemann, M. et al. 2012, A\&A, 540, L11

\bibitem[Skrutskie et al.(2006)]{skrutskie06}
Skrutskie, M. F., Cutri, R. M., Stiening, R., et al. 2006, AJ, 131, 1163

\bibitem[Stanek et al.(1998)]{stanek98}
Stanek K. Z., Zaritsky D., Harris J., 1998, ApJ, 500, L141

\bibitem[Tang et al.(2014)]{tang14}
Tang, J., Bressan, A., Rosenfield, P., et al. 2014, MNRAS, 445, 4287

\bibitem[Townsley et al.(2006)]{townsley06}
Townsley, L.K., Broos, P.S., Feigelson, E.D., Garmire, G.P., Getman, Konstantin V. 2006, AJ, 131, 2140

\bibitem[Udalski et al.(1999a)]{udalski99a}
Udalski, A., Soszyn\`ski, I., Szymanski, M., et al. 1999a, Acta Astron., 49, 223

\bibitem[Udalski et al.(1999b)]{udalski99b}
Udalski, A., Soszyn\`ski, I., Szymanski, M., et al. 1999b, Acta Astron., 49, 437

\bibitem[Walborn et al.(2013)]{walborn12}
Walborn. N.R., Barb\`a, R.H., Sewilo, M.M. 2013, AJ, 145, 98 

\bibitem[Walborn et al.(2002)]{walborn02}
 Walborn, N. R., Ma\'iz Apell\'aniz, J., \& Barb\'a, R. H. 2002, AJ, 124, 1601 

\bibitem[Walborn et al.(1999)]{walborn99}
Walborn, N. R., Barb\'a, R. H., Brandner, W., et al. 1999, AJ, 117, 225

\bibitem[Walborn et al.(1995)]{walborn95}
Walborn, N.R., MacKenty, J.W., Saha, A., White, R.L., Parkerm J.W. 1995, ApJ, 439, L47 

\bibitem[Walborn \& Blades(1997)]{walborn97}
Walborn, N.R., \& Bladesm J.C. 1997, ApJS, 112, 457

\bibitem[Wang \& Helfand(1991)]{wang91}
Wang, Q., Helfand, D. J. 1991, ApJ, 370, 541

\bibitem[Zaritsky et al.(2004)]{zaritsky04}
Zaritsky, D., Harris, J., Tomphson, I.B., Grebel, E.K. 2004, AJ, 128, 1606
\end{thebibliography}
\end{document}